\documentclass[twoside,11pt]{article}

\usepackage{amsmath,amsfonts}
\usepackage{amssymb,latexsym}
\usepackage{amsxtra}
\usepackage{upref}
\usepackage{exscale}
\usepackage[mathscr]{eucal}

\newcommand{\be}{\begin{equation}}
\newcommand{\ee}{\end{equation}}
\newcommand{\beq}{\begin{eqnarray}}
\newcommand{\eeq}{\end{eqnarray}}
\newcommand{\no}{\nonumber}
\newcommand{\bea}{\begin{array}}
\newcommand{\eea}{\end{array}}
\newcommand{\lb}{\label}

\newcommand{\mscr}{\mathscr}
\newcommand{\mfrak}{\mathfrak}

\newcommand{\ts}{\textstyle}
\newcommand{\pp}{\partial}
\newcommand{\im}{\imath}
\newcommand{\ppr}{^{\boldsymbol{\prime}}}
\newcommand{\bprime}{\boldsymbol{\prime}}

\newcommand{\pdag}{^{\dagger}}
\newcommand{\wt}{\widetilde}

\newcommand{\ph}{\phantom}
\newcommand{\scr}{\scriptstyle}
\newcommand{\scrscr}{\scriptscriptstyle}
\newcommand{\scz}{\scriptsize}

\numberwithin{equation}{section}
\allowdisplaybreaks[4]

\begin{document}

\begin{center}
{\large\bf Extension of the integrable, (1+1) Gross-Pitaevskii equation} \vspace*{0.1cm}\\
{\large\bf to chaotic behaviour and arbitrary dimensions}  \vspace*{0.3cm}\\
{\bf Bernhard Mieck}\footnote{e-mail: "bjmeppstein@arcor.de";\newline freelance activity during 2007-2009;
current location : Zum Kohlwaldfeld 16, D-65817 Eppstein, Germany.}
\end{center}

\begin{abstract}
The integrable, (1+1) Gross-Pitaevskii (GP-) equation with hermitian property is extended to
chaotic behaviour as part of general complex fields within the $\mbox{sl}(2,\mathbb{C})$ algebra
for Lax pairs. Furthermore, we prove the involution property of conserved quantities in the case of
GP-type equations with an arbitrary external potential. We solve the corresponding
zero-curvature condition, following from compatibility of mixed partial derivatives with
Lax pair matrices, in terms of the '$\mfrak{ad}$'-operators which are derived from a Cartan-Weyl basis
of the general $\mbox{sl}(n,\mathbb{C})$ algebra or one of its sub-algebras. A gauge invariance of the
Lax pairs and its zero-curvature relation is proven so that one can reduce the total Cartan-Weyl basis
for the spatial Lax matrix to the maximal commuting Cartan sub-algebra. This allows to attest the
involution property of conserved quantities under very general conditions apart from the well-known,
classical '$\mfrak{r}$'-matrix approach. We generalize the approach of Lax pair matrices to arbitrary
spacetime dimensions and conclude for the type of nonlinear equations from the structure constants
of the underlying algebra. One can also calculate conserved quantities from loops within the (N-1)
dimensional base space and the mapping to the manifold of the general $\mbox{SL}(n,\mathbb{C})$
group or a sub-group, provided that the resulting fibre space is of nontrivial homotopic kind.
This condition is of crucial importance for proving the corresponding involution property and
avoids possible contraction of the considered loops to trivial point mappings.\newline

\vspace*{0.1cm}

\end{abstract}

\newpage

\tableofcontents

\newpage

\section{Introduction}\lb{s1}

In the study of nonlinear field equations there frequently occur various Lax pairs which allow to construct conserved
quantities as part of an integrable system. Although it is generally difficult to determine for a given nonlinear
field equation a corresponding Lax pair, the zero-curvature condition enables deep insight into integrability and
soliton solutions by the inverse scattering and B\"acklund transformations for several, physically relevant
equations, as e.g. the Gross-Pitaevskii (GP)-equation, the Korteweg-de Vries (KdV)-type
or the Sine-Gordon-like equations \cite{Abl1}-\cite{Gram2}.
However, this paper is initiated by a different notion of the integrable (1+1) GP-equation
which can be summarized by the questions \cite{Grah1,Grah2,Grah3} :
\begin{itemize}
\item Is really any kind of the (1+1) GP-equation (\ref{s1_1}), e.g. with an external potential $V(x,t)$,
integrable in the sense of an infinite number of independent, conserved quantities ? (cf. the definition of
Liouville integrability in Ref. \cite{Manton})
\item Which conclusions can be achieved from the corresponding Lax pair with respect to the involution of the
conserved quantities and the classical $\mfrak{r}$-matrix ?
\end{itemize}
\be\lb{s1_1}
\im\:\big(\pp_{t}\psi(x,t)\,\big)=-\big(\pp_{x}\pp_{x}\psi(x,t)\,\big)-2\:V(x,t)\:\psi(x,t)
-2\:\big|\psi(x,t)\big|^{2}\:\psi(x,t) \;.
\ee
In particular we examine the (1+1) nonlinear GP-equation which has been intensively computed with an external
potential, a so-called kick- or delta-function potential, for chaotic behaviour on a periodic circle
\(\psi(x,t)=\psi(x+2\pi,t)\) \cite{Raizen1,Y2K1}
\be\lb{s1_2}
\im\:\big(\pp_{t}\psi(x,t)\,\big)=-\big(\pp_{x}\pp_{x}\psi(x,t)\,\big)+
K\;\cos(x)\sum_{n=-\infty}^{+\infty}\delta(t-n)\;\psi(x,t)+g\:\big|\psi(x,t)\big|^{2}\:\psi(x,t)\;.
\ee
In advance, we mention that the obvious, hermitian discretization (\ref{s1_3}) of the continuous (1+1)
GP-equations (\ref{s1_1},\ref{s1_2}) fails to give a corresponding Lax pair whereas the peculiar, discrete,
non-hermitian form (\ref{s1_4}) can be obtained from a Lax pair \cite{Abl1,Gram2,Suris1}
\beq \lb{s1_3}
\im\:\big(\pp_{t}\psi(x_{n},t)\,\big) &=&-\;\frac{1}{(\Delta x)^{2}}
\Big(\psi(x_{n+1},t)-2\:\psi(x_{n},t)+\psi(x_{n-1},t)\Big) + \\ \no &-&2\:V(x_{n},t)\:\psi(x_{n},t)-
2\:\big|\psi(x_{n},t)\big|^{2}\:\psi(x_{n},t)   \;;  \\ \lb{s1_4}
\im\:\big(\pp_{t}\psi(x_{n},t)\,\big) &=&-\;\frac{1}{(\Delta x)^{2}}
\Big(\psi(x_{n+1},t)-2\:\psi(x_{n},t)+\psi(x_{n-1},t)\Big) +  \\ \no &-&
\big(V(x_{n+1},t)+V(x_{n},t)\,\big)\:\psi(x_{n},t)-
\big|\psi(x_{n},t)\big|^{2}\:\big(\psi(x_{n+1},t)+\psi(x_{n-1},t)\,\big)   \;.
\eeq
This suggests the preference of the discrete, non-hermitian kind (\ref{s1_4}) instead of the discrete, hermitian
form (\ref{s1_3}) in order to prove an infinite number of conserved quantities from Lax pairs.

According to a mathematical view, it is necessary to apply the discrete, non-hermitian form (\ref{s1_4}) of the
GP-equation in order to preserve exactly the Liouville integrability from the Lax pairs of the continuous
kind (\ref{s1_1},\ref{s1_2}) of the GP-equation. Apart from the involution of the conserved quantities which is
proven for more general cases in later sections, this can explain the numerically observed chaos of the kicked
GP-equation which may be caused by a false representation of discrete grid points for the numerical integration
despite of its hermitian form. Therefore, one should use the discrete kind (\ref{s1_4}) with non-hermitian property in order
to keep the Liouville integrability from the Lax pairs for the
continuous, kicked GP-equation (\ref{s1_1},\ref{s1_2}) without any chaos.

From a physical point of view, one has to remark that any physical problem has a time-, length-scale or
energy-, momentum-scale so that the application of delta-function-like time kicks within the (1+1) GP-equation
has to be regarded as problematic because these time kicks involve any energies (from highly excited Rydberg-atoms to
cosmic dimensions) \cite{Dancer1}. Furthermore, it is not astonishing that different, discrete realizations of continuous
nonlinear equations can cause different integrable or chaotic behaviour because of varying numerical properties;
however, in order to maintain the Liouville integrability from Lax pairs of a equation as the kicked, continuous
GP-equation (\ref{s1_2}), one has simply to introduce a {\it finite} time-,length-scale or broadened delta-function of time
for the external kick-potential and has even more to include "{\it many more}" (sufficient), discrete
spacetime steps in the numerical integration so that one can tract the time development for the particular scale
of the external potential without artifacts coming from below the physical scale.

In this paper we take the latter point or physical view and furthermore seek for a chaotic behaviour of the (1+1)
GP-equations. Chaotic behaviour seems to be completely excluded
because one can easily conclude for non-chaotic properties from the hermiticity and
finite norm in any case of a (physical) external potential on a spatial, periodically confined circle.
Therefore, any kind of GP-equation only seems
to allow for integrable properties. However, we can also extend to a possible chaotic behaviour , as one considers the
\(\mbox{su}(2)\)- or \(\mbox{sp}(2,\mathbb{R})\)-Lax pairs (for an attractive or repulsive interaction) as
sub-algebras of the general \(\mbox{sl}(2,\mathbb{C})\) algebra and Lax pairs. We specify the derivation of a Lax
pair for an arbitrary external potential $V(x,t)$ of the general \(\mbox{sl}(2,\mathbb{C})\) case so that this Lax
pair can be used for the construction and involution of the conserved quantities following from the monodromy matrix.

In section \ref{s3} we further analyze the point that the construction of a Lax pair, as e.g. for the GP-equation, appears
to be accidental as one chooses two matrix potentials \(\mfrak{L}(x,t)\), \(\mfrak{M}(x,t)\) for a
zero-curvature condition which finally determines the nonlinear equations. In appendix \ref{sa} we therefore
demonstrate the reduction of the very general \(n\times n\), \(\mbox{gl}(n,\mathbb{C})\) Lax pair algebra
to its nontrivial, traceless parts \(\mbox{sl}(n,\mathbb{C})\); furthermore, we prove a gauge invariance
of the Lax pairs and their zero-curvature condition in section \ref{s32}, provided that a Maurer-Cartan equation holds for
the gauge matrices. The gauge invariance
of the zero-curvature condition underlines that a Lax pair should not be assigned to a single, definite
nonlinear equation (as e.g. the characteristic GP-case), but to a whole set of equivalent nonlinear equations.
We can solve the zero-curvature condition for general \(\mbox{sl}(n,\mathbb{C})\) Lax pairs (or reduced to one
of its sub-algebras as \(\mbox{su}(n)\), etc.) in terms of the '$\mfrak{ad}$'- operator
\be\lb{s1_5}
\overrightarrow{\boldsymbol{[}k_{i}\:\hat{H}^{i}+\mscr{P}_{\alpha}(x,t)\:\hat{E}^{\alpha}_{+}+
\mscr{Q}_{\alpha}(x,t)\:\hat{E}_{-}^{\alpha}\:\boldsymbol{,}\:
\ldots\boldsymbol{]_{-}}}\ldots \;,
\ee
for the Cartan-Weyl basis of the prevailing generators of a chosen, closed algebra \cite{Schweig1}; this generally attests
that the spatial- and time-development of (seemingly accidental) nonlinear equations as part of a whole set
of equivalent equations is confined to the development within group manifolds, either by the general
\(\mbox{sl}(n,\mathbb{C})\) Lax pair algebra or one of its possible sub-algebras as the \(\mbox{su}(n)\)-case.
According to this property, it is more appropriate to classify the whole set of equivalent nonlinear equations
as the GP-like case by the underlying, chosen sub-algebras \(\subseteq\mbox{sl}(n,\mathbb{C})\) for the
Lax pairs.

Since the spatial and time development of nonlinear equations from Lax pairs is limited to that within
group manifolds, it is not astonishing that one can confirm involution properties of the conserved quantities from the
monodromy matrix under rather general assumptions by performing the possible gauge transformations
to a diagonal, spatial Lax matrix \(\mfrak{L}(x,t)\rightarrow\mfrak{L}_{{\scrscr\hat{H}}}(x,t)\). This gauge transformation
allows to detect conditions for the general independence of conserved quantities derived from the Lax pair
and its zero-curvature relation apart from the introduction of classical $\mfrak{r}$-matrices. Nevertheless, we also
investigate the calculation of the classical $\mfrak{r}$-matrix in the case of spatially-ultralocal Lax matrices
\(\mfrak{L}(x,t)\) from the tensor product within canonical Poisson brackets (cf. section \ref{s42} and appendix \ref{sb})
\be\lb{s1_6}
\boldsymbol{\big\{}\mfrak{L}(x_{1},t)\otimes\hat{1}\stackrel{\otimes}{\boldsymbol{,}}\hat{1}\otimes
\mfrak{L}(x_{2},t)\boldsymbol{\big\}}\;\;,
\ee
which can be separated into an eigenvalue part and a commutator of the Lax matrices
\(\mfrak{L}_{1}(x_{1},t)=\mfrak{L}(x_{1},t)\otimes\hat{1}\),
\(\mfrak{L}_{2}(x_{2},t)=\hat{1}\otimes\mfrak{L}(x_{2},t)\)  with the \(\mfrak{r}_{12}\)-,\(\mfrak{r}_{21}\)-
matrices (concerning the general definitions of the tensor product space and the tensor product within Poisson brackets, etc. ,
see Refs. \cite{Chowd1,Manton}, appendix A and chap. 2.5, respectively).

In section \ref{s5} we extend the zero-curvature relation of Lax pairs beyond (1+1) dimensions and determine
the nonlinear field equations which are also confined to the spacetime development within group manifolds
by the '$\mfrak{ad}$'- operators of the \(\mbox{sl}(n,\mathbb{C})\) algebra or one of its sub-algebras.
The calculation of conserved quantities and their involution as in the (1+1) dimensional case can be
straightforwardly generalized for the corresponding Lax pairs within arbitrary spacetime dimensions; however,
it has to be taken into account the following additional point : \newline
As we specify closed loops within group manifolds for determining independent, conserved
quantities, it has to be ascertained that these loops, contained within the mapping of n-1 spatial coordinates
to the fields within the considered group manifold, possess a nontrivial homotopy in order to prevent
possible contractions to trivial point mappings.

\section{Lax pairs for GP-type equations from the $\boldsymbol{\mbox{sl}(2,\mathbb{C})}$ algebra} \lb{s2}

\subsection{Lie algebras of $\mbox{su}(2)$-, $\mbox{sp}(2,\mathbb{R})$- and
$\mbox{sl}(2,\mathbb{C})$-Lax pairs for integrable and chaotic behaviour}\lb{s21}

In this section we follow Refs. \cite{Abl1,Abl2}, but additionally emphasize the algebraic \(\mbox{sl}(2,\mathbb{C})\) properties
of the Lax pairs for the GP-equation. One starts out from two traceless \(2\times 2\) matrices
\(\mfrak{L}_{\mbox{\scz sl}(2,\mathbb{C})}(\mscr{Q},\mscr{P};k)\), \(\mfrak{M}_{\mbox{\scz sl}(2,\mathbb{C})}(\mscr{Q},\mscr{P};k)\) (\ref{s2_4},\ref{s2_5}) with complex-valued
entries \(\mscr{Q}(x,t)\), \(\mscr{P}(x,t)\) for the physical wavefunctions and a complex-valued spectral parameter \(k\) (\ref{s2_3})
which is included to obtain general \(\mbox{sl}(2,\mathbb{C})\) matrices. In the context of eqs. (\ref{s2_1},\ref{s2_2}), the
Lax matrices with the physical fields \(\mscr{Q}(x,t)\), \(\mscr{P}(x,t)\) appear to be accidental, but separately determine the spatial
and time evolution of a two component, complex-valued, (non-physical), auxiliary vector field
\(\Xi(x,t)=(\Xi_{1}(x,t)\:,\:\Xi_{2}(x,t)\,)^{T}\)
\beq \lb{s2_1}
\Xi_{x} &=& \bigg(\bea{cc} -\im\,k & \mscr{P} \\ \mscr{Q} & \im\,k \eea\bigg)\;\Xi(x,t) = \mfrak{L}_{\mbox{\scz sl}(2,\mathbb{C})}(\mscr{Q},\mscr{P};k)\;\Xi(x,t)\;; \\ \lb{s2_2}
\Xi_{t} &=& \bigg(\bea{cc} 2\im\,k^{2}+\im\:\mscr{P}\;\mscr{Q} & -2\,k\:\mscr{P}-\im\:\mscr{P}_{x} \\
-2\,k\:\mscr{Q}+\im\:\mscr{Q}_{x} & -2\im\,k^{2}-\im\:\mscr{P}\;\mscr{Q} \eea\bigg)\;\Xi(x,t)=\mfrak{M}_{\mbox{\scz sl}(2,\mathbb{C})}(\mscr{Q},\mscr{P};k)\;\Xi(x,t) \;; \\  \lb{s2_3}
\Xi &=& \Xi(x,t)\in\mathbb{C}\;;\;\;\;\mscr{P}=\mscr{P}(x,t)\in\mathbb{C}\;;\;\;\;\mscr{Q}=\mscr{Q}(x,t)\in\mathbb{C}\;;  \;\;\;k\in\mathbb{C}\;;\\  \lb{s2_4}
\mfrak{L}_{\mbox{\scz sl}(2,\mathbb{C})}(\mscr{Q},\mscr{P};k) &=& \!\!\!-(\im\,k)\; \big(\hat{\tau}_{3}\big) + \mscr{P}(x,t)\;\big(\hat{\tau}_{+}\big)+ \mscr{Q}(x,t)\;
\big(\hat{\tau}_{-}\big)
\;\Longrightarrow\in\mbox{sl}(2,\mathbb{C}) \;; \\  \lb{s2_5}
\mfrak{M}_{\mbox{\scz sl}(2,\mathbb{C})}(\mscr{Q},\mscr{P};k) &=&\!\!\! (2\im\,k^{2}+\im\:\mscr{P}\;\mscr{Q})\; \big(\hat{\tau}_{3}\big) \!+\!
(-2\,k\:\mscr{P}-\im\:\mscr{P}_{x})\;\big(\hat{\tau}_{+}\big)\!+\! (-2\,k\:\mscr{Q}+\im\:\mscr{Q}_{x})\;\big(\hat{\tau}_{-}\big)
\!\Longrightarrow\in\!\mbox{sl}(2,\!\mathbb{C}\!) ;  \\ \no &;&\mbox{ladder operators }\;\big(\hat{\tau}_{+}\big),\;\big(\hat{\tau}_{-}\big)
\mbox{ of Pauli matrices }\;\big(\hat{\tau}_{1}\big),\;\big(\hat{\tau}_{2}\big),\;\big(\hat{\tau}_{3}\big)\;\;.
\eeq
It is the separate evolution of space and time (\ref{s2_1},\ref{s2_2}) for the auxiliary field \(\Xi(x,t)\) which enforces
the pairwise equivalence of the mixed and exchanged spacetime derivatives (\ref{s2_6},\ref{s2_7})
for sufficiently continuous, (non-multivalued!) functions. After substitution of \((\pp_{x}\Xi)\) and   \((\pp_{t}\Xi)\)
by (\ref{s2_1},\ref{s2_2}), one transfers the equivalence of mixed and exchanged derivatives of the auxiliary field \(\Xi(x,t)\)
to the Lax matrices \(\mfrak{L}_{\mbox{\scz sl}(2,\mathbb{C})}(\mscr{Q},\mscr{P};k)\),
\(\mfrak{M}_{\mbox{\scz sl}(2,\mathbb{C})}(\mscr{Q},\mscr{P};k)\) and attains the so-called
zero-curvature relation (\ref{s2_8}) which restricts the physical fields \(\mscr{Q}(x,t)\), \(\mscr{P}(x,t)\) to a nonlinear field equation
\beq\no
&&\mbox{\bf zero-curvature condition}  \\ \lb{s2_6}
\big(\pp_{t}\pp_{x}\Xi(x,t)\big) &=& \big(\pp_{t}\mfrak{L}_{\mbox{\scz sl}(2,\mathbb{C})}(\mscr{Q},\mscr{P};k)\big)\;\Xi(x,t) +
\mfrak{L}_{\mbox{\scz sl}(2,\mathbb{C})}(\mscr{Q},\mscr{P};k)\;\big(\pp_{t}\Xi(x,t)\big) \\ \no &=&
\big(\pp_{t}\mfrak{L}_{\mbox{\scz sl}(2,\mathbb{C})}(\mscr{Q},\mscr{P};k)\big)\;\Xi(x,t) +
\mfrak{L}_{\mbox{\scz sl}(2,\mathbb{C})}(\mscr{Q},\mscr{P};k)\;\mfrak{M}_{\mbox{\scz sl}(2,\mathbb{C})}(\mscr{Q},\mscr{P};k)\;\Xi(x,t) \;;\\ \lb{s2_7}
\big(\pp_{x}\pp_{t}\Xi(x,t)\big) &=& \big(\pp_{x}\mfrak{M}_{\mbox{\scz sl}(2,\mathbb{C})}(\mscr{Q},\mscr{P};k)\big)\;\Xi(x,t) +
\mfrak{M}_{\mbox{\scz sl}(2,\mathbb{C})}(\mscr{Q},\mscr{P};k)\;\big(\pp_{x}\Xi(x,t)\big) \\ \no &=&
\big(\pp_{x}\mfrak{M}_{\mbox{\scz sl}(2,\mathbb{C})}(\mscr{Q},\mscr{P};k)\big)\;\Xi(x,t) +
\mfrak{M}_{\mbox{\scz sl}(2,\mathbb{C})}(\mscr{Q},\mscr{P};k)\;\mfrak{L}_{\mbox{\scz sl}(2,\mathbb{C})}(\mscr{Q},\mscr{P};k)\;\Xi(x,t) \;;\\  \lb{s2_8}
\Longrightarrow \; 0&=& \big(\pp_{t}\mfrak{L}_{\mbox{\scz sl}(2,\mathbb{C})}(\mscr{Q},\mscr{P};k)\big)-\big(\pp_{x}
\mfrak{M}_{\mbox{\scz sl}(2,\mathbb{C})}(\mscr{Q},\mscr{P};k)\big)+
\boldsymbol{\big[}\mfrak{L}_{\mbox{\scz sl}(2,\mathbb{C})}(\mscr{Q},\mscr{P};k)\;\boldsymbol{,}\;
\mfrak{M}_{\mbox{\scz sl}(2,\mathbb{C})}(\mscr{Q},\mscr{P};k)\,\boldsymbol{\big]_{-}}.
\eeq
As we calculate every matrix element of the \((2\times 2)\) zero-curvature condition (\ref{s2_9}),
we (seemingly accidental) achieve
two coupled, nonlinear equations (\ref{s2_10},\ref{s2_11}) of \(\mscr{Q}(x,t)\), \(\mscr{P}(x,t)\) from the
off-diagonal matrix entries of (\ref{s2_8}) whereas
the diagonal part completely adds to zero without a physical meaning as a wave equation. Since the trace of a commutator vanishes
for finite dimensional matrices, the zero-curvature condition with the commutator between
\(\mfrak{L}_{sl(2,\mathbb{c})}(\mscr{Q},\mscr{P};k)\) and \(\mfrak{M}_{sl(2,\mathbb{c})}(\mscr{Q},\mscr{P};k)\) always remains within the complex-valued,
traceless Lie algebra generators \(\mbox{sl}(2,\mathbb{C})\). (In appendix \ref{sa} we demonstrate how to separate a trivial,
diagonal, complex-valued unit part from the most general \(\mbox{gl}(n,\mathbb{C})\) Lax pair ansatz so that one has only to
consider the traceless, \(\mbox{sl}(n,\mathbb{C})\) parts of Lax pairs and their zero-curvature relation for physical field
equations which are indeed nonlinear and nontrivial.) It is possible to derive a continuity equation (\ref{s2_12}) of the
complex fields \(\mscr{Q}(x,t)\), \(\mscr{P}(x,t)\) from the two field equations (\ref{s2_10},\ref{s2_11}) which allows to conclude for
a time-like conserved quantity with complex constant \(c_{0}\) as we require spatially periodic boundary conditions
(\ref{s2_10},\ref{s2_11}) of the physical fields
\beq\lb{s2_9}
\lefteqn{\hspace*{-6.3cm}\overbrace{\bigg(\bea{cc} 0 & \mscr{P}_{t} \\ \mscr{Q}_{t} & 0 \eea\bigg)}^{(\pp_{t}\mfrak{L})}-
\overbrace{\bigg(\bea{cc} \im\,(\mscr{P}\,\mscr{Q})_{x} & -2k\:\mscr{P}_{x}-\im\:\mscr{P}_{xx} \\ -2k\:\mscr{Q}_{x}+\im\:\mscr{Q}_{xx}
& -\im\,(\mscr{P}\,\mscr{Q})_{x} \eea\bigg)}^{(\pp_{x}\mfrak{M})} +
\overbrace{\bigg(\bea{cc} \im\,(\mscr{P}\,\mscr{Q})_{x} & -2k\:\mscr{P}_{x}-2\im\:\mscr{P}^{2}\:\mscr{Q} \\
-2k\:\mscr{Q}_{x}+2\im\:\mscr{P}\:\mscr{Q}^{2} & -\im\,(\mscr{P}\,\mscr{Q})_{x} \eea\bigg)}^{[\mfrak{L}\,,\,\mfrak{M}]_{-}} =}  \\  \no &=&
\bigg( \bea{cc} 0 & \mscr{P}_{t}+\im\,\mscr{P}_{xx}-2\im\:\mscr{P}^{2}\:\mscr{Q} \\
\mscr{Q}_{t}-\im\,\mscr{Q}_{xx}+2\im\:\mscr{P}\:\mscr{Q}^{2} & 0 \eea\bigg)\equiv 0 \;;  \\ \lb{s2_10}
\im\,\big(\pp_{t}\mscr{P}\big) &=& \big(\pp_{x}\pp_{x}\mscr{P}\big)-
2\:\mscr{Q}\:\mscr{P}^{2} \;;\;\;\;\mscr{P}(x=0,t)=\mscr{P}(x=2\pi,t)\;; \\ \lb{s2_11}
-\im\,\big(\pp_{t}\mscr{Q}\big) &=& \big(\pp_{x}\pp_{x}\mscr{Q}\big)-
2\:\mscr{P}\:\mscr{Q}^{2} \;;\;\;\; \mscr{Q}(x=0,t)=\mscr{Q}(x=2\pi,t)\;;  \\ \lb{s2_12}
\Longrightarrow \;\;\pp_{t}\big(\mscr{P}\:\mscr{Q}\big) &=&-\im\:\pp_{x}\big(\mscr{Q}\:(\pp_{x}\mscr{P})-
\mscr{P}\:(\pp_{x}\mscr{Q})\,\big) \;; \\  \lb{s2_13}
\Longrightarrow \;\;\pp_{t}\int_{0}^{2\pi}dx\;\big(\mscr{P}(x,t)\;\mscr{Q}(x,t)\,\big) &=&\hspace*{-0.2cm} -\im\hspace*{-0.1cm}
\int_{0}^{2\pi} \hspace*{-0.2cm} dx\;\pp_{x}\big(\mscr{Q}(x,t)\:(\pp_{x}\mscr{P}(x,t)\,)-\mscr{P}(x,t)\:(\pp_{x}\mscr{Q}(x,t)\,)\,\big)=0 \;; \\ \lb{s2_14}
\Longrightarrow\;\; \int_{0}^{2\pi}dx\;\big(\mscr{P}(x,t)\;\mscr{Q}(x,t)\,\big) &=&\mbox{const.}=c_{0}\in\mathbb{C}\;.
\eeq
The conserved, complex-valued quantity (\ref{s2_14}) refers to a generalized norm of \(\mscr{Q}(x,t)\), \(\mscr{P}(x,t)\), compared
to the standard GP-equations, but lacks of its positive definiteness. It is the non-positive property of the generalized norm
(\ref{s2_14}) which will allow us to accomplish a possible chaotic behaviour within \(\mbox{sl}(2,\mathbb{C})\) Lax pairs
of physical fields. The restriction of the \(\mbox{sl}(2,\mathbb{C})\) generators \(\mfrak{L}_{\mbox{\scz sl}(2,\mathbb{C})}\),
\(\mfrak{M}_{\mbox{\scz sl}(2,\mathbb{C})}\) to the sub-algebras \(\mbox{su}(2)\) and \(\mbox{sp}(2,\mathbb{R})\) leads to the
well-known, (1+1) GP-equations (\ref{s2_15},\ref{s2_16}) for the attractive and repulsive interaction case. As we set in (\ref{s2_15})
\((\mscr{Q}(x,t)=-\mscr{P}^{*}(x,t)=\psi(x,t)\,)\) to the physical wavefunction with periodic boundary conditions and with a {\it real} spectral
parameter $k$, we acquire anti-hermitian Lax pairs \(\mfrak{L}_{\mbox{\scz su}(2)}(\mscr{Q},\mscr{P};k)\), \(\mfrak{M}_{\mbox{\scz su}(2)}(\mscr{Q},\mscr{P};k)\)
whose zero-curvature condition determines the attractive case of interaction of the (1+1) GP-equation
\beq \lb{s2_15}
\mbox{I)} &:& \mbox{attractive case},\;\;\mbox{su}(2)\;\mbox{sub-algebra of }\mbox{sl}(2,\mathbb{C}) \\ \no
\mscr{Q}(x,t) &=&-\mscr{P}^{*}(x,t) =\psi(x,t)\;\in\mathbb{C}\;;\;\;\;k\in\mathbb{R} \;; \\ \no
\im\:\big(\pp_{t}\psi) &=&-\big(\pp_{x}\pp_{x}\psi)-2\:|\psi|^{2}\:\psi \;;\;\;\;\psi(x=0,t)=\psi(x=2\pi,t)\;; \\ \no
\mfrak{L}_{\mbox{\scz su}(2)}(\mscr{Q},\mscr{P};k) &=& \bigg(\bea{cc} -\im\,k & -\psi^{*} \\ \psi & \im\,k \eea\bigg)\;;\;\;\;
\mfrak{M}_{\mbox{\scz su}(2)}(\mscr{Q},\mscr{P};k) =
\bigg(\bea{cc} 2\im\,k^{2}-\im\:|\psi|^{2} & +2\,k\:\psi^{*}+\im\:\psi_{x}^{*} \\
-2\,k\:\psi+\im\:\psi_{x} & -2\im\,k^{2}+\im\:|\psi|^{2} \eea\bigg)\;; \\ \no
\mfrak{L}_{\mbox{\scz su}(2)}(\mscr{Q},\mscr{P};k) &=&-\big(\mfrak{L}_{\mbox{\scz su}(2)}(\mscr{Q},\mscr{P};k)\,\big)\pdag\;;\;\;\;
\mfrak{M}_{\mbox{\scz su}(2)}(\mscr{Q},\mscr{P};k) =-\big(\mfrak{M}_{\mbox{\scz su}(2)}(\mscr{Q},\mscr{P};k)\,\big)\pdag\;.
\eeq
Similarly, one can choose the combination \((\mscr{Q}(x,t)=+\mscr{P}^{*}(x,t)=\psi(x,t)\,)\) for the physical wavefunction with a {\it real}
spectral parameter $k$ in order to achieve Lax matrices within the symplectic algebra \(\mbox{sp}(2,\mathbb{R})\), having
the real parameters \(\Re(\psi(x,t)\,)\), \(\Im(\psi(x,t)\,)\), \(k\in\mathbb{R}\); the zero-curvature relation then results
in the GP-equation with a repulsive interaction
\beq\lb{s2_16}
\mbox{II)}&:& \mbox{repulsive case},\;\;\mbox{sp}(2,\mathbb{R})\;\mbox{sub-algebra of }\mbox{sl}(2,\mathbb{C}) \\ \no
\mscr{Q}(x,t) &=&+\mscr{P}^{*}(x,t) =\psi(x,t)\;\in\mathbb{C}\;;\;\;\; (\im\:k\rightarrow) k\in\mathbb{R} \;; \\ \no
\im\:\big(\pp_{t}\psi) &=&-\big(\pp_{x}\pp_{x}\psi)+2\:|\psi|^{2}\:\psi \;;\;\;\;\psi(x=0,t)=\psi(x=2\pi,t)\;; \\ \no
\mfrak{L}_{\mbox{\scz sp}(2,\mathbb{R})}(\mscr{Q},\mscr{P};k) &=& \bigg(\bea{cc} -k & \psi^{*} \\ \psi & k \eea\bigg)\;;\;\;\;
\mfrak{M}_{\mbox{\scz sp}(2,\mathbb{R})}(\mscr{Q},\mscr{P};k) =
\bigg(\bea{cc} -2\im\,k^{2}+\im\:|\psi|^{2} & 2\,\im\,k\:\psi^{*}-\im\:\psi_{x}^{*} \\
2\,\im\,k\:\psi+\im\:\psi_{x} & 2\im\,k^{2}-\im\:|\psi|^{2} \eea\bigg)\;;  \\  \no
0 &\equiv&
\mfrak{L}_{\mbox{\scz sp}(2,\mathbb{R})}(\mscr{Q},\mscr{P};k) \;\bigg(\bea{cc} 0 & -1 \\ 1 & 0 \eea\bigg)+
\bigg(\bea{cc} 0 & -1 \\ 1 & 0 \eea\bigg)\;\big(\mfrak{L}_{\mbox{\scz sp}(2,\mathbb{R})}(\mscr{Q},\mscr{P};k)\,\big)^{T}\;;  \\ \no
0 &\equiv& \mfrak{M}_{\mbox{\scz sp}(2,\mathbb{R})}(\mscr{Q},\mscr{P};k) \;\bigg(\bea{cc} 0 & -1 \\ 1 & 0 \eea\bigg)+
\bigg(\bea{cc} 0 & -1 \\ 1 & 0 \eea\bigg)\;\big(\mfrak{M}_{\mbox{\scz sp}(2,\mathbb{R})}(\mscr{Q},\mscr{P};k)\,\big)^{T}\;.
\eeq
Both cases, the \(\mbox{su}(2)\)- and \(\mbox{sp}(2,\mathbb{R})\)-sub-algebras for the attractive and repulsive interaction, have the
identical continuity equation (\ref{s2_17}), due to the hermitian property of the corresponding Hamilton operator. The physical
norm of the periodic fields on a circle is always positive definite and is set to one
(or e.g. to a constant particle number $N_{0}$) for an interpretation of a true probability density. Apart from the
normalization of wavefunctions at arbitrary time, we can apply the positive norm of constant value to estimate the maximal
achievable value for the norm of the difference of two wavefunctions at different times. Since the absolute value of any
wavefunction is restricted by the maximal probability density one at arbitrary times, the maximal achievable difference
of the absolute values of two wavefunctions at two different time points is bounded by the value \(8\,\pi\), due to the
restricted spatial coordinate \(x\in[0,\,2\pi)\). Therefore, the hermitian property of Hamilton operators in the
attractive \(\mbox{su}(2)\)- and repulsive \(\mbox{sp}(2,\mathbb{R})\)-cases only allows to assign a non-chaotic
behaviour to the GP-equations (\ref{s2_15},\ref{s2_16}),
due to the absence of an exponential-(or other type-)divergence at two arbitrary times
\beq\lb{s2_17}
\Longrightarrow \;\;\pp_{t}\big(\psi^{*}\:\psi\big) &=&\im\:\pp_{x}\big(\psi^{*}\:(\pp_{x}\psi)-
\psi\:(\pp_{x}\psi^{*})\,\big) \;; \\  \lb{s2_18}
\Longrightarrow \;\;\pp_{t}\int_{0}^{2\pi}\!\!\!dx\;\big(\psi^{*}(x,t)\;\psi(x,t)\,\big) &=&\im
\int_{0}^{2\pi}\!\!\!dx\;
\pp_{x}\big(\psi^{*}(x,t)\:(\pp_{x}\psi(x,t)\,)-\psi(x,t)\:(\pp_{x}\psi^{*}(x,t)\,)\,\big)=0 ; \\  \lb{s2_19}
\Longrightarrow\;\; \int_{0}^{2\pi}dx\;\big(\psi^{*}(x,t)\;\psi(x,t)\,\big) &=&\mbox{const.}=r_{0}\in\mathbb{R}\;;\;\;\;
r_{0}:=1\;; \\  \lb{s2_20}
\lefteqn{\hspace*{-5.5cm}\Longrightarrow \;\;\int_{0}^{2\pi}dx\;\big(\psi^{*}(x,t)-\psi^{*}(x,t=0)\,\big)\;
\big(\psi(x,t)-\psi(x,t=0)\,\big) = } \\ \no
\lefteqn{\hspace*{-5.5cm}=\int_{0}^{2\pi}dx\;\Big(\big|\psi(x,t)\big|^{2}+\big|\psi(x,t=0)\big|^{2}
-2\:\cos\Big(\angle\big(\psi^{*}(x,t)\big|\psi(x,t=0)\big)\Big)\;\;\big|\psi(x,t)\big|\;\;\big|\psi(x,t=0)\big| \Big)\leq } \\ \no &\leq&
\int_{0}^{2\pi}dx\;\big(1+1-2\cdot(-1)\cdot 1 \cdot 1\big)=\int_{0}^{2\pi}dx\:4=8\:\pi\;.
\eeq
In section \ref{s4} we also prove the time-like conservation of traces from arbitrary powers of related
monodromy matrices (\ref{s2_21}-\ref{s2_23})
which even fulfill the involution property of the Poisson brackets. According to our separate consideration of the
three algebraic cases, one can derive conserved quantities from the \(\mfrak{T}_{\mbox{\scz sl}(2,\mathbb{C})}(x,t;k)\) matrices for the
rather general \(\mbox{sl}(2,\mathbb{C})\) case with independent fields \(\mscr{Q}(x,t)\), \(\mscr{P}(x,t)\) or for the two special
\(\mbox{su}(2)\)-, \(\mbox{sp}(2,\mathbb{R})\)-sub-algebras with the exponential matrices
\(\mfrak{T}_{\mbox{\scz su}(2)}(x,t;k)\), \(\mfrak{T}_{\mbox{\scz sp}(2,\mathbb{R})}(x,t;k)\) of the spatial Lax generators
\(\mfrak{L}_{\mbox{\scz su}(2)}(\xi,t;k)\), \(\mfrak{L}_{\mbox{\scz sp}(2,\mathbb{R})}(\xi,t;k)\).
The inequality (\ref{s2_20}), which restricts
the wavefunctions to non-chaotic behaviour, complies with the infinite number of independently conserved quantities derived from
the \(\mfrak{T}_{\mbox{\scz su}(2)}(x,t;k)\), \(\mfrak{T}_{\mbox{\scz sp}(2,\mathbb{R})}(x,t;k)\) matrices for the attractive and repulsive
case, respectively (cf. section \ref{s4} and appendix \ref{sb}). According to the inequality (\ref{s2_20}) and the independently conserved quantities from the traces of
powers of \(\mfrak{T}_{\mbox{\scz su}(2)}(x,t;k)\), \(\mfrak{T}_{\mbox{\scz sp}(2,\mathbb{R})}(x,t;k)\), it seems to be impossible to obtain
any chaotic behaviour from the \(\mbox{su}(2)\)-, \(\mbox{sp}(2,\mathbb{R})\)- Lax pairs and their zero-curvature conditions. Chaotic
behaviour from Lax pairs appears to be impossible in particular because the inequality and the independently conserved quantities
from \(\mfrak{T}_{\mbox{\scz su}(2)}(x,t;k)\), \(\mfrak{T}_{\mbox{\scz sp}(2,\mathbb{R})}(x,t;k)\) can be simply generalized to the case with
an arbitrary external potential in the GP-equations with attractive or repulsive interaction
\beq \lb{s2_21}
\mfrak{T}_{\mbox{\scz sl}(2,\mathbb{C})}(x,t;k) &=& \overleftarrow{\exp}\Big\{\int_{0}^{x}d\xi\;
\mfrak{L}_{\mbox{\scz sl}(2,\mathbb{C})}(\xi,t;k)\Big\}\Longrightarrow
C_{sl(2,\mathbb{C})}^{(n)}(t;k)=\mbox{Tr}\Big[\big(\mfrak{T}_{\mbox{\scz sl}(2,\mathbb{C})}(x=2\pi,t;k)\big)^{n}\Big]\;; \\ \no
1 &=&\mfrak{T}_{\mbox{\scz sl}(2,\mathbb{C})}^{-1}(x,t;k)\;\:\mfrak{T}_{\mbox{\scz sl}(2,\mathbb{C})}(x,t;k)\;;\;\;\;\;
\mfrak{T}_{\mbox{\scz sl}(2,\mathbb{C})}^{-1}(x,t;k)=\overrightarrow{\exp}\Big\{-\int_{0}^{x}d\xi\;
\mfrak{L}_{\mbox{\scz sl}(2,\mathbb{C})}(\xi,t;k)\Big\} \;; \\ \lb{s2_22}
\mfrak{T}_{\mbox{\scz su}(2)}(x,t;k) &=& \overleftarrow{\exp}\Big\{\int_{0}^{x}d\xi\;
\mfrak{L}_{\mbox{\scz su}(2)}(\xi,t;k)\Big\}\Longrightarrow
C_{su(2)}^{(n)}(t;k)=\mbox{Tr}\Big[\big(\mfrak{T}_{\mbox{\scz su}(2)}(x=2\pi,t;k)\big)^{n}\Big]\;; \\ \no
1 &=&\mfrak{T}_{\mbox{\scz su}(2)}\pdag(x,t;k)\;\:\mfrak{T}_{\mbox{\scz su}(2)}(x,t;k)\;;\;\;\;\;
\mfrak{T}_{\mbox{\scz su}(2)}\pdag(x,t;k)=\overrightarrow{\exp}\Big\{-\int_{0}^{x}d\xi\;
\mfrak{L}_{\mbox{\scz su}(2)}(\xi,t;k)\Big\} \;; \\    \lb{s2_23}
\mfrak{T}_{\mbox{\scz sp}(2,\mathbb{R})}(x,t;k) &=& \overleftarrow{\exp}\Big\{\int_{0}^{x}d\xi\;
\mfrak{L}_{\mbox{\scz sp}(2,\mathbb{R})}(\xi,t;k)\Big\}\Longrightarrow
C_{sp(2,\mathbb{R})}^{(n)}(t;k)=\mbox{Tr}\Big[\big(\mfrak{T}_{\mbox{\scz sp}(2,\mathbb{R})}(x=2\pi,t;k)\big)^{n}\Big]\;; \\ \no
\bigg(\bea{cc} 0 &-1 \\ 1 & 0\eea\bigg) &=&\mfrak{T}_{\mbox{\scz sp}(2,\mathbb{R})}^{T}(x,t;k)\;
\bigg(\bea{cc} 0 &-1 \\ 1 & 0\eea\bigg)\;\mfrak{T}_{\mbox{\scz sp}(2,\mathbb{R})}(x,t;k)\;; \\ \no
\mfrak{T}_{\mbox{\scz sp}(2,\mathbb{R})}^{T}(x,t;k)&=&\overrightarrow{\exp}\Big\{\int_{0}^{x}d\xi\;
\mfrak{L}_{\mbox{\scz sp}(2,\mathbb{R})}^{T}(\xi,t;k)\Big\} \;.
\eeq
Nevertheless, it is possible to compose the fields \(\mscr{Q}(x,t)\), \(\mscr{P}(x,t)\) of the rather general \(\mbox{sl}(2,\mathbb{C})\) Lax pair
algebra by the combination (\ref{s2_25},\ref{s2_26}) of the physical wavefunctions \(\psi_{\mbox{\scz su}(2)}(x,t)\),
\(\psi_{\mbox{\scz sp}(2,\mathbb{R})}(x,t)\) so that probability of \(\mbox{su}(2)\)- and \(\mbox{sp}(2,\mathbb{R})\)- sub-algebra
matrices with their corresponding wavefunctions as parameters is allowed to flow and change within the total \(\mbox{sl}(2,\mathbb{C})\)
Lax pair algebra; in consequence chaotic behaviour becomes possible despite of the formulation in terms of Lax pairs and
their zero-curvature relations
\beq \lb{s2_24}
c_{0}(\in\mathbb{C})&=&\int_{0}^{2\pi}dx\;\big(\mscr{P}(x,t)\;\mscr{Q}(x,t)\,\big) \;;  \\ \lb{s2_25}
\mscr{Q}(x,t) &=&\psi_{\mbox{\scz su}(2)}(x,t)+\psi_{\mbox{\scz sp}(2,\mathbb{R})}(x,t) \;; \\  \lb{s2_26}
\mscr{P}(x,t) &=&-\psi_{\mbox{\scz su}(2)}^{*}(x,t)+\psi_{\mbox{\scz sp}(2,\mathbb{R})}^{*}(x,t) \;.
\eeq
We illustrate this possibility from the {\it generalized} conserved norm (\ref{s2_24}) of fields \(\mscr{Q}(x,t)\), \(\mscr{P}(x,t)\) with the
complex constant $c_{0}$. One has to compose the fields \(\mscr{Q}(x,t)\), \(\mscr{P}(x,t)\) of \(\mbox{sl}(2,\mathbb{C})\) by the wavefunctions
\(\psi_{\mbox{\scz su}(2)}(x,t)\), \(\psi_{\mbox{\scz sp}(2,\mathbb{R})}(x,t)\) of the attractive and repulsive interaction case of GP-equations
according to eqs. (\ref{s2_25},\ref{s2_26}) so that one can resolve from the real "\(\Re(c_{0})\)" and imaginary part "\(\Im(c_{0})\)"
of the generalized norm (\ref{s2_24}) two independent conserved relations (\ref{s2_29},\ref{s2_30})
in terms of the independent physical wavefunctions
\(\psi_{\mbox{\scz su}(2)}(x,t)\), \(\psi_{\mbox{\scz sp}(2,\mathbb{R})}(x,t)\). In correspondence to the non-compact property of
\(\mbox{SL}(2,\mathbb{C})\), we attain unbounded norms of \(\psi_{\mbox{\scz su}(2)}(x,t)\), \(\psi_{\mbox{\scz sp}(2,\mathbb{R})}(x,t)\),
which only differ by the constant \(\Re(c_{0})\), and also acquire a restriction for a {\it generalized angle} between the
{\it wavefunction vectors} of \(\psi_{\mbox{\scz su}(2)}(x,t)\), \(\psi_{\mbox{\scz sp}(2,\mathbb{R})}(x,t)\) from the imaginary
part (as one transfers the spatial argument "$x$" of wavefunctions to a discrete index). The decomposition of \(\mscr{Q}(x,t)\), \(\mscr{P}(x,t)\)
to the parameters \(\psi_{\mbox{\scz su}(2)}(x,t)\), \(\psi_{\mbox{\scz sp}(2,\mathbb{R})}(x,t)\) of the sub-algebras
\(\mbox{su}(2)\), \(\mbox{sp}(2,\mathbb{R})\) within \(\mbox{sl}(2,\mathbb{C})\) therefore removes the restriction to
completely integrable behaviour (even for the cases with arbitrary external potential \(V(x,t)\))
\beq\lb{s2_27}
c_{0}(\in\mathbb{C})&=&\int_{0}^{2\pi}dx\;\Big(-\psi_{\mbox{\scz su}(2)}^{*}(x,t)+\psi_{\mbox{\scz sp}(2,\mathbb{R})}^{*}(x,t)\Big)\;
\Big(\psi_{\mbox{\scz su}(2)}(x,t)+\psi_{\mbox{\scz sp}(2,\mathbb{R})}(x,t)\Big)\;; \\  \lb{s2_28}
\int_{0}^{2\pi}dx\;\big|\psi_{\mbox{\scz sp}(2,\mathbb{R})}(x,t)\big|^{2} &=&
\int_{0}^{2\pi}dx\;\big|\psi_{\mbox{\scz su}(2)}(x,t)\big|^{2} +\Re(c_{0}) + \\ \no  &-&\im
\bigg[\Im(c_{0})-2\int_{0}^{2\pi}dx\;\Im\Big(\psi_{\mbox{\scz su}(2)}^{*}(x,t)\;\psi_{\mbox{\scz sp}(2,\mathbb{R})}(x,t)\Big)\bigg] \;;
\eeq
\beq\lb{s2_29}
\int_{0}^{2\pi}dx\;\big|\psi_{\mbox{\scz sp}(2,\mathbb{R})}(x,t)\big|^{2} &=&
\int_{0}^{2\pi}dx\;\big|\psi_{\mbox{\scz su}(2)}(x,t)\big|^{2} +\Re(c_{0}) \;;  \\ \lb{s2_30}
\int_{0}^{2\pi}dx\;\Im\Big(\psi_{\mbox{\scz su}(2)}^{*}(x,t)\;\psi_{\mbox{\scz sp}(2,\mathbb{R})}(x,t)\Big)&=&{\ts\frac{1}{2}}\;\Im(c_{0})\;.
\eeq
After replacing the fields \(\mscr{Q}(x,t)\), \(\mscr{P}(x,t)\) by (\ref{s2_25},\ref{s2_26}) in the \(\mbox{sl}(2,\mathbb{C})\) zero-curvature relation
and field equations (\ref{s2_31}), we attain two coupled GP-type equations with the attractive case of the \(\mbox{su}(2)\)-field
\(\psi_{\mbox{\scz su}(2)}(x,t)\) and with the repulsive case of the \(\mbox{sp}(2,\mathbb{R})\)-field \(\psi_{\mbox{\scz sp}(2,\mathbb{R})}(x,t)\).
Aside from the cubic, attractive interaction '\(-2\:|\psi_{\mbox{\scz su}(2)}(x,t)|^{2}\:\psi_{\mbox{\scz su}(2)}(x,t)\)' and the
cubic, repulsive interaction '\(+2\:|\psi_{\mbox{\scz sp}(2,\mathbb{R})}(x,t)|^{2}\:\psi_{\mbox{\scz sp}(2,\mathbb{R})}(x,t)\), there appear
the parts '\(+4\:|\psi_{\mbox{\scz sp}(2,\mathbb{R})}(x,t)|^{2}\:\psi_{\mbox{\scz su}(2)}(x,t)\)' and
'\(-4\:|\psi_{\mbox{\scz su}(2)}(x,t)|^{2}\:\psi_{\mbox{\scz sp}(2,\mathbb{R})}(x,t)\)' acting as a repulsive and attractive external
potential, respectively. Moreover, one has the two incoherent terms
'\(-2\:\psi_{\mbox{\scz su}(2)}^{*}(x,t)\:\psi_{\mbox{\scz sp}(2,\mathbb{R})}^{2}(x,t)\)' and
'\(+2\:\psi_{\mbox{\scz sp}^{*}(2,\mathbb{R})}(x,t)\:\psi_{\mbox{\scz su}(2)}^{2}(x,t)\)' which can give rise to a change of norm and integrated,
total probability
\beq \lb{s2_31}
\lefteqn{\bea{rcl}\im\,(\pp_{t}\mscr{P}) &=& \mscr{P}_{xx} -2\:r\;\mscr{P}^{2} \\
-\im\,(\pp_{t}r) &=& \mscr{Q}_{xx}-2\:\mscr{P}\:r^{2} \eea\bigg\}\Longrightarrow } \\ \lb{s2_32}
\im\,\big(\pp_{t}\psi_{\mbox{\scz su}(2)}\,\big) &=& -\big(\pp_{x}\pp_{x}\psi_{\mbox{\scz su}(2)}\big)
-2\,\Big(\big|\psi_{\mbox{\scz su}(2)}\big|^{2}-2\big|\psi_{\mbox{\scz sp}(2,\mathbb{R})}\big|^{2}\Big)\:\psi_{\mbox{\scz su}(2)}
-2\:\psi_{\mbox{\scz su}(2)}^{*}\;\big(\psi_{\mbox{\scz sp}(2,\mathbb{R})}\big)^{2}  \;;  \\  \lb{s2_33}
\im\,\big(\pp_{t}\psi_{\mbox{\scz sp}(2,\mathbb{R})}\,\big) &=& -\big(\pp_{x}\pp_{x}\psi_{\mbox{\scz sp}(2,\mathbb{R})}\big)
+2\,\Big(\big|\psi_{\mbox{\scz sp}(2,\mathbb{R})}\big|^{2}-2\big|\psi_{\mbox{\scz su}(2)}\big|^{2}\Big)\:\psi_{\mbox{\scz sp}(2,\mathbb{R})}
+2\:\psi_{\mbox{\scz sp}(2,\mathbb{R})}^{*}\;\big(\psi_{\mbox{\scz su}(2)}\big)^{2}  \;.
\eeq
This becomes obvious as we derive continuity equations with the corresponding source parts
\beq \lb{s2_34}
\big(\pp_{t}|\psi_{\mbox{\scz su}(2)}(x,t)|^{2}\big) &=& \hspace*{-0.3cm}-2\;\pp_{x}\Im\Big[\big(\psi_{\mbox{\scz su}(2)}^{*}(x,t)\;
\big(\pp_{x}\psi_{\mbox{\scz su}(2)}(x,t)\,\big)\,\big)\Big] -4\;|\psi_{\mbox{\scz su}(2)}(x,t)|^{2}\;
\Im\big(\psi_{\mbox{\scz sp}(2,\mathbb{R})}^{2}(x,t)\,\big)  \;; \\ \lb{s2_35}
\big(\pp_{t}|\psi_{\mbox{\scz sp}(2,\mathbb{R})}(x,t)|^{2}\big) &=&\hspace*{-0.3cm} -2\;\pp_{x}\Im\Big[\big(\psi_{\mbox{\scz sp}(2,\mathbb{R})}^{*}(x,t)\;
\big(\pp_{x}\psi_{\mbox{\scz sp}(2,\mathbb{R})}(x,t)\,\big)\,\big)\Big] \!+\!4\;|\psi_{\mbox{\scz sp}(2,\mathbb{R})}(x,t)|^{2}\;
\Im\big(\psi_{\mbox{\scz su}(2)}^{2}(x,t)\,\big)_{\!\mbox{.}}
\eeq
According to the coupled GP-type equations (\ref{s2_32},\ref{s2_33}), one can distinguish three different, physical cases. If we set the
two wavefunctions \(\psi_{\mbox{\scz su}(2)}(x,t)\) , \(\psi_{\mbox{\scz sp}(2,\mathbb{R})}(x,t)\) to an identical value
\(\psi_{0}(x,t)\), one just considers the trivial free propagation with the kinetic term '\(-(\pp_{x}\pp_{x}\psi_{0}(x,t)\,)\)' (\ref{s2_36}).
However, we can also take the attractive or repulsive case with the wavefunctions
\(\psi_{\mbox{\scz su}(2)}(x,t)\) , \(\psi_{\mbox{\scz sp}(2,\mathbb{R})}(x,t)\) as the dominant term \(\psi_{0}(x,t)\) in the two, coupled,
nonlinear equations (\ref{s2_32},\ref{s2_33}) and can then regard the other \(\psi_{\mbox{\scz sp}(2,\mathbb{R})}(x,t)\) or
\(\psi_{\mbox{\scz su}(2)}(x,t)\) wave component as the corresponding perturbation \(\Delta\psi(x,t)\) in order to compute for a
possible chaotic behaviour of the \(\psi_{\mbox{\scz su}(2)}(x,t)\) and \(\psi_{\mbox{\scz sp}(2,\mathbb{R})}(x,t)\)
wavefunctions, respectively
\beq\lb{s2_36}
1)\;\mbox{ if } &:&\psi_{\mbox{\scz su}(2)}=\psi_{\mbox{\scz sp}(2,\mathbb{R})}=\psi_{0}\;;\;\;\;\Delta\psi\equiv0\;\Longrightarrow\;
\mbox{free propagation}\;;\;\;\; \im\,\big(\pp_{t}\psi_{0}\big)=-\big(\pp_{x}\pp_{x}\psi_{0}\big) \;; \\ \lb{s2_37}
2)\;\mbox{ if } &:&\psi_{\mbox{\scz su}(2)}=\psi_{0}\;;\;\;\;
\psi_{\mbox{\scz sp}(2,\mathbb{R})}=\Delta\psi\;(\mbox{perturbation})\;\Longrightarrow\;
\mbox{chaotic behaviour of }\;\psi_{\mbox{\scz su}(2)}\;;  \\ \lb{s2_38}
3)\;\mbox{ if } &:&\psi_{\mbox{\scz sp}(2,\mathbb{R})}=\psi_{0}\;;\;\;\;
\psi_{\mbox{\scz su}(2)}=\Delta\psi\;(\mbox{perturbation})\;\Longrightarrow\;
\mbox{chaotic behaviour of }\;\psi_{\mbox{\scz sp}(2,\mathbb{R})}\;.
\eeq
The latter two cases (\ref{s2_37},\ref{s2_38}) as sub-algebras of \(\mbox{sl}(2,\mathbb{C})\) have an infinite number of
independent, exactly conserved quantities which are derived from  traces of powers of
\(\mfrak{T}_{\mbox{\scz sl}(2,\mathbb{C})}(x,t;k)\) monodromy matrices. These exactly conserved quantities from
\(\mfrak{T}_{\mbox{\scz sl}(2,\mathbb{C})}(x,t;k)\) contain both sub-groups, the \(\mbox{SU}(2)\)-group with
\(\mfrak{T}_{\mbox{\scz su}(2)}(x,t;k)\) matrices and the \(\mbox{Sp}(2,\mathbb{R})\)-group with the
\(\mfrak{T}_{\mbox{\scz sp}(2,\mathbb{R})}(x,t;k)\) matrices, so that one can consider perturbations of the dominantly
conserved parts, which are given as traces of powers of \(\mfrak{T}_{\mbox{\scz su}(2)}(x,t;k)\) or
\(\mfrak{T}_{\mbox{\scz sp}(2,\mathbb{R})}(x,t;k)\)
with corresponding perturbative terms of \(\mfrak{L}_{\mbox{\scz sp}(2,\mathbb{R})}(x,t;k)\) or
\(\mfrak{L}_{\mbox{\scz su}(2)}(x,t;k)\), respectively.

In this section we have limited discussion to \(\mbox{sl}(2,\mathbb{C})\) Lax pairs and zero-curvature relations
with the \(\mbox{su}(2)\), \(\mbox{sp}(2,\mathbb{R})\) sub-algebras of the corresponding integrable, attractive or repulsive
GP-equations. However, the coupled equations (\ref{s2_32},\ref{s2_33}) of
\(\psi_{\mbox{\scz su}(2)}(x,t)\) and \(\psi_{\mbox{\scz sp}(2,\mathbb{R})}(x,t)\) wavefunctions within the non-compact
\(\mbox{sl}(2,\mathbb{C})\) algebra can be generalized to arbitrary algebras \(\mbox{sl}(n,\mathbb{C})\) with a chosen sub-algebra
\(\mbox{su}(n)\) so that one can achieve a possible chaotic behaviour for field equations following from \(\mbox{sl}(n,\mathbb{C})\)
Lax pairs and zero-curvature conditions. The rather general \(\mbox{sl}(n,\mathbb{C})\) Lax pair has to be decomposed into sub-algebras
as the compact \(\mbox{su}(n)\) case of anti-hermitian Lax pairs and remaining algebra parts so that the norm of wavefunctions is not bounded
by real, positive constants. This extension to chaotic behaviour, which follows from taking compact sub-algebras of the rather
general \(\mbox{sl}(n,\mathbb{C})\) algebra, is also possible for Lax pairs beyond (1+1) dimensions.

\subsection{Lax pairs of $\boldsymbol{\mbox{sl}(2,\mathbb{C})}$ for an external potential} \lb{s22}

The derivations, concerning the integrable and chaotic behaviour in the previous section \ref{s21}, can even be transferred to the case
with an arbitrary external potential \(V(x,t)\) in the attractive \(\mbox{su}(2)\) and repulsive \(\mbox{sp}(2,\mathbb{R})\) interaction
case of GP-equations. In the following we construct \(\mbox{sl}(2,\mathbb{C})\) Lax matrices \(\mfrak{L}_{\mbox{\scz sl}(2,\mathbb{C})}(\mscr{Q},\mscr{P};k)\),
\(\mfrak{M}_{\mbox{\scz sl}(2,\mathbb{C})}(A,B,C)\) so that the nonlinear field equations for \(\mscr{Q}(x,t)\), \(\mscr{P}(x,t)\) of the zero-curvature
relation allow under restriction to \(\psi_{\mbox{\scz su}(2)}(x,t)\), \(\psi_{\mbox{\scz sp}(2,\mathbb{R})}(x,t)\) fields and sub-algebras
the GP-equations with the external potential \(V(x,t)\). It suffices to derive the nonlinear equations
\beq\lb{s2_39}
\im\,\mscr{P}_{t} &=& \mscr{P}_{xx} -2\,\big(\mscr{Q}\:\mscr{P}-V(x,t)\,\big)\,\mscr{P}  \;;  \\  \lb{s2_40}
-\im\,\mscr{Q}_{t} &=& \mscr{Q}_{xx} -2\,\big(\mscr{P}\:\mscr{Q}-V(x,t)\,\big)\,\mscr{Q}  \;,
\eeq
so that the whole discussion of section \ref{s21} can be conveyed to the case with an external potential, especially the conclusions
for integrable and chaotic behaviour. We point out that the spatial Lax matrix \(\mfrak{L}_{\mbox{\scz sl}(2,\mathbb{C})}(\mscr{Q},\mscr{P};k)\)
does not change aside from a possible spacetime dependence of the complex spectral parameter \(k\rightarrow k(x,t)\in\mathbb{C}\).
Therefore, the conserved quantities, following from traces of powers of the monodromy matrix \(\mfrak{T}_{\mbox{\scz sl}(2,\mathbb{C})}(x,t;k)\),
are not essentially altered and the canonical variables \(\mscr{Q}(x,t)\), \(\mscr{P}(x,t)\) within the Poisson brackets do not cause a different
\(\mfrak{r}\)-matrix which specifies the involution of the independently conserved quantities. (In the sequel one is even allowed
to choose a constant spectral parameter $k\in\mathbb{C}$, concerning the involution property, because the complex-valued, auxiliary field
$W$ can always be adapted in such a manner so that an unwanted appearance of \(V_{x}(x,t)\) can be absorbed by suitable dependence of
\(W_{x}\) in (\ref{s2_59}).)

In analogy to eqs. (\ref{s2_1}-\ref{s2_5}), we start out from traceless matrices \(\mfrak{L}_{\mbox{\scz sl}(2,\mathbb{C})}(\mscr{Q},\mscr{P};k)\),
\(\mfrak{M}_{\mbox{\scz sl}(2,\mathbb{C})}(A,B,C)\) where the latter matrix for the time evolution of the auxiliary field
\(\Xi(x,t)\) is determined by an ansatz with the three complex-valued fields \(A(\mscr{P},\mscr{Q};k)\), \(B(\mscr{P},\mscr{Q};k)\), \(C(\mscr{P},\mscr{Q};k)\)
\beq\lb{s2_41}
\Xi_{x} &=& \bigg(\bea{cc} -\im\,k & \mscr{P} \\ \mscr{Q} & \im\,k \eea\bigg)\;\Xi(x,t) = \mfrak{L}_{\mbox{\scz sl}(2,\mathbb{C})}(\mscr{Q},\mscr{P};k)\;\Xi(x,t) \\ \lb{s2_42}
\Xi_{t} &=& \bigg(\bea{cc} A(\mscr{P},\mscr{Q};k) & B(\mscr{P},\mscr{Q};k) \\
C(\mscr{P},\mscr{Q};k) & -A(\mscr{P},\mscr{Q};k)  \eea\bigg)\;\Xi(x,t)=\mfrak{M}_{\mbox{\scz sl}(2,\mathbb{C})}(A,B,C)\;\Xi(x,t) \;; \\ \lb{s2_43}
\Xi &=& \Xi(x,t)\in\mathbb{C}\;;\;\;\;\mscr{P}=\mscr{P}(x,t)\in\mathbb{C}\;;\;\;\;\mscr{Q}=\mscr{Q}(x,t)\in\mathbb{C}\;;  \;\;\;k=k(x,t)\in\mathbb{C}\;;\\ \lb{s2_44}
A &=& A(\mscr{P},\mscr{Q};k)\in\mathbb{C}\;;\;\;\;B=B(\mscr{P},\mscr{Q};k)\in\mathbb{C}\;;\;\;\;C=C(\mscr{P},\mscr{Q};k)\in\mathbb{C}\;; \\  \lb{s2_45}
\mfrak{L}_{\mbox{\scz sl}(2,\mathbb{C})}(\mscr{Q},\mscr{P};k) &=& -(\im\,k)\; \big(\hat{\tau}_{3}\big) + \mscr{P}(x,t)\;\big(\hat{\tau}_{+}\big)+ \mscr{Q}(x,t)\;\big(\hat{\tau}_{-}\big) \;\Longrightarrow\in\mbox{sl}(2,\mathbb{C}) \;; \\  \lb{s2_46}
\mfrak{M}_{\mbox{\scz sl}(2,\mathbb{C})}(A,B,C) &=&A(\mscr{P},\mscr{Q};k) \; \big(\hat{\tau}_{3}\big) +
B(\mscr{P},\mscr{Q};k)\;\big(\hat{\tau}_{+}\big)+C(\mscr{P},\mscr{Q};k)\;\big(\hat{\tau}_{-}\big)
\;\Longrightarrow\in\mbox{sl}(2,\mathbb{C}) \;.
\eeq
The zero-curvature condition (\ref{s2_47}) results into the field equations (\ref{s2_49},\ref{s2_50}) and additionally into the spectral
parameter equation (\ref{s2_51}) which follow from the off-diagonal and diagonal matrix entries in (\ref{s2_48}).
A modified ansatz (\ref{s2_52}-\ref{s2_54}) for the
coefficients \(A(\mscr{P},\mscr{Q};k)\), \(B(\mscr{P},\mscr{Q};k)\), \(C(\mscr{P},\mscr{Q};k)\), similar to the case in previous section \ref{s21}, leads to the equations
(\ref{s2_57},\ref{s2_58}) of the fields \(\mscr{Q}(x,t)\), \(\mscr{P}(x,t)\). Aside from the external potential
in the coefficient \(A(\mscr{P},\mscr{Q};k)\) (\ref{s2_52}),
the ansatz (\ref{s2_52}-\ref{s2_54}) is supplied with a sufficient number of complex-valued fields
\(W(\mscr{P},\mscr{Q};k;V)\), \(Y(\mscr{P},\mscr{Q};k;V)\), \(Z(\mscr{P},\mscr{Q};k;V)\) (\ref{s2_55}) which have to be determined in such a manner that the nonlinear field equations
(\ref{s2_57},\ref{s2_58}) of \(\mscr{Q}(x,t)\), \(\mscr{P}(x,t)\) only change by the external potential \(V(x,t)\). These conditions
(cf. the two braces in eqs. (\ref{s2_57},\ref{s2_58})) specify first order partial differential equations of
\(Y_{x}(\mscr{P},\mscr{Q};k;V)\), \(Z_{x}(\mscr{P},\mscr{Q};k;V)\) whereas the additional equation (\ref{s2_59})
of a (possibly chosen spacetime dependent !) spectral parameter
\(k(x,t)\) results into a differential equation with the partial derivative \(W_{x}(\mscr{P},\mscr{Q};k;V)\)
\beq\no
&&\mbox{\bf zero-curvature condition}  \\ \lb{s2_47}
\Longrightarrow 0&=& \big(\pp_{t}\mfrak{L}_{\mbox{\scz sl}(2,\mathbb{C})}(\mscr{Q},\mscr{P};k)\big)-
\big(\pp_{x}\mfrak{M}_{\mbox{\scz sl}(2,\mathbb{C})}(A,B,C)\big)+
\boldsymbol{\big[}\mfrak{L}_{\mbox{\scz sl}(2,\mathbb{C})}(\mscr{Q},\mscr{P};k)\;\boldsymbol{,}\;
\mfrak{M}_{\mbox{\scz sl}(2,\mathbb{C})}(A,B,C)\,\boldsymbol{\big]_{-}} ;
\\ \lb{s2_48} \lefteqn{\hspace*{-0.6cm}\bigg(\bea{cc} -\im\,k_{t} & \mscr{P}_{t} \\ \mscr{Q}_{t} & \im\,k_{t} \eea\bigg)-
\bigg(\bea{cc} A_{x} & B_{x} \\ C_{x} & -A_{x} \eea\bigg) +
\bigg(\bea{cc} \mscr{P}\:C-\mscr{Q}\:B &  -2\,\mscr{P}\:A-2\im\,k\:B \\
2\,\mscr{Q}\:A+2\im\,k\:C & -(\mscr{P}\:C-\mscr{Q}\:B) \eea\bigg)\equiv 0 \;;}  \\ \lb{s2_49}
\mscr{P}_{t} &=& B_{x}+2\,\mscr{P}\:A+2\im\,k\:B \;;  \\ \lb{s2_50}
\mscr{Q}_{t} &=& C_{x}-2\,\mscr{Q}\:A-2\im\,k\:C \;;  \\ \lb{s2_51}
-\im\,k_{t} &=& A_{x} - \mscr{P}\:C + \mscr{Q}\:B \;;  \\   \lb{s2_52}
A &=& \im\,\mscr{P}\:\mscr{Q} + 2\im\,\:k^{2} -\im\:V(x,t) + W \;; \\  \lb{s2_53}
B &=& -\im\,\mscr{P}_{x} -2\,k\:\mscr{P} +Y \;;  \\  \lb{s2_54}
C &=& \im\,\mscr{Q}_{x} -2\,k\:\mscr{Q} + Z \;;  \\ \lb{s2_55}
W &=& W(\mscr{P},\mscr{Q};k;V)\in\mathbb{C}\;;\;\;\;Y=Y(\mscr{P},\mscr{Q};k;V)\in\mathbb{C}\;;\;\;\;Z=Z(\mscr{P},\mscr{Q};k;V)\in\mathbb{C}\;;  \\   \lb{s2_56}
k&=&k(x,t)\in\mathbb{C}\;;\;\;\;V=V(x,t)\in\mathbb{C}\;;    \\   \lb{s2_57}
\im\:\mscr{P}_{t} &=& \mscr{P}_{xx} -2\,(\mscr{Q}\:\mscr{P}-V)\,\mscr{P}\overbrace{+\im\,Y_{x}-2\im\,(k_{x}-W)\,\mscr{P}-2\,k\:Y}^{\equiv 0}  \;;  \\  \lb{s2_58}
-\im\:\mscr{Q}_{t} &=& \mscr{Q}_{xx} -2\,(\mscr{P}\:\mscr{Q}-V)\,\mscr{Q}\underbrace{-\im\,Z_{x}+2\im\,(k_{x}+W)\,\mscr{Q}-2\,k\:Z}_{\equiv 0} \;;  \\   \lb{s2_59}
-\im\:k_{t} &=& \im\,(2\,(k^{2})_{x} - V_{x}) +W_{x} +Y\:\mscr{Q} -Z\:\mscr{P}  \;.
\eeq
We combine the restrictions of the auxiliary fields \(W(\mscr{P},\mscr{Q};k;V)\), \(Y(\mscr{P},\mscr{Q};k;V)\), \(Z(\mscr{P},\mscr{Q};k;V)\) (\ref{s2_55}) of the modified ansatz
with coefficients \(A(\mscr{P},\mscr{Q};k)\), \(B(\mscr{P},\mscr{Q};k)\), \(C(\mscr{P},\mscr{Q};k)\) from relations (\ref{s2_47}-\ref{s2_59}) and require the nonlinear
field equations (\ref{s2_60},\ref{s2_61}) with the external potential \(V(x,t)\) for the physical fields
\(\mscr{Q}(x,t)\), \(\mscr{P}(x,t)\). The nonlinear field equations (\ref{s2_60},\ref{s2_61}) only follow from the zero-curvature condition
with modified Lax matrix \(\mfrak{M}_{\mbox{\scz sl}(2,\mathbb{C})}(A,B,C)\), provided that the additional fields
\(W(\mscr{P},\mscr{Q};k;V)\), \(Y(\mscr{P},\mscr{Q};k;V)\), \(Z(\mscr{P},\mscr{Q};k;V)\) (\ref{s2_55}) fulfill the first order, spatial evolution equation (\ref{s2_62}).
However, this evolution equation (\ref{s2_62}) in the spatial variable $x$ can be solved under rather general conditions with the appropriate
ordering of exponential step operators and initial value \(\vec{\mfrak{w}}(x=0,t)\) (\ref{s2_63},\ref{s2_64}).
Apart from the homogenous solution with latter field vector \(\vec{\mfrak{w}}(x=0,t)\), one has a particular solution with vector field
\(\vec{\mfrak{v}}(\mscr{P},\mscr{Q};k;V)\) which also contains the spatial derivative of the external potential \(V(x,t)\)
\beq \lb{s2_60}
\im\,\mscr{P}_{t} &=& \mscr{P}_{xx} -2\,\big(\mscr{Q}\:\mscr{P}-V(x,t)\,\big)\,\mscr{P}  \;;  \\  \lb{s2_61}
-\im\,\mscr{Q}_{t} &=& \mscr{Q}_{xx} -2\,\big(\mscr{P}\:\mscr{Q}-V(x,t)\,\big)\,\mscr{Q}  \;;  \\  \lb{s2_62}
\pp_{x}\left(\bea{c} W \\ Y \\ Z \eea\right) &=&
\left(\bea{ccc}  0 & -\mscr{Q} & \mscr{P} \\ -2\,\mscr{P} & -2\im\,k & 0 \\
2\,\mscr{Q} & 0 & 2\im\,k \eea\right)\left(\bea{c} W \\ Y \\ Z \eea\right) +
\left(\bea{c} \im\big(V_{x}-k_{t}-2\,(k^{2})_{x}\big) \\ 2\,k_{x}\,\mscr{P} \\ 2\,k_{x}\,\mscr{Q} \eea\right) \;; \\ \lb{s2_63}
\pp_{x}\vec{\mfrak{w}} &=& \hat{\mscr{W}}(\mscr{P},\mscr{Q};k)\:\vec{\mfrak{w}} + \vec{\mfrak{v}}(\mscr{P},\mscr{Q};k;V) \;;  \\ \no
\vec{\mfrak{w}} &=&\big(W\,,\,Y\,,\,Z\big)^{T}\;;\;\;\;
\vec{\mfrak{v}}(\mscr{P},\mscr{Q};k;V)=\Big(\im\big(V_{x}-k_{t}-2\,(k^{2})_{x}\big)\:,\: 2\,k_{x}\,\mscr{P} \:,\: 2\,k_{x}\,\mscr{Q} \Big)^{T}  \;;  \\   \lb{s2_64}
\vec{\mfrak{w}}(x,t) &=& \overleftarrow{exp}\Big\{\int_{0}^{x}d\xi\;\hat{\mscr{W}}(\xi,t)\Big\}\;\vec{\mfrak{w}}(x=0,t) +
\int_{0}^{x}dy\;\overleftarrow{exp}\Big\{\int_{y}^{x}d\xi\;\hat{\mscr{W}}(\xi,t)\Big\}\;\vec{\mfrak{v}}(y,t) \;.
\eeq
Therefore, one can construct Lax matrices for generalized GP-type equations (\ref{s2_60},\ref{s2_61}) with the additional
fields \(W(\mscr{P},\mscr{Q};k;V)\), \(Y(\mscr{P},\mscr{Q};k;V)\), \(Z(\mscr{P},\mscr{Q};k;V)\) (\ref{s2_55},\ref{s2_62}-\ref{s2_64}) in such a manner, that the
zero-curvature relation (\ref{s2_47},\ref{s2_48}) confines to the cases with an external potential \(V(x,t)\).
As we have already mentioned at the beginning of this section \ref{s22}, one can directly repeat considerations of section \ref{s21}
in order to conclude for integrable or a possible chaotic behaviour by regarding the corresponding cases of
\(\mbox{sl}(2,\mathbb{C})\) Lax pairs with sub-algebras \(\mbox{su}(2)\) and \(\mbox{sp}(2,\mathbb{R})\).

\section{The general $\boldsymbol{n\times n}$ Lax pair matrices as $\boldsymbol{\mbox{sl}(n,\mathbb{C})}$ algebras} \lb{s3}

\subsection{Solution of the zero-curvature condition} \lb{s31}

The most general matrices for the Lax pairs \(\mfrak{L}(x,t)\), \(\mfrak{M}(x,t)\) are given by the complex-valued \(\mbox{gl}(n,\mathbb{C})\)
algebra with \(2\:n^{2}\) real parameters; however, due to the occurrence of the commutator
\(\boldsymbol{[}\mfrak{L}(x,t)\:\boldsymbol{,}\:\mfrak{M}(x,t)\boldsymbol{]_{-}}\),
one can split a trivial, diagonal unity part from this most general ansatz
with two complex-valued fields consisting of the sum of the diagonal entries from the \(\mfrak{L}(x,t)\), \(\mfrak{M}(x,t)\) matrices
within \(\mbox{gl}(n,\mathbb{C})\) (cf. appendix \ref{sa}). Therefore, the \(\mbox{sl}(n,\mathbb{C})\) algebra with \(2\:n^{2}-2\) real
parameters is only taken as the most general ansatz for Lax pairs \(\mfrak{L}(x,t)\), \(\mfrak{M}(x,t)\) so that, indeed, field equations
remain with nontrivial, nonlinear properties.

In the following, we assume that the spatial Lax matrix \(\mfrak{L}(x,t)\) is chosen within a \(\mbox{sl}(n,\mathbb{C})\) algebra or one
of its sub-algebras as e.\ g.\ \(\mbox{su}(n)\), etc.\ . Therefore, one can view the zero-curvature relation as a spatial evolution equation
(\ref{s3_1}) of \((\pp_{x}\mfrak{M})\) with the \(\mfrak{ad}\)-operator
{\small\(\overrightarrow{\boldsymbol{[}\mfrak{L}(\xi,t)\:\boldsymbol{,}\:\ldots\boldsymbol{]_{-}}}\)}
of a closed algebra acting onto \(\mfrak{M}(x,t)\)
and an inhomogeneity \((\pp_{t}\mfrak{L}(x,t)\,)\). The spatial evolution equation (\ref{s3_1}) has a homogenous matrix solution
\(\mfrak{M}_{hom}(x,t)\) (\ref{s3_2}) with 'ini(tial)' matrix \(\mfrak{M}_{ini(tial)}(x=0,t)\) at the spatial origin
which develops with spatial steps
{\small\({\scr \Delta x}\cdot \overrightarrow{\boldsymbol{[}\mfrak{L}(\xi,t)\:\boldsymbol{,}\:\ldots\boldsymbol{]_{-}}}\)}
in exponential operators with appropriate spatial ordering from right to left
'\(\overleftarrow{\exp}\{\ldots\}\)'. The action of the \(\mfrak{ad}\)-operator
{\small\(\overrightarrow{\boldsymbol{[}\mfrak{L}(\xi,t)\:\boldsymbol{,}\:\ldots\boldsymbol{]_{-}}}\)}
within the spatially ordered exponentials has the effect
of left and right propagation for the initial matrix \(\mfrak{M}_{ini}(x=0,t)\) with exponents of
\(\pm {\scr \Delta x}\cdot\mfrak{L}(\xi,t)\); this is reminiscent of the Heisenberg equation of motion in quantum mechanics or of the total
development operator for the von-Neumann equation in statistical mechanics
\beq \lb{s3_1}
\big(\pp_{x}\mfrak{M}\big) &=&\boldsymbol{\big[}\mfrak{L}\:\boldsymbol{,}\:\mfrak{M}\boldsymbol{\big]_{-}}+
\big(\pp_{t}\mfrak{L}\big) = \overrightarrow{\boldsymbol{[}\mfrak{L}\:\boldsymbol{,}\:\ldots\boldsymbol{]_{-}}}\mfrak{M}+
\big(\pp_{t}\mfrak{L}\big)  \;;  \\ \no \mbox{homogenous solution} &:&
\mfrak{M}_{hom}(x,t) \;; \\ \lb{s3_2}
\mfrak{M}_{hom}(x,t) &=& \overleftarrow{\exp}\Big\{\int_{0}^{x}d\xi\:
\overrightarrow{\boldsymbol{[}\mfrak{L}(\xi,t)\:\boldsymbol{,}\:\ldots\boldsymbol{]_{-}}}\Big\}\mfrak{M}_{ini(tial)}(x=0,t) \;;  \\  \lb{s3_3}
\lefteqn{\hspace*{-6.0cm}\mfrak{M}_{hom}(x,t) =
\exp\big\{{\scr \Delta x}\,\mfrak{L}(x,t)\big\}\:\exp\big\{{\scr \Delta x}\,\mfrak{L}(x-{\scr \Delta x},t)\big\}\:
\exp\big\{{\scr \Delta x}\,\mfrak{L}(x-2{\scr \Delta x},t)\big\}\:\ldots \;\times }  \\ \no \lefteqn{\hspace*{-6.0cm}\times\ldots\:
\exp\big\{{\scr \Delta x}\,\mfrak{L}(2{\scr \Delta x},t)\big\}\:\exp\big\{{\scr \Delta x}\,\mfrak{L}({\scr \Delta x},t)\big\}\:\mfrak{M}_{ini}(x=0,t)\:
\exp\big\{-{\scr \Delta x}\,\mfrak{L}({\scr \Delta x},t)\big\}\:\exp\big\{-{\scr \Delta x}\,\mfrak{L}(2{\scr \Delta x},t)\big\}\:\ldots \;\times} \\ \no
\lefteqn{\hspace*{-6.0cm}\times\ldots\:
\exp\big\{-{\scr \Delta x}\,\mfrak{L}(x-2{\scr \Delta x},t)\big\}\:\exp\big\{-{\scr \Delta x}\,\mfrak{L}(x-{\scr \Delta x},t)\big\}\:
\exp\big\{-{\scr \Delta x}\,\mfrak{L}(x,t)\big\}\;.}
\eeq
The particular solution of (\ref{s3_1}) similarly results as in the case of an ordinary, first order differential equation by a variational
matrix-ansatz \(\mfrak{N}(x,t)\) (\ref{s3_4}) instead of the initial matrix \(\mfrak{M}_{ini}(x=0,t)\). Straightforward transformations
(\ref{s3_5},\ref{s3_6}) lead to the general solution (\ref{s3_7}) which consists of the homogenous part with matrix \(\mfrak{M}_{ini}(x=0,t)\)
and the particular part with the inhomogeneity \((\pp_{t}\mfrak{L}(y,t)\,)\)
\beq \no
\mbox{Variational ansatz for initial matrix } \mfrak{M}_{ini}(x=0,t) \mbox{ with } \mfrak{N}(x,t) && \\ \lb{s3_4}
\lefteqn{\hspace*{-13.6cm}\mfrak{M}(x,t) = \overleftarrow{\exp}\Big\{\int_{0}^{x}d\xi\:
\overrightarrow{\boldsymbol{[}\mfrak{L}(\xi,t)\:\boldsymbol{,}\:\ldots\boldsymbol{]_{-}}}\Big\}\mfrak{M}_{ini}(x=0,t) +
\overleftarrow{\exp}\Big\{\int_{0}^{x}d\xi\:
\overrightarrow{\boldsymbol{[}\mfrak{L}(\xi,t)\:\boldsymbol{,}\:\ldots\boldsymbol{]_{-}}}\Big\}\mfrak{N}(x,t)  \;; } \\  \no
\mbox{inserted into zero-curvature condition} && \\  \lb{s3_5}
\lefteqn{\hspace*{-13.6cm}\overleftarrow{\exp}\Big\{\int_{0}^{x}d\xi\:
\overrightarrow{\boldsymbol{[}\mfrak{L}(\xi,t)\:\boldsymbol{,}\:\ldots\boldsymbol{]_{-}}}\Big\}\:
\big(\pp_{x}\mfrak{N}(x,t)\,\big) =\big(\pp_{t}\mfrak{L}(x,t)\,\big) \;; }    \\   \lb{s3_6}
\lefteqn{\hspace*{-13.6cm}\mfrak{N}(x,t)=\int_{0}^{x}dy\:\overrightarrow{\exp}\Big\{-\int_{0}^{y}d\xi\:
\overrightarrow{\boldsymbol{[}\mfrak{L}(\xi,t)\:\boldsymbol{,}\:\ldots\boldsymbol{]_{-}}}\Big\}\:
\big(\pp_{t}\mfrak{L}(y,t)\big)\;;} \\  \lb{s3_7}
\lefteqn{\hspace*{-13.6cm}\mfrak{M}(x,t)=\overleftarrow{\exp}\Big\{\int_{0}^{x}d\xi\:
\overrightarrow{\boldsymbol{[}\mfrak{L}(\xi,t)\:\boldsymbol{,}\:\ldots\boldsymbol{]_{-}}}\Big\}\mfrak{M}_{ini}(x=0,t)+
\int_{0}^{x}dy\:\overleftarrow{\exp}\Big\{\int_{y}^{x}d\xi\:
\overrightarrow{\boldsymbol{[}\mfrak{L}(\xi,t)\:\boldsymbol{,}\:\ldots\boldsymbol{]_{-}}}\Big\}\:\big(\pp_{t}\mfrak{L}(y,t)\big).}
\eeq
The general solution (\ref{s3_7}) of the zero-curvature relation demonstrates that it suffices to choose a spatial Lax matrix
\(\mfrak{L}(\xi,t)\) from a closed algebra as the general \(\mbox{sl}(n,\mathbb{C})\) case or one of its sub-algebras as \(\mbox{su}(n)\)
in order to determine the matrix \(\mfrak{M}(x,t)\) or the physical, nonlinear equations wave equations following from it. After choosing
the spatial matrix \(\mfrak{L}(\xi,t)\) and the initial matrix \(\mfrak{M}_{ini}(x=0,t)\)
within a closed Lie algebra of a Cartan-Weyl basis (\ref{s3_8}),
one is constrained to the spectral development within the corresponding, closed Lie group, due to the action of the \(\mfrak{ad}\)-operator
{\small\(\overrightarrow{\boldsymbol{[}\mfrak{L}(\xi,t)\:\boldsymbol{,}\:\ldots\boldsymbol{]_{-}}}\)} of the closed Lie algebra.
This points to a classification of nonlinear equations by the underlying closed Lie algebra of Lax pairs (as the general \(\mbox{sl}(n,\mathbb{C})\)
case with its sub-algebras as e.g. \(\mbox{su}(n)\)) instead by the precise form of the physical, nonlinear equations (as e.g. the
typical GP-equations). This becomes even more obvious as we can prove a general gauge invariance from a closed algebra of Lax pairs
so that one has to regard a whole set of Lax pairs for a certain type of physical, nonlinear equations. Let us consider
the Cartan-Weyl basis (\ref{s3_8}) for a closed Lie algebra \cite{Schweig1}
\be \lb{s3_8}
\bea{rclrcl}
\boldsymbol{\big[}\hat{H}^{i}\:\boldsymbol{,}\:\hat{H}^{j}\boldsymbol{\big]_{-}}&=&0\;;&
\boldsymbol{\big[}\hat{E}^{\alpha}\:\boldsymbol{,}\:\hat{E}^{\beta}\boldsymbol{\big]_{-}}&=&N^{\alpha\beta}\;
\hat{E}^{\alpha+\beta}\;;\;\;\;(\alpha+\beta\neq0)\;; \\
\boldsymbol{\big[}\hat{H}^{j}\:\boldsymbol{,}\:\hat{E}^{\alpha}\boldsymbol{\big]_{-}}&=&\alpha^{j}\;\hat{E}^{\alpha}\;; &
\boldsymbol{\big[}\hat{E}^{\alpha}\:\boldsymbol{,}\:\hat{E}^{-\alpha}\boldsymbol{\big]_{-}}&=&\alpha_{j}\;\hat{H}^{j}\;,
\eea
\ee
so that the spatial Lax matrix has the general, ultralocal form (\ref{s3_9}) with the fields
\(\phi_{\alpha}(x,t)=(\mscr{Q}_{\alpha}(x,t)\:,\:\mscr{P}_{\alpha}(x,t)\,)\) for the ladder operators
\(\hat{E}^{\alpha}=(\hat{E}_{-}^{\alpha}\:,\:\hat{E}_{+}^{\alpha})\) and with the general, spectral parameters \(k_{i}(x,t)\)
for the (maximal commuting, traceless) Cartan sub-algebra. According to the gauge invariance of Lax pairs (cf. next section \ref{s32}),
one can transform the general ansatz of \(\mfrak{L}(x,t;k)\) (\ref{s3_9})  with the ladder operators \(\hat{E}_{\pm}^{\alpha}\) to the
(maximal commuting, traceless) Cartan sub-algebra generators \(\hat{H}^{i}\) where the corresponding fields \(\lambda_{i}^{{\scrscr(\hat{H}^{i})}}(x,t;k)\) (\ref{s3_10}) follow from the fields \(\phi_{\alpha}(x,t)\) and the spectral parameters
\beq\lb{s3_9}
\mfrak{L}(x,t;k)&=&\hat{H}^{i}\;k_{i}(x,t)+\hat{E}^{\alpha}\;\phi_{\alpha}(x,t)\;; \\ \no
\hat{E}^{\alpha}\:\phi_{\alpha}(x,t)&=&\hat{E}_{-}^{\alpha}\:\mscr{Q}_{\alpha}(x,t)+\hat{E}_{+}^{\alpha}\:\mscr{P}_{\alpha}(x,t)\;;\;\;\;
\phi_{\alpha}(x,t)=(\mscr{Q}_{\alpha}(x,t)\:,\:\mscr{P}_{\alpha}(x,t)\,)\;;  \\ \lb{s3_10}
\mfrak{L}(x,t;k)&\rightarrow& \mfrak{L}_{{\scrscr\hat{H}}}(x,t;k)=\hat{H}^{i}\;\lambda_{i}^{{\scrscr(\hat{H}^{i})}}\big(\phi_{\alpha};k_{j}\big)=
\hat{H}^{i}\;\lambda_{i}^{{\scrscr(\hat{H}^{i})}}(x,t;k)\;; \mbox{(by a gauge transformation)} ;  \\  \lb{s3_11}
\mfrak{M}_{ini}(x=0,t) &=&
\hat{H}^{i}\;A_{i}^{(ini)}(x=0,t)+\hat{E}^{\alpha}\;B_{\alpha}^{(ini)}(x=0,t) \;;  \\ \no
\hat{E}^{\alpha}\:B_{\alpha}^{(ini)}(x,t) &=&\hat{E}_{-}^{\alpha}\:B_{-,\alpha}^{(ini)}(x,t)+
\hat{E}_{+}^{\alpha}\:B_{+,\alpha}^{(ini)}(x,t)\;;   \\ \lb{s3_12}
\mfrak{M}(x,t)&:=&\hat{H}^{i}\;a_{i}\big(\phi_{\beta}(x,t);k_{j}\big)+
\hat{E}^{\alpha}\;b_{\alpha}\big(\phi_{\beta}(x,t);k_{j}\big)=
\hat{H}^{i}\;a_{i}(x,t)+\hat{E}^{\alpha}\;b_{\alpha}(x,t) \;; \\  \no
\hat{E}^{\alpha}\;b_{\alpha}(x,t) &=& \hat{E}_{-}^{\alpha}\;b_{-,\alpha}(x,t) +
\hat{E}_{+}^{\alpha}\;b_{+,\alpha}(x,t)  \;; \\ \lb{s3_13}
\mfrak{M}(x,t)&=&  \int_{0}^{x}dy\:\:\overleftarrow{\exp}\Big\{\int_{y}^{x}d\xi\;\lambda_{i}^{{\scrscr(\hat{H}^{i})}}(\xi,t)\;
\overrightarrow{\boldsymbol{[}\hat{H}^{i}\:\boldsymbol{,}\:\ldots\boldsymbol{]_{-}}}\Big\}\:\times \\ \no &\times&
\bigg[\hat{H}^{j}\;\big(\pp_{t}\lambda_{j}^{{\scrscr(\hat{H}^{j})}}(y,t)\big)+\delta(y)\;
\Big(\hat{H}^{i}\;A_{i}^{(ini)}(y,t)+\hat{E}^{\alpha}\;B_{\alpha}^{(ini)}(y,t)\Big)\bigg]\;; \\ \lb{s3_14}
\mfrak{M}(x,t)&=&\hat{H}^{i}\;a_{i}\big(\phi_{\beta}(x,t);k_{j}\big)+
\hat{E}^{\alpha}\;b_{\alpha}\big(\phi_{\beta}(x,t);k_{j}\big)=
\hat{H}^{i}\;a_{i}(x,t)+\hat{E}^{\alpha}\;b_{\alpha}(x,t) \\ \no &=&
\hat{H}^{j}\;\Big(A_{j}^{(ini)}(x=0,t)+
\int_{0}^{x}dy\;\big(\pp_{t}\lambda_{j}^{{\scrscr(\hat{H}^{j})}}(y,t)\big)\Big) +   \\ \no &+&
\overleftarrow{\exp}\Big\{\int_{0}^{x}d\xi\;\lambda_{i}^{{\scrscr(\hat{H}^{i})}}(\xi,t)
\overrightarrow{\boldsymbol{[}\hat{H}^{i}\:\boldsymbol{,}\:\ldots\boldsymbol{]_{-}}}\Big\}\:
\hat{E}^{\alpha}\;B_{\alpha}^{(ini)}(x=0,t) \;.
\eeq
Recursive application of the exponential operators with the \(\mfrak{ad}\)-operator
{\small\(\overrightarrow{\boldsymbol{[}\hat{H}^{i}\:\boldsymbol{,}\:\ldots\boldsymbol{]_{-}}}\)} of the Cartan
sub-algebra in (\ref{s3_14}) specifies the form of the time-like Lax matrix \(\mfrak{M}(x,t)\)
by using the components \(\alpha^{j}\) of the
root vectors \(\vec{\alpha}\) of corresponding ladder operators \(\hat{E}_{\pm}^{\alpha}\)
\beq \lb{s3_15}
\lefteqn{\hspace*{-4.8cm}\mfrak{M}(x,t):=
\hat{H}^{i}\;a_{i}(x,t)+\hat{E}^{\alpha}\;b_{\alpha}(x,t)= }  \\ \no \lefteqn{\hspace*{-4.2cm}=
\hat{H}^{j}\;\Big(A_{j}^{(ini)}(x=0,t)+
\int_{0}^{x}dy\;\big(\pp_{t}\lambda_{j}^{{\scrscr(\hat{H}^{j})}}(y,t)\big)\Big) +
\sum_{\alpha\in\mbox{\scz root}}\overleftarrow{\exp}\Big\{\int_{0}^{x}d\xi\;\lambda_{i}^{{\scrscr(\hat{H}^{i})}}(\xi,t)\;\alpha^{i}\Big\}\:
\hat{E}^{\alpha}\;B_{\alpha}^{(ini)}(x=0,t) \;;} \\ \lb{s3_16}
a_{i}\big(\phi_{\beta}(x,t);k_{j}\big) &=&a_{i}(x,t)=A_{i}^{(ini)}(x=0,t)+
\int_{0}^{x}dy\;\big(\pp_{t}\lambda_{i}^{{\scrscr(\hat{H}^{i})}}(y,t)\big) \;; \\ \lb{s3_17}
b_{\alpha}\big(\phi_{\beta}(x,t);k_{j}\big)&=&b_{\alpha}(x,t)=B_{\alpha}^{(ini)}(x=0,t)\;\times\;
\overleftarrow{\exp}\Big\{\int_{0}^{x}d\xi\;\lambda_{i}^{{\scrscr(\hat{H}^{i})}}(\xi,t)\;\alpha^{i}\Big\} \;.
\eeq
We summarize the algebraic properties of the zero-curvature condition with following additional notes :\newline
If \(\mfrak{L}(x,t)\) belongs to the generators of a closed algebra with the physical fields being
the parameters or angles of the corresponding Lie group, one can generate \(\mfrak{M}(x,t)\) for the
zero-curvature condition by (\ref{s3_14},\ref{s3_15}) which then determines the nonlinear equations
(\ref{s3_16},\ref{s3_17}) as GP-type equations, or other types corresponding to the chosen algebra of
\(\mfrak{L}(x,t)\subseteq\mbox{sl}(n,\mathbb{C})\). One has to take into account that one does not obtain
an overdetermined system which finally results with the achieved matrix \(\mfrak{M}(x,t)\) (\ref{s3_14},\ref{s3_15})
and its chosen parametric dependence \(a_{i}(\phi_{\beta}(x,t);k_{j})\),
\(b_{\alpha}(\phi_{\beta}(x,t);k_{j})\) (\ref{s3_12})
into contradictory equations constraining the solutions from the zero-curvature condition to fixed time-
and spatial-terms without a physical time development.

\subsection{Gauge invariance of the Lax pair and the zero-curvature condition} \lb{s32}

It has already been stated that the Lax pair \(\mfrak{L}(x,t)\), \(\mfrak{M}(x,t)\) and its zero-curvature condition is by no means unique.
If one restricts to transformations (\ref{s3_18}) with a solely time dependent gauge matrix \(\mscr{G}_{0}=\mscr{G}(t)\), one can
immediately verify the invariance (\ref{s3_20}) of the zero-curvature relation under the solely time dependent gauge transformation (\ref{s3_19})
\beq  \lb{s3_18}
\mscr{G}_{0}:=\mscr{G}(t) &;& \big(\pp_{x}\mscr{G}_{0}\big)=\big(\pp_{x}\mscr{G}(t)\,\big)\equiv 0\;; \\  \lb{s3_19}
\mfrak{L}\rightarrow\mfrak{L}\ppr=\mscr{G}_{0}\;\mfrak{L}\;\mscr{G}_{0}^{-1} &;&
\mfrak{M}\rightarrow\mfrak{M}\ppr=\mscr{G}_{0}\;\mfrak{M}\;\mscr{G}_{0}^{-1}+
\big(\pp_{t}\mscr{G}_{0}\big)\:\mscr{G}_{0}^{-1} \;; \\   \lb{s3_20}
\big(\pp_{t}\mfrak{L}\big)-\big(\pp_{x}\mfrak{M}\big)+\boldsymbol{\big[}\mfrak{L}\;\boldsymbol{,}\;\mfrak{M}\boldsymbol{\big]_{-}}=0
&\Longleftrightarrow&
\big(\pp_{t}\mfrak{L}\ppr\big)-\big(\pp_{x}\mfrak{M}\ppr\big)+\boldsymbol{\big[}\mfrak{L}\ppr\;\boldsymbol{,}\;
\mfrak{M}\ppr\boldsymbol{\big]_{-}}=0 \;.
\eeq
The gauge invariance with \(\mscr{G}_{0}=\mscr{G}(x,t)\) (\ref{s3_21}-\ref{s3_28}) can also be extended to a general spacetime dependence
within the corresponding Lie group, following from the closed algebra \(\mfrak{L}(x,t)\), as e.\ g.\ in the Cartan-Weyl basis. As we begin
with the transformations (\ref{s3_23}-\ref{s3_25}) of \(\mscr{G}_{0}=\mscr{G}(x,t)\), acting onto the auxiliary field \(\Xi(x,t)\) and
Lax pair \(\mfrak{L}(x,t)\), \(\mfrak{M}(x,t)\), we attain the invariance of the Lax pair equations (\ref{s3_26}) and its
zero-curvature relation (\ref{s3_27}), provided that the Maurer-Cartan relation (\ref{s3_28}) is
fulfilled by the gauge matrix \(\mscr{G}_{0}=\mscr{G}(x,t)\)
\beq\no
\mbox{Gauge invariance }\;\mscr{G}_{0}&=&\mscr{G}(x,t)\; \mbox{ with Lie group of the closed algebra from }\;\mfrak{L}(x,t)\;; \\ \lb{s3_21}
 &&\big(\pp_{x}\Xi\big)=\mfrak{L}\;\Xi\;;\;\;\;\big(\pp_{t}\Xi\big)=\mfrak{M}\;\Xi\;; \\ \lb{s3_22} \Longrightarrow &&
\big(\pp_{t}\mfrak{L}\big)-\big(\pp_{x}\mfrak{M}\big)+
\boldsymbol{\big[}\mfrak{L}\;\boldsymbol{,}\;\mfrak{M}\boldsymbol{\big]_{-}}=0 \;; \\  \lb{s3_23}
\Xi &\rightarrow& \Xi\ppr=\mscr{G}_{0}\;\Xi \;; \\   \lb{s3_24}
\mfrak{L} &\rightarrow&\mfrak{L}\ppr=\mscr{G}_{0}\;\mfrak{L}\;\mscr{G}_{0}^{-1}+
\big(\pp_{x}\mscr{G}_{0}\big)\;\mscr{G}_{0}^{-1}\;; \\   \lb{s3_25}
\mfrak{M} &\rightarrow&\mfrak{M}\ppr=\mscr{G}_{0}\;\mfrak{M}\;\mscr{G}_{0}^{-1}+
\big(\pp_{t}\mscr{G}_{0}\big)\;\mscr{G}_{0}^{-1}\;; \\  \lb{s3_26}  \Longrightarrow &&
\big(\pp_{x}\Xi\ppr\big)=\mfrak{L}\ppr\;\Xi\ppr\;;\;\;\;\big(\pp_{t}\Xi\ppr\big)=\mfrak{M}\ppr\;\Xi\ppr\;; \\  \lb{s3_27}
\Longrightarrow &&  \big(\pp_{t}\mfrak{L}\ppr\big)-\big(\pp_{x}\mfrak{M}\ppr\big)+
\boldsymbol{\big[}\mfrak{L}\ppr\;\boldsymbol{,}\;\mfrak{M}\ppr\boldsymbol{\big]_{-}}=0 \;; \\ \lb{s3_28}  \mbox{provided that} &:&
\pp_{t}\big((\pp_{x}\mscr{G}_{0})\,\mscr{G}_{0}^{-1}\big)-\pp_{x}\big((\pp_{t}\mscr{G}_{0})\,\mscr{G}_{0}^{-1}\big)+
\boldsymbol{\big[}(\pp_{x}\mscr{G}_{0})\,\mscr{G}_{0}^{-1}\;\boldsymbol{,}\;(\pp_{t}\mscr{G}_{0})\,\mscr{G}_{0}^{-1}\boldsymbol{\big]_{-}}=0\;.
\eeq
On the condition of the general Maurer-Cartan relation (\ref{s3_28}), we can conclude for a
whole set of equivalent Lax matrices (\ref{s3_23}-\ref{s3_26}) and their zero-curvature relations (\ref{s3_27}). However, it is in general
even possible to transform the spatial Lax matrix \(\mfrak{L}(x,t)\) with a suitable gauge matrix \(\mscr{G}(x,t)\) to a diagonal
form \(\mfrak{L}_{{\scrscr\hat{H}}}(x,t)=\hat{H}^{i}\;\lambda_{i}^{{\scrscr(\hat{H}^{i})}}(x,t)\) constrained to the commuting
Cartan sub-algebra within a Cartan-Weyl basis of a Lie algebra. The diagonal form \(\mfrak{L}_{{\scrscr\hat{H}}}(x,t)\) of the
Lax matrix \(\mfrak{L}(x,t)\) has been used in section \ref{s31} to derive a general solution of the zero-curvature condition for
the corresponding set of physical, nonlinear equations. In the following we construct the equations for the gauge matrix \(\mscr{G}(x,t)\)
and its algebra \(\mfrak{g}^{j}\) with 'rotation' angles \(\vartheta_{j}(x,t)\) which transform the general Lax matrix \(\mfrak{L}(x,t)\)
to a diagonal form \(\mfrak{L}_{{\scrscr\hat{H}}}(x,t)=\hat{H}^{i}\;\lambda_{i}^{{\scrscr(\hat{H}^{i})}}(x,t)\)
\beq  \lb{s3_29}
\mfrak{L}_{{\scrscr\hat{H}}}(x,t) &=& \mscr{G}(x,t)\;\mfrak{L}(x,t)\;\mscr{G}^{-1}(x,t)+
\big(\pp_{x}\mscr{G}(x,t)\,\big)\;\mscr{G}^{-1}(x,t)\;; \\  \lb{s3_30}
\mfrak{g}(x,t) &=& \mfrak{g}^{j}\;\vartheta_{j}(x,t)\;;\;\;\;\mscr{G}(x,t)=\exp\big\{\mfrak{g}(x,t)\big\}\;.
\eeq
In order to simplify the Lie group current \((\pp_{x}\mscr{G}(x,t)\,)\;\mscr{G}^{-1}(x,t)\) in (\ref{s3_29}), we apply relation (\ref{s3_31})
which allows to resolve the current into the action of the \(\mfrak{ad}\)-operator
{\small\(\overrightarrow{\boldsymbol{[}\mfrak{g}(x,t)\;\boldsymbol{,}\;\ldots\boldsymbol{]_{-}}}\)} onto
\(\pp\mfrak{g}(x,t)\,/\,\pp x\) with the function \((e^{x}-1)/x\)
\beq \lb{s3_31}
\lefteqn{\hspace*{-1.0cm}\big(\pp_{x}\mscr{G}(x,t)\,\big)\;\mscr{G}^{-1}(x,t) = \Big(\pp_{x}\exp\big\{\mfrak{g}(x,t)\big\}\Big)\;
\exp\big\{-\mfrak{g}(x,t)\big\}= } \\ \no &=& \int_{0}^{1}dv\;\:\exp\big\{v\;\mfrak{g}(x,t)\big\}\;\;\frac{\pp\mfrak{g}(x,t)}{\pp x}\;\;
\exp\big\{-v\;\mfrak{g}(x,t)\big\} =
\int_{0}^{1}dv\;\:\exp\Big\{v\;\overrightarrow{\boldsymbol{[}\mfrak{g}(x,t)\;\boldsymbol{,}\;\ldots\boldsymbol{]_{-}}}\Big\}\;
\frac{\pp\mfrak{g}(x,t)}{\pp x} \\ \no &=&
\bigg(\frac{\exp\{\overrightarrow{\boldsymbol{[}\mfrak{g}(x,t)\;\boldsymbol{,}\;\ldots\boldsymbol{]_{-}}}\}-1}
{\overrightarrow{\boldsymbol{[}\mfrak{g}(x,t)\;\boldsymbol{,}\;\ldots\boldsymbol{]_{-}}}}\;\frac{\pp\mfrak{g}(x,t)}{\pp x}\bigg)\;.
\eeq
After insertion of (\ref{s3_31}) into (\ref{s3_29}), we achieve eq. (\ref{s3_32}) for the transformation of
\(\mfrak{L}(x,t)\rightarrow\mfrak{L}_{{\scrscr\hat{H}}}(x,t)\) and \(\pp\mfrak{g}(x,t)\,/\,\pp x\). Under the assumption of an invertible
function (\ref{s3_34}) for the \(\mfrak{ad}\)-operator
{\small\(\overrightarrow{\boldsymbol{[}\mfrak{g}(x,t)\;\boldsymbol{,}\;\ldots\boldsymbol{]_{-}}}\)}, one can finally derive a first
order differential equation (\ref{s3_33}) for the Lie algebra \(\mfrak{g}(x,t)=\mfrak{g}^{j}\;\vartheta_{j}(x,t)\) of the gauge matrix
\(\mscr{G}(x,t)\) so that the general spatial Lax matrix \(\mfrak{L}(x,t)\) can be converted to a diagonal form
\(\mfrak{L}_{{\scrscr\hat{H}}}(x,t)=\hat{H}^{i}\;\lambda_{i}^{{\scrscr(\hat{H}^{i})}}(x,t)\) with a suitable angular dependence
\(\vartheta_{j}(x,t)\) within the gauge transformation
\beq \lb{s3_32}
\mfrak{L}_{{\scrscr\hat{H}}}(x,t) &=&
\Big(\exp\big\{\overrightarrow{\boldsymbol{\big[}\mfrak{g}(x,t)\;\boldsymbol{,}\;\ldots
\boldsymbol{\big]_{-}}}\big\}\;\mfrak{L}(x,t)\Big) +
\bigg(\frac{\exp\{\overrightarrow{\boldsymbol{\big[}\mfrak{g}(x,t)\;\boldsymbol{,}\;\ldots\boldsymbol{\big]_{-}}}\}-1}
{\overrightarrow{\boldsymbol{\big[}\mfrak{g}(x,t)\;\boldsymbol{,}\;\ldots
\boldsymbol{\big]_{-}}}}\;\frac{\pp\mfrak{g}(x,t)}{\pp x}\bigg) \;; \\  \lb{s3_33}
\frac{\pp\mfrak{g}(x,t)}{\pp x} &=&\bigg[\bigg(\frac{\exp\{\overrightarrow{
\boldsymbol{\big[}\mfrak{g}(x,t)\;\boldsymbol{,}\;\ldots\boldsymbol{\big]_{-}}}\}-1}
{\overrightarrow{\boldsymbol{\big[}\mfrak{g}(x,t)\;\boldsymbol{,}\;\ldots
\boldsymbol{\big]_{-}}}}\bigg)^{\boldsymbol{-1}}
\Big[\mfrak{L}_{{\scrscr\hat{H}}}(x,t)-\Big(\exp\big\{\overrightarrow{
\boldsymbol{\big[}\mfrak{g}(x,t)\;\boldsymbol{,}\;\ldots
\boldsymbol{\big]_{-}}}\big\}\;\mfrak{L}(x,t)\Big)\Big] \bigg]\;;
\\ \lb{s3_34} \frac{e^{x}-1}{x} &\rightarrow& \Big(\frac{e^{x}-1}{x}\Big)^{-1} \;.
\eeq

\section{Determination and independence of conserved quantities}\lb{s4}

\subsection{Calculation of conserved quantities from the Lax pair}\lb{s41}

The Lax pair \(\mfrak{L}(x,t)\), \(\mfrak{M}(x,t)\) and its zero-curvature condition, which specifies the nonlinear equations of
physical fields, is accompanied by conserved, time-independent quantities. Let us consider the matrix \(\mfrak{T}(x,t;k)\) (\ref{s4_1})
or the monodromy matrix \(\mfrak{T}(t;k)\) (\ref{s4_2}) with periodic, spatial boundary conditions on a circle \(x\in[0,2\pi)\)
\beq \lb{s4_1}
\mfrak{T}(x,t;k) &=& \overleftarrow{\exp}\Big\{\int_{0}^{x}d\xi\;\mfrak{L}(\xi,t;k)\Big\}\;; \\ \lb{s4_2}
\mfrak{T}(t;k) &=& \overleftarrow{\exp}\Big\{\int_{0}^{2\pi}d\xi\;\mfrak{L}(\xi,t;k)\Big\}\;,
\eeq
then the supposition, that all fields are periodic in $x$ with period \(2\pi\), implies that traces of powers of the
monodromy matrix generate conserved quantities \(C^{(n)}(k)=C^{(n)}(t;k)\) independent of time
\be \lb{s4_3}
C^{(n)}(t;k)=\mbox{Tr}\big[\big(\mfrak{T}(t;k)\,\big)^{n}\big]\;.
\ee
In order to attest this statement, we outline the defining, spatial ordering (\ref{s4_4}) of the exponential with generator
\(\mfrak{L}(\xi,t;k)\) of the spatial Lax matrix
\beq \lb{s4_4}
\mfrak{T}(t;k)&=&\overleftarrow{\exp}\Big\{\int_{0}^{2\pi}d\xi\;\mfrak{L}(\xi,t;k)\Big\}   =
\exp\big\{{\scr \Delta x}\:\mfrak{L}(\xi=2\pi,t;k)\big\}\:\exp\big\{{\scr \Delta x}\:\mfrak{L}(\xi=2\pi-{\scr \Delta x},t;k)\big\}\:\ldots\times
\\  \no &\times& \ldots\:
\exp\big\{{\scr \Delta x}\:\mfrak{L}(\xi=2\,{\scr \Delta x},t;k)\big\}\:\exp\big\{{\scr \Delta x}\:\mfrak{L}(\xi={\scr \Delta x},t;k)\big\}\;.
\eeq
The time-like derivative \((\pp_{t}\mfrak{T}(t;k)\,)\) (\ref{s4_5}) of the monodromy matrix (\ref{s4_2}) involves the product rule
with \((\pp_{t}\mfrak{L}(y,t;k)\,)\) according to the spatial ordering of the exponentials (\ref{s4_4}). After substitution of
\((\pp_{t}\mfrak{L}(y,t;k)\,)\) by the zero-curvature condition, the integrand reduces to a total, spatial derivative
\(\pp_{y}(\ldots)\) (\ref{s4_5}) within the integration boundaries \(y\in[0,2\pi)\)
of a circle. Hence, we can perform the spatial \(dy\)-integration
along the circle and acquire a commutator (\ref{s4_7}) between the time-like Lax matrix \(\mfrak{M}(x=0,t;k)\) at the origin and the
monodromy matrix \(\mfrak{T}(t;k)\), due to the presupposed periodicity (\ref{s4_6}) on a spatial circle
\beq \lb{s4_5}
\big(\pp_{t}\mfrak{T}(t;k)\,\big)  &=&\int_{0}^{2\pi}dy\;
\exp\Big\{\int_{y}^{2\pi}d\xi_{2}\;\mfrak{L}(\xi_{2},t;k)\Big\}\;
\big(\pp_{t}\mfrak{L}(y,t;k)\big)\;\exp\Big\{\int_{0}^{y}d\xi_{1}\;\mfrak{L}(\xi_{1},t;k)\Big\}  \\ \no &=&
\int_{0}^{2\pi}dy\;\exp\Big\{\int_{y}^{2\pi}d\xi_{2}\;\mfrak{L}(\xi_{2},t;k)\Big\}\;
\Big[\big(\pp_{y}\mfrak{M}(y,t;k)\big)+
\boldsymbol{\big[}\mfrak{M}(y,t;k)\:\boldsymbol{,}\:\mfrak{L}(y,t;k)\boldsymbol{\big]_{-}}\Big]\;\times \\ \no &\times&
\exp\Big\{\int_{0}^{y}d\xi_{1}\;\mfrak{L}(\xi_{1},t;k)\Big\}  \\ \no &=&
\int_{0}^{2\pi}dy\;\pp_{y}\bigg(\exp\Big\{\int_{y}^{2\pi}d\xi_{2}\;\mfrak{L}(\xi_{2},t;k)\Big\}\;\mfrak{M}(y,t;k)\;
\exp\Big\{\int_{0}^{y}d\xi_{1}\;\mfrak{L}(\xi_{1},t;k)\Big\} \bigg)  \\ \no &=&
\exp\Big\{\int_{y}^{2\pi}d\xi_{2}\;\mfrak{L}(\xi_{2},t;k)\Big\}\;\mfrak{M}(y,t;k)\;
\exp\Big\{\int_{0}^{y}d\xi_{1}\;\mfrak{L}(\xi_{1},t;k)\Big\}\bigg|_{y=0}^{y=2\pi}   \\ \no &=&
\mfrak{M}(x=2\pi,t;k)\;\mfrak{T}(t;k)-\mfrak{T}(t;k)\;\mfrak{M}(x=0,t;k) \;;  \\ \lb{s4_6}
\mbox{circle} &:&\Longrightarrow\;\mfrak{M}(x=2\pi,t;k)=\mfrak{M}(x=0,t;k)\;;  \\ \lb{s4_7}
\big(\pp_{t}\mfrak{T}(t;k)\,\big) &=&\boldsymbol{\big[}\mfrak{M}(x=0,t;k)\;\boldsymbol{,}\;
\mfrak{T}(t;k)\boldsymbol{\big]_{-}}\;.
\eeq
The conversion (\ref{s4_7}) of the time-like derivative \((\pp_{t}\mfrak{T}(t;k)\,)\) to a commutator allows to demonstrate
the time independence of traces of arbitrary powers of the monodromy matrix \(\mfrak{T}(t;k)\) (\ref{s4_2})
\beq\no
\pp_{t}\mbox{Tr}\Big[\big(\mfrak{T}(t;k)\,\big)^{n}\Big] &=& n\;\mbox{Tr}\Big[\big(\pp_{t}\mfrak{T}(t;k)\,\big)\;
\big(\mfrak{T}(t;k)\,\big)^{n-1}\Big] = n\;\mbox{Tr}\Big[\boldsymbol{\big[}\mfrak{M}(x=0,t;k)\;\boldsymbol{,}
\;\mfrak{T}(t;k)\boldsymbol{\big]_{-}}\;\big(\mfrak{T}(t;k)\,\big)^{n-1}\Big]\equiv0 \;;  \\ \lb{s4_8}
\pp_{t}C^{(n)}(t;k) &=& \pp_{t}\mbox{Tr}\Big[\big(\mfrak{T}(t;k)\,\big)^{n}\Big] \equiv 0\;;\;\;\;
\Longrightarrow \;C^{(n)}(t;k)=C^{(n)}(k)\;.
\eeq
Therefore, the general time independence of \(C^{(n)}(t;k)=\mbox{Tr}[(\mfrak{T}(t;k)\,)^{n}]\) is confirmed for arbitrary powers
\(n\in\mathbb{N}\) so that the \(C^{(n)}(k)=C^{(n)}(t;k)\) (\ref{s4_8}) have to be regarded as the conserved quantities within
a Liouville integrability  of the corresponding nonlinear equations for the physical fields. Instead of the label '$n$' for powers
of \(\mfrak{T}(t;k)\) (\ref{s4_2}), one can also conduct a power series expansion with the spectral parameters \(k_{j}\) in order
to generate conserved quantities from (orthogonal) polynomials of \(k_{j}\). Note that in the described cases of
the (1+1) GP-equations for the attractive and repulsive interactions with the \(\mbox{su}(2)\) and \(\mbox{sp}(2,\mathbb{R})\) algebra, respectively,
\be \lb{s4_9}
\mfrak{L}_{\mbox{\scz su}(2)}(\mscr{Q},\mscr{P};k) = \bigg(\bea{cc} -\im\,k & -\psi^{*} \\ \psi & \im\,k \eea\bigg)\;;\;\;\;(k\in\mathbb{R})\;;\;\;\;
\mfrak{L}_{\mbox{\scz sp}(2,\mathbb{R})}(\mscr{Q},\mscr{P};k) = \bigg(\bea{cc} -k & \psi^{*} \\ \psi & k \eea\bigg)\;;\;\;\;(k\in\mathbb{R})\;.
\ee
the spatial Lax matrices essentially stay unaltered as one introduces the external potential \(V(x,t)\). One can even choose a completely
constant spectral parameter \(k(x,t):=k_{0}\) so that the conserved quantities \(C^{(n)}(t;k)\) are not affected by the inclusion
of an external potential \(V(x,t)\) into the GP-equations
with a solely attractive ('\(\mbox{su}(2)\)') or repulsive ('\(\mbox{sp}(2,\mathbb{R})\)') interaction. Therefore, the independence
of conserved quantities \(C^{(n)}(t;k)\) can be directly transferred from the well-known case of the integrable, (1+1) GP-equations
without any external potential to the more general GP-types with an arbitrary external potential \(V(x,t)\).

\subsection{Involution of conserved quantities and the classical $\boldsymbol{\mfrak{r}}$-matrix}  \lb{s42}

In order to prove the independence of the derived, conserved quantities \(C^{(n)}(t;k)\) of previous section \ref{s41}, one has to verify
the involution from the Poisson brackets of the canonical fields (\ref{s4_10}).
These have to be taken into account because the conserved quantities
\(C^{(n)}(t;k)\) are generators for first order evolution equations with corresponding parameters \(t_{n}\) (\ref{s4_11})
and have furthermore to fulfill the Jacobi identity (\ref{s4_12},\ref{s4_13}), following from the Poisson brackets of canonical fields
\(\phi_{\alpha}(x,t)=(\mscr{Q}_{\alpha}(x,t)\:,\:\mscr{P}_{\alpha}(x,t)\,)\)
\beq \lb{s4_10}
0&\stackrel{!}{=}&
\boldsymbol{\big\{}C^{(n_{1})}(t;k)\:\boldsymbol{,}\:
C^{(n_{2})}(t;k)\boldsymbol{\big\}}\;;\;\;\;(n_{1},\,n_{2}\in\mathbb{N})\;; \\  \lb{s4_11}
\frac{\pp}{\pp t_{n_{1}}}\ldots &=& \boldsymbol{\big\{}C^{(n_{1})}(t;k)\:\boldsymbol{,}\:\ldots\boldsymbol{\big\}} \;; \\ \lb{s4_12}
0 &=& \boldsymbol{\big\{}C^{(n_{1})}(t;k)\:\boldsymbol{,}\:
\boldsymbol{\big\{}C^{(n_{2})}(t;k)\:\boldsymbol{,}\:\ldots\boldsymbol{\big\}}\boldsymbol{\big\}}+ \\ \no &+&
\boldsymbol{\big\{}C^{(n_{2})}(t;k)\:\boldsymbol{,}\:\boldsymbol{\big\{}\ldots\:\boldsymbol{,}\:
 C^{(n_{1})}(t;k)\boldsymbol{\big\}}\boldsymbol{\big\}}+
\boldsymbol{\big\{}\ldots\:\boldsymbol{,}\:
\boldsymbol{\big\{}C^{(n_{1})}(t;k)\:\boldsymbol{,}\:\boldsymbol{\big\{}C^{(n_{2})}(t;k)\boldsymbol{\big\}}\boldsymbol{\big\}} \;; \\ \lb{s4_13}
\Longleftrightarrow 0 &=& \underbrace{\bigg(\frac{\pp}{\pp t_{n_{1}}}\frac{\pp}{\pp t_{n_{2}}}\ldots-
\frac{\pp}{\pp t_{n_{2}}}\frac{\pp}{\pp t_{n_{1}}}\ldots\bigg)}_{\stackrel{!}{=}0} \; -\;
\boldsymbol{\big\{}\underbrace{\boldsymbol{\big\{}C^{(n_{1})}(t;k)\:\boldsymbol{,}\:
C^{(n_{2})}(t;k)\boldsymbol{\big\}}}_{\Longrightarrow =0\;!}
\:\boldsymbol{,}\:\ldots\boldsymbol{\big\}}\;.
\eeq
The involution of traces of powers of the monodromy matrix
\be  \lb{s4_14}
\boldsymbol{\big\{}C^{(n_{1})}(t;k)\:\boldsymbol{,}\:C^{(n_{2})}(t;k)\boldsymbol{\big\}}=0\;,
\ee
is usually investigated by the so-called \(\mfrak{r}\)-matrix approach where one assumes  the validity of a '{\it fundamental
Poisson bracket relation}' (\ref{s4_15}) between tensor products \(\mfrak{L}_{1}(x_{1},t;k_{1}) \), \(\mfrak{L}_{2}(x_{2},t;k_{2}) \)
of spatial Lax matrices (cf. appendix A in \cite{Chowd1} and chap.\ 2.5 in \cite{Manton})
\beq \lb{s4_15}
\boldsymbol{\big\{}\mfrak{L}_{1}(\xi_{1},t;k_{1})\;\stackrel{\otimes}{\boldsymbol{,}}\;\mfrak{L}_{2}(\xi_{2},t;k_{2})\boldsymbol{\big\}}& =&
\boldsymbol{\big[}\mfrak{r}_{12}(k_{1},k_{2};t)\;\boldsymbol{,}\;\mfrak{L}_{1}(x_{1},t;k_{1})+\mfrak{L}_{2}(x_{2},t;k_{2})
\,\boldsymbol{\big]_{-}}\;\delta(\xi_{1}-\xi_{2})\;; \\ \no \mfrak{r}_{12}(k_{1},k_{2};t) &=&-\mfrak{r}_{21}(k_{2},k_{1};t) \;.
\eeq
The Poisson bracket \(\boldsymbol{\{}\mfrak{L}_{1}(\xi_{1},t;k_{1})\;
\stackrel{\otimes}{\boldsymbol{,}}\;\mfrak{L}_{2}(\xi_{2},t;k_{2})\boldsymbol{\}}\) (\ref{s4_15}) is replaced by the commutator with the
\(\mfrak{r}_{12}(k_{1},k_{2};t) \)-matrix where we further require the '{\it ultralocal}' form with the spatial delta function \(\delta(\xi_{1}-\xi_{2})\).
(The 'ultralocal' condition is in general obtained from spatial Lax matrices which only depend on the physical fields
\(\mscr{Q}_{\alpha}(x,t)\), \(\mscr{P}_{\alpha}(x,t)\) without any derivatives of these. This condition is hence fulfilled for the (1+1) GP-equations.)
As we consider more general types of \(\mfrak{T}(x,y;t;k)\) matrices (\ref{s4_16}) with initial coordinate $y$ and end point $x$
for the monodromy matrix (\ref{s4_2}) instead of \(\mfrak{T}(x;t;k)\) (\ref{s4_1})
\be \lb{s4_16}
\mfrak{T}(x,y;t;k) = \overleftarrow{\exp}\Big\{\int_{y}^{x}d\xi\;\mfrak{L}(\xi,t;k)\Big\}\;,
\ee
we can start out from the general, tensorial Poisson bracket relation (\ref{s4_17})
whose right-hand side results from the validity of the Leibniz product rule
for the Poisson bracket operation \(\boldsymbol{\{}\ldots\;\stackrel{\otimes}{\boldsymbol{,}}\;\ldots\boldsymbol{\}}\)
\beq  \lb{s4_17}
\big\{\mfrak{T}_{1}(x_{1},y_{1};t;k_{1})\;\stackrel{\otimes}{\boldsymbol{,}}\;
\mfrak{T}_{2}(x_{2},y_{2};t;k_{2})\big\} &=&
\int_{y_{1}}^{x_{1}}d\xi_{1}\int_{y_{2}}^{x_{2}}d\xi_{2}\;\;\mfrak{T}_{1}(x_{1},\xi_{1};t;k_{1})\;\;
\mfrak{T}_{2}(x_{2},\xi_{2};t;k_{2})\;\;\times  \\ \no &\times&
\boldsymbol{\big\{}\mfrak{L}_{1}(\xi_{1},t;k_{1})\;\stackrel{\otimes}{\boldsymbol{,}}\;
\mfrak{L}_{2}(\xi_{2},t;k_{2})\boldsymbol{\big\}}\;\;
\mfrak{T}_{1}(\xi_{1},y_{1};t;k_{1})\;\;\mfrak{T}_{2}(\xi_{2},y_{2};t;k_{2})\;.
\eeq
In appendix \ref{sb}, we demonstrate according to Ref. \cite{Manton} how to achieve the involution
\(\boldsymbol{\{}C^{(n_{1})}(t;k)\:\boldsymbol{,}\:C^{(n_{2})}(t;k)\boldsymbol{\}}=0\)
from the generally valid relation (\ref{s4_17}) of extended
monodromy matrices \(\mfrak{T}(x,y;t;k)\)  under the special assumption of the 'fundamental Poisson bracket relation' (\ref{s4_15}) with
a {\it spatially constant} \(\mfrak{r}_{12}(k_{1},k_{2};t)\)-matrix. Since the spatial Lax matrices
\(\mfrak{L}_{\mbox{\scz su}(2)}(\mscr{Q},\mscr{P};k)\), \(\mfrak{L}_{\mbox{\scz sp}(2,\mathbb{R})}(\mscr{Q},\mscr{P};k)\)
of the integrable, (1+1) GP-equations do not change under the inclusion of an arbitrary external potential \(V(x,t)\)
(apart from a possibly chosen spacetime dependence of the spectral parameter \(k\rightarrow k(x,t)\)), the whole derivation of the involution
of conserved quantities can be directly conveyed from the case without an external potential to the case with an arbitrary potential \(V(x,t)\).

The given derivation of the involution property (\ref{s4_10})
of conserved quantities \(C^{(n)}(t;k)\) in appendix \ref{sb} depends on the assumed fundamental Poisson
bracket relation (\ref{s4_15})
with a {\it spatially constant} \(\mfrak{r}_{12}(k_{1},k_{2};t)\)-matrix. In the following we attain more general
statements as we begin from relation (\ref{s4_17}),
caused by the Leibniz product rule of the tensorial Poisson brackets, and apply the gauge invariance of the
Lax matrices which allows to transform to the diagonal, commuting Cartan sub-algebra.
Further transformations of above relation (\ref{s4_17})
rely on simplification of the tensorial Poisson bracket \(\boldsymbol{\{}\ldots\;\stackrel{\otimes}{\boldsymbol{,}}\;\ldots\boldsymbol{\}}\)
of the spatial Lax matrices \(\mfrak{L}_{1}(x_{1},t;k_{1})\), \(\mfrak{L}_{2}(x_{2},t;k_{2})\). As we suppose a closed Lie algebra
for \(\mfrak{L}(x,t;k)\), we can diagonalize or parametrize to \(\hat{\Lambda}(x,t;k)\) (of  the commuting Cartan sub-algebra)
with the invertible 'eigenvector' matrix \(\mfrak{U}(x,t;k)\) which also contains the ladder operators \(\hat{E}_{\pm}^{\alpha}\). This kind of
parameters is taken within each part of the tensor product space \(\mfrak{L}_{1}(x_{1},t;k_{1})\), \(\mfrak{L}_{2}(x_{2},t;k_{2})\)
of the spatial Lax matrix \(\mfrak{L}(x,t;k)\)
\beq \lb{s4_18}
\mfrak{L}(x,t;k) &=& \mfrak{U}(x,t;k)\;\hat{\Lambda}(x,t;k)\;\mfrak{U}^{-1}(x,t;k) \;; \;\;\;
\hat{\Lambda}(x,t;k) = \hat{H}^{i}\;\lambda_{i}^{{\scrscr(\hat{H}^{i})}}(x,t;k)\;; \\  \lb{s4_19}
\mfrak{L}_{1}(x_{1},t;k_{1}) &=& \mfrak{U}_{1}(x_{1},t;k_{1})\;\hat{\Lambda}_{1}(x_{1},t;k_{1})\;
\mfrak{U}_{1}^{-1}(x_{1},t;k_{1}) \;; \\ \lb{s4_20}
\mfrak{L}_{2}(x_{2},t;k_{2}) &=& \mfrak{U}_{2}(x_{2},t;k_{2})\;\hat{\Lambda}_{2}(x_{2},t;k_{2})\;\mfrak{U}_{2}^{-1}(x_{2},t;k_{2}) \;.
\eeq
On the condition of the "ultralocal" case of spatial Lax matrices, one can transfer the tensorial Poisson bracket (\ref{s4_21})
\(\boldsymbol{\{}\ldots\;\stackrel{\otimes}{\boldsymbol{,}}\;\ldots\boldsymbol{\}}\) of Lax matrices
\(\mfrak{L}_{1}(x_{1},t;k_{1})\), \(\mfrak{L}_{2}(x_{2},t;k_{2}) \) (\ref{s4_19},\ref{s4_20})
to terms with their eigenvalues \(\hat{\Lambda}_{1}(x_{1},t;k_{1})\),
\(\hat{\Lambda}_{2}(x_{2},t;k_{2})\) and 'eigenvector' matrices \(\mfrak{U}_{1}(x_{1},t;k_{1})\),
\(\mfrak{U}_{2}(x_{2},t;k_{2})\) and a further part consisting of general spacetime dependent
\(\mfrak{r}_{12}(x_{1},k_{1};x_{2},k_{2};t)\)-, \(\mfrak{r}_{21}(x_{2},k_{2};x_{1},k_{1};t)\)-matrices within commutators of
\(\mfrak{L}_{1}(x_{1},t;k_{1})\), \(\mfrak{L}_{2}(x_{2},t;k_{2})\), respectively. The
\(\mfrak{r}_{12}(x_{1},k_{1};x_{2},k_{2};t)\)-, \(\mfrak{r}_{21}(x_{2},k_{2};x_{1},k_{1};t)\)-matrices (\ref{s4_22})
are composed of further parts \(\mfrak{K}_{12}\), \(\mfrak{O}_{12} \) (\ref{s4_23},\ref{s4_24})
 which contain terms with the Poisson brackets among the eigenvector matrices
\(\mfrak{U}_{1}(x_{1},t;k_{1})\), \(\mfrak{U}_{2}(x_{2},t;k_{2})\) and between the eigenvector matrix
\(\mfrak{U}_{1}(x_{1},t;k_{1})\) and the diagonal Cartan sub-algebra part \(\hat{\Lambda}_{2}(x_{2},t;k_{2})\).
In summary the basic nine terms in (\ref{s4_21}) are achieved by the Leibniz product rule of the tensorial
Poisson bracket of canonical fields and can be grouped into a first part with the Poisson bracket of the eigenvalues
\(\boldsymbol{\{}\hat{\Lambda}_{1}(x_{1},t;k_{1}) \;\stackrel{\otimes}{\boldsymbol{,}}\;\hat{\Lambda}_{2}(x_{2},t;k_{2})\boldsymbol{\}}\)
and a second part with commutators between the \(\mfrak{r}_{12}(x_{1},k_{1};x_{2},k_{2};t)\)-, \(\mfrak{r}_{21}(x_{2},k_{2};x_{1},k_{1};t)\)-matrices and the corresponding spatial Lax matrices \(\mfrak{L}_{1}(x_{1},t;k_{1})\), \(\mfrak{L}_{2}(x_{2},t;k_{2}) \)
(in the ultralocal case '\(\delta(x_{1}-x_{2})\)')
\beq \lb{s4_21}
\lefteqn{\boldsymbol{\big\{}\mfrak{L}_{1}(x_{1},t;k_{1})\;\stackrel{\otimes}{\boldsymbol{,}}\;
\mfrak{L}_{2}(x_{2},t;k_{2})\boldsymbol{\big\}} =} \\ \no &=&
\mfrak{U}_{1}(x_{1},t;k_{1})\:\mfrak{U}_{2}(x_{2},t;k_{2})\;\boldsymbol{\big\{}\hat{\Lambda}_{1}(x_{1},t;k_{1}) \;
\stackrel{\otimes}{\boldsymbol{,}}\;\hat{\Lambda}_{2}(x_{2},t;k_{2})\boldsymbol{\big\}}\;\mfrak{U}_{1}^{-1}(x_{1},t;k_{1})\;
\mfrak{U}_{2}^{-1}(x_{2},t;k_{2}) + \\ \no &+&
\Big(\boldsymbol{\big[}\mfrak{r}_{12}(x_{1},k_{1};x_{2},k_{2};t)\;\boldsymbol{,}\;
\mfrak{L}_{1}(x_{1},t;k_{1})\,\boldsymbol{\big]_{-}}-
\boldsymbol{\big[}\mfrak{r}_{21}(x_{2},k_{2};x_{1},k_{1};t)\;\boldsymbol{,}\;
\mfrak{L}_{2}(x_{2},t;k_{2})\,\boldsymbol{\big]_{-}}\Big)\;\delta(x_{1}-x_{2})\;; \\  \lb{s4_22}
\lefteqn{\mfrak{r}_{12}(x_{1},k_{1};x_{2},k_{2};t) = \mfrak{O}_{12} +\frac{1}{2}\;\boldsymbol{\big[}\mfrak{K}_{12}\;\boldsymbol{,}\;
\mfrak{L}_{2}(x_{2},t;k_{2})\,\boldsymbol{\big]_{-}} \;;}   \\ \lb{s4_23} \mfrak{K}_{12} &=&
\boldsymbol{\big\{} \mfrak{U}_{1}(x_{1},t;k_{1}) \;\stackrel{\otimes}{\boldsymbol{,}}\;
\mfrak{U}_{2}(x_{2},t;k_{2}) \boldsymbol{\big\}}\;\mfrak{U}_{1}^{-1}(x_{1},t;k_{1}) \;
\mfrak{U}_{2}^{-1}(x_{2},t;k_{2}) \;;  \\  \lb{s4_24}
\mfrak{O}_{12}  &=& \mfrak{U}_{2}(x_{2},t;k_{2})
\boldsymbol{\big\{} \mfrak{U}_{1}(x_{1},t;k_{1}) \;\stackrel{\otimes}{\boldsymbol{,}}\;
\hat{\Lambda}_{2}(x_{2},t;k_{2}) \boldsymbol{\big\}}\;
\mfrak{U}_{1}^{-1}(x_{1},t;k_{1}) \; \mfrak{U}_{2}^{-1}(x_{2},t;k_{2}) \;.
\eeq
The (spatially constant !) \(\mfrak{r}\)-matrix approach, given in appendix \ref{sb}, only allows to conclude for the involution of the
conserved quantities \(C^{(n)}(t;k)\), provided that the eigenvalues
\(\hat{\Lambda}_{1}(x_{1},t;k_{1})\), \(\hat{\Lambda}_{2}(x_{2},t;k_{2})\)
of the spatial Lax matrices are in involution (or their tensorial Poisson
brackets vanish completely) (\ref{s4_25}). This can be accomplished for a symmetrical dependence on canonical fields
\(\phi_{\alpha}(x,t)=(\mscr{Q}_{\alpha}(x,t)\:,\:\mscr{P}_{\alpha}(x,t)\,)\) within the eigenvalues
\(\lambda_{i}^{{\scrscr(\hat{H}^{i})}}(\mscr{Q}_{\alpha},\mscr{P}_{\alpha};k)\stackrel{!}{=}\lambda_{i}^{{\scrscr(\hat{H}^{i})}}(\mscr{P}_{\alpha},\mscr{Q}_{\alpha};k)\)
which may originate from symmetrically chosen parameters \(\phi_{\alpha}(x,t)=(\mscr{Q}_{\alpha}(x,t)\:,\:\mscr{P}_{\alpha}(x,t)\,)\)
within the original Lax matrix \(\mfrak{L}(\mscr{Q}_{\alpha},\mscr{P}_{\alpha};k_{j})\) of a closed algebra determined by
a Cartan-Weyl basis
of ladder operators (e.\ g.\ \(\phi_{\alpha}\;E^{\alpha}=\mscr{Q}_{\alpha}\;E_{-}^{\alpha}+\mscr{P}_{\alpha}\;E_{+}^{\alpha}\))
\beq \lb{s4_25}
\lefteqn{\boldsymbol{\big\{}\hat{\Lambda}_{1}(x_{1},t;k_{1}) \;
\stackrel{\otimes}{\boldsymbol{,}}\;
\hat{\Lambda}_{2}(x_{2},t;k_{2})\boldsymbol{\big\}} = 0\;;\;\;\;\Longrightarrow \Big(\mbox{symmetric dependence on} \;}   \\ \no &&\hspace*{-0.6cm}
\mbox{canonical fields } \; \phi_{\alpha}=(\mscr{Q}_{\alpha}(x,t)\:,\:\mscr{P}_{\alpha}(x,t)\,)\;
\mbox{ as e.g. }\;\lambda_{j}^{{\scrscr(\hat{H}^{j})}}(x,t)=\sqrt{k_{j}\:k^{j}+\mscr{Q}_{\alpha}\cdot \mscr{P}^{\alpha}}\Big)\;.
\eeq
Under the assumption of a symmetrical dependence on the canonical fields \(\mscr{Q}_{\alpha}(x,t)\:,\:\mscr{P}_{\alpha}(x,t)\) (\ref{s4_25}),
the general tensorial Poisson bracket (\ref{s4_21}) reduces to the commutator (\ref{s4_26}) between the
\(\mfrak{r}_{12}(x_{1},k_{1};x_{2},k_{2};t)\)-, \(\mfrak{r}_{21}(x_{2},k_{2};x_{1},k_{1};t)\)-matrices
and the spatial Lax matrices \(\mfrak{L}_{1}(x_{1},t;k_{1})\), \(\mfrak{L}_{2}(x_{2},t;k_{2})\)
\beq \lb{s4_26}
\lefteqn{\Longrightarrow\;
\boldsymbol{\big\{}\mfrak{L}_{1}(x_{1},t;k_{1})\;\stackrel{\otimes}{\boldsymbol{,}}\;
\mfrak{L}_{2}(x_{2},t;k_{2})\boldsymbol{\big\}} = }  \\ \no &=&
\Big(\boldsymbol{\big[}\mfrak{r}_{12}(x_{1},k_{1};x_{2},k_{2};t)\;\boldsymbol{,}\;
\mfrak{L}_{1}(x_{1},t;k_{1})\,\boldsymbol{\big]_{-}}-
\boldsymbol{\big[}\mfrak{r}_{21}(x_{2},k_{2};x_{1},k_{1};t)\;\boldsymbol{,}\;
\mfrak{L}_{2}(x_{2},t;k_{2})\,\boldsymbol{\big]_{-}}\Big)\;\delta(x_{1}-x_{2})\;.
\eeq
Relation (\ref{s4_26}) is similar to the already assumed fundamental Poisson bracket eq. (\ref{s4_15}); however, the given
\(\mfrak{r}_{12}(x_{1},k_{1};x_{2},k_{2};t)\)-, \(\mfrak{r}_{21}(x_{2},k_{2};x_{1},k_{1};t)\)-matrices  have a priori no special symmetries,
neither anti-symmetric nor symmetric. In order to be applicable for the derivation in appendix \ref{sb}, we transform by a commutator
between the spatial Lax matrix \(\mfrak{L}_{1}(x_{1},t;k_{1})\), \(\mfrak{L}_{2}(x_{2},t;k_{2})\) and an additional
symmetric matrix \(\hat{\sigma}_{12}=\hat{\sigma}_{21}\) (\ref{s4_27}-\ref{s4_29})
which retains eq. (\ref{s4_26}) invariant. This allows to assign the anti-symmetry to the
transformed \(\mfrak{r}_{12}=-\mfrak{r}_{21}\) matrix (\ref{s4_30},\ref{s4_31})
so that the derivation of appendix \ref{sb} can be performed with the further
restriction to a {\it spatially constant} \(\mfrak{r}_{12}(k_{1},k_{2};t)=-\mfrak{r}_{21}(k_{2},k_{1};t)\) matrix (\ref{s4_32})
\beq \lb{s4_27}
\mfrak{r}_{12}(x_{1},k_{1};x_{2},k_{2};t) &\rightarrow&\mfrak{r}_{12}(x_{1},k_{1};x_{2},k_{2};t)+
\boldsymbol{\big[}\hat{\sigma}_{12}(x_{1},k_{1};x_{2},k_{2};t)\;\boldsymbol{,}\;
\mfrak{L}_{2}(x_{2},t;k_{2})\boldsymbol{\big]_{-}} \;;
\\  \lb{s4_28} \mfrak{r}_{21}(x_{2},k_{2};x_{1},k_{1};t) &\rightarrow&\mfrak{r}_{21}(x_{2},k_{2};x_{1},k_{1};t)+
\boldsymbol{\big[}\hat{\sigma}_{21}(x_{2},k_{2};x_{1},k_{1};t)\;\boldsymbol{,}\;
\mfrak{L}_{1}(x_{1},t;k_{1})\boldsymbol{\big]_{-}} \;; \\ \lb{s4_29}
\hat{\sigma}_{12}(x_{1},k_{1};x_{2},k_{2};t) &=& \hat{\sigma}_{21}(x_{2},k_{2};x_{1},k_{1};t)\;;\;\;\;
(\mbox{symmetric matrix !})\;;
\eeq
\beq\lb{s4_30}
\mfrak{r}_{12}(\sharp 1;\sharp 2;t)+
\boldsymbol{\big[}\hat{\sigma}_{12}(\sharp 1;\sharp 2;t)\;\boldsymbol{,}\;\mfrak{L}_{2}(x_{2},t;k_{2})\boldsymbol{\big]_{-}}
 &=&\!\!\!-\Big(\mfrak{r}_{21}(\sharp 2;\sharp 1;t)+
\boldsymbol{\big[}\hat{\sigma}_{21}(\sharp 2;\sharp 1;t)\;\boldsymbol{,}\;\mfrak{L}_{1}(x_{1},t;k_{1})\boldsymbol{\big]_{-}}\Big);
\\ \lb{s4_31} 0\neq \mfrak{r}_{12}(\sharp 1;\sharp 2;t) + \mfrak{r}_{21}(\sharp 2;\sharp 1;t) &=&\!\!\!
-\boldsymbol{\big[}\hat{\sigma}_{12}(\sharp 1;\sharp 2;t)\;\boldsymbol{,}\;\mfrak{L}_{1}(x_{1},t;k_{1}) +
\mfrak{L}_{2}(x_{2},t;k_{2})\boldsymbol{\big]_{-}} \;; \\  \lb{s4_32}
\mbox{Problem of spatial dependence of} &:&
\mfrak{r}_{12}(x_{1},k_{1};x_{2},k_{2};t)\rightarrow
\mfrak{r}_{12}(k_{1},k_{2};t)\;.
\eeq
Apart from the involution of conserved quantities according to the so-called
\(\mfrak{r}_{12}\)-matrix approach in appendix \ref{sb} and \cite{Manton},
we also suggest a different proof which relies on a gauge transformation of the spatial Lax matrix \(\mfrak{L}(x,t;k)\) to a completely
diagonal form \(\mfrak{L}_{{\scrscr \hat{H}}}(x,t;k)=\hat{\Lambda}(x,t;k)\)
given in section \ref{s32} and (\ref{s4_38},\ref{s4_39}). This results in vanishing
\(\mfrak{r}_{12}\)-, \(\mfrak{r}_{21}\)-matrices (\ref{s4_33},\ref{s4_34})
within the basic tensorial Poisson bracket relation (\ref{s4_21}-\ref{s4_24}) so that the Poisson bracket
of the diagonal Cartan sub-algebra matrices \(\mfrak{L}_{{\scrscr \hat{H}},1}(x,t;k)=\hat{\Lambda}_{1}(x,t;k)\),
\(\mfrak{L}_{{\scrscr \hat{H}},2}(x,t;k)=\hat{\Lambda}_{2}(x,t;k)\) only remain
because the diagonalizing matrices \(\mfrak{U}_{1}(x_{2},t;k_{2})\),
\( \mfrak{U}_{2}(x_{1},t;k_{1})\) reduce to tensorial unit matrices (\ref{s4_35}).
We remark again that \(\mfrak{L}(x,t;k)\) and its gauge
transformed, diagonal form \(\mfrak{L}_{{\scrscr \hat{H}}}(x,t;k)=\hat{\Lambda}(x,t;k)\) represent the same nonlinear equations of fields
\(\phi_{\alpha}(x,t)=(\mscr{Q}_{\alpha}(x,t)\:,\:\mscr{P}_{\alpha}(x,t)\,)\) within the general \(\mbox{sl}(n,\mathbb{C})\) algebra or within one of
its sub-algebras as e.\ g.\ \(\mbox{su}(n)\)
\beq \lb{s4_33}
\mbox{if }\;\mfrak{L}(x,t;k) &\rightarrow& \mfrak{L}_{{\scrscr\hat{H}}}(x,t;k)=\hat{\Lambda}(x,t;k)\;\mbox{by a gauge transformation} \\  \lb{s4_34}
\Longrightarrow\;\mfrak{r}_{12}(\sharp 1;\sharp 2;t)\equiv0  &\wedge&
\mfrak{r}_{21}(\sharp 2;\sharp 1;t)\equiv0\;; \\  \lb{s4_35}
\lefteqn{\hspace*{-2.0cm}\Longrightarrow \;
\big\{\mfrak{L}_{1}(x_{1},t;k_{1})\;\stackrel{\otimes}{\boldsymbol{,}}\;\mfrak{L}_{2}(x_{2},t;k_{2})\big\} \;
\Longrightarrow }  \\ \no &\Longrightarrow&
\overbrace{\boldsymbol{\big\{}\mfrak{L}_{{\scrscr \hat{H}},1}(x_{1},t;k_{1})\;\stackrel{\otimes}{\boldsymbol{,}}\;
\mfrak{L}_{{\scrscr\hat{H}},2}(x_{2},t;k_{2})\boldsymbol{\big\}}}^{=\boldsymbol{\big\{}\hat{\Lambda}_{1}(x_{1},t;k_{1}) \;
\stackrel{\otimes}{\boldsymbol{,}}\;\hat{\Lambda}_{2}(x_{2},t;k_{2})\boldsymbol{\big\}}}+
\\ \no &\hspace*{-1.0cm}+&\hspace*{-1.0cm}
\Big(\boldsymbol{\big[}\underbrace{\mfrak{r}_{12}(\sharp 1;\sharp 2;t)}_{\equiv0}\;\boldsymbol{,}\;
\mfrak{L}_{{\scrscr\hat{H}},1}(x_{1},t;k_{1})\,\boldsymbol{\big]_{-}}-
\boldsymbol{\big[}\underbrace{\mfrak{r}_{21}(\sharp 2;\sharp 1;t)}_{\equiv0}\;\boldsymbol{,}\;
\mfrak{L}_{{\scrscr\hat{H}},2}(x_{2},t;k_{2})\,\boldsymbol{\big]_{-}}\Big)\;
\delta(x_{1}-x_{2})\;.
\eeq
As we repeat to choose symmetrical dependences (\ref{s4_36})
on canonical fields \(\phi_{\alpha}(x,t)=(\mscr{Q}_{\alpha}(x,t)\:,\:\mscr{P}_{\alpha}(x,t)\,)\)
with diagonal, traceless Cartan sub-algebra matrices
\(\mfrak{L}_{{\scrscr \hat{H}}}(x,t;k) = \hat{H}^{i}\;\lambda_{i}^{{\scrscr(\hat{H}^{i})}}(\mscr{Q}_{\alpha},\mscr{P}_{\alpha};k)=
\hat{\Lambda}(\mscr{Q}_{\alpha},\mscr{P}_{\alpha};k)\)
\beq \lb{s4_36}
\lambda_{i}^{{\scrscr(\hat{H}^{i})}}(\mscr{Q}_{\alpha}(x,t)\:,\:\mscr{P}_{\alpha}(x,t)\,) &\Longrightarrow& \mbox{symmetrical dependences on }\;
 \mscr{Q}_{\alpha}(x,t)\:,\:\mscr{P}_{\alpha}(x,t)\;! \;; \\  \no
\Longrightarrow\;\lambda_{i}^{{\scrscr(\hat{H}^{i})}}(\mscr{Q}_{\alpha}(x,t)\:,\:\mscr{P}_{\alpha}(x,t)\,) &=&
\lambda_{i}^{{\scrscr(\hat{H}^{i})}}(\mscr{P}_{\alpha}(x,t)\:,\:\mscr{Q}_{\alpha}(x,t)\,) \;; \\  \lb{s4_37} &\Longrightarrow&
\boldsymbol{\big\{}\mfrak{L}_{{\scrscr\hat{H}},1}(x_{1},t;k_{1})\;\stackrel{\otimes}{\boldsymbol{,}}\;
\mfrak{L}_{{\scrscr\hat{H}},2}(x_{2},t;k_{2})\boldsymbol{\big\}}\equiv0\;,
\eeq
we resolve the involution of the conserved quantities following from (\ref{s4_21},\ref{s4_17})
by the remaining, vanishing tensorial Poisson bracket (\ref{s4_37}) of
the diagonal Lax matrices \(\mfrak{L}_{{\scrscr \hat{H}},1}(\mscr{Q}_{\alpha},\mscr{P}_{\alpha};k)\),
\(\mfrak{L}_{{\scrscr \hat{H}},2}(\mscr{Q}_{\alpha},\mscr{P}_{\alpha};k)\).
In section \ref{s32} it has already been exemplified that the gauge transformation (\ref{s4_38},\ref{s4_39}),
which constrains the spatial Lax matrix
\(\mfrak{L}(x,t;k)\) to the Cartan sub-algebra
\(\mfrak{L}_{{\scrscr \hat{H}}}(x,t;k) = \hat{H}^{i}\;\lambda_{i}^{{\scrscr(\hat{H}^{i})}}(\mscr{Q}_{\alpha},\mscr{P}_{\alpha};k)\),
should exist for very general conditions of the physical fields
\beq \lb{s4_38}
\frac{\pp\mfrak{g}(x,t)}{\pp x} &=&
\bigg[\bigg(\frac{\exp\{\overrightarrow{\boldsymbol{\big[}\mfrak{g}(x,t)\;\boldsymbol{,}\;\ldots\boldsymbol{\big]_{-}}}\}-1}
{\overrightarrow{\boldsymbol{\big[}\mfrak{g}(x,t)\;\boldsymbol{,}\;\ldots\boldsymbol{\big]_{-}}}}\bigg)^{\boldsymbol{-1}}
\Big[\hat{H}^{i}\;\lambda_{i}^{{\scrscr(\hat{H}^{i})}}(\mscr{Q}_{\alpha}(x,t)\:,\:\mscr{P}_{\alpha}(x,t)\,) +  \\ \no &-&
\Big(\exp\big\{\overrightarrow{\boldsymbol{\big[}\mfrak{g}(x,t)\;\boldsymbol{,}\;\ldots\boldsymbol{\big]_{-}}}\big\}\;
\big(\hat{H}^{i}\;k_{i}+\hat{E}_{+}^{\alpha}\;\mscr{P}_{\alpha}(x,t)+\hat{E}_{-}^{\alpha}\;\mscr{Q}_{\alpha}(x,t)\,\big)\Big)\Big]
\bigg]\;; \\ \lb{s4_39} \Longrightarrow\;\mscr{G}(x,t)&=&\exp\big\{\mfrak{g}(x,t)\}\;\mbox{ gauge transformstion should exist !}\;.
\eeq

\section{Extension of the zero-curvature condition beyond (1+1)-dimensions}  \lb{s5}

\subsection{Determination of the nonlinear equations for the fields of relevant, physical observables}\lb{s51}

In previous sections \ref{s2}-\ref{s4} we have emphasized the algebraic properties of Lax pairs in (1+1) dimensions and have also
considered cases where one can obtain chaotic behaviour within the general, non-compact \(\mbox{sl}(n,\mathbb{C})\) algebra
by separating into sub-algebra parts as e.\ g.\ \(\mbox{su}(n)\) or \(\mbox{sp}(n,\mathbb{R})\), etc.\ . In this section we point out
a possible extension to ((N-1)+1) Euclidean dimensions \(x_{\nu}=(\vec{x},t)\), \(x^{\nu}=(\vec{x},t)\) for Lax matrices \(\mfrak{A}^{\nu}(\vec{x},t)\) taking values within the \(\mbox{sl}(n,\mathbb{C})\)- or one of its sub-algebras.
Instead of two equations for the (1+1) dimensions,
one has to regard \(N\) equations (\ref{s5_1}) with Lax matrices \(\mfrak{A}^{\nu}(\vec{x},t)\) acting onto the
\(n\)-component, auxiliary field \(\Xi(\vec{x},t)\). In analogy to a \(N\) dimensional electromagnetic theory, the Lax matrices \(\mfrak{A}^{\nu}(\vec{x},t)\) are termed as matrix-potentials \(\mfrak{A}^{\nu}(\vec{x},t)\) which depend on canonical fields
\(\phi_{\alpha}^{\nu}(x,t)=(\mscr{Q}_{\alpha}^{\nu}(x,t)\:,\:\mscr{P}_{\alpha}^{\nu}(x,t)\,)\) and spectral parameters
\(k_{i}^{\nu}(\vec{x},t)\) within a Cartan-Weyl basis (\ref{s5_2},\ref{s3_8})
of the general \(\mbox{sl}(n,\mathbb{C})\) algebra or a chosen sub-algebra as e.\ g.\ \(\mbox{su}(n)\)
\beq \lb{s5_1}
\pp^{\nu}\Xi(\vec{x},t)&=&\mfrak{A}^{\nu}(\vec{x},t)\;\Xi(\vec{x},t)\;;\;\;\;\;
(\kappa,\,\lambda,\,\mu,\,\nu=0,\:(1,\ldots,N-1)\,)\;; \\ \lb{s5_2}
\mfrak{A}^{\nu}(\vec{x},t) &=&\hat{H}^{i}\;k_{i}^{\nu}(\vec{x},t)+
\hat{E}^{\alpha}\;\phi_{\alpha}^{\nu}(\vec{x},t)=\hat{H}^{i}\;k_{i}^{\nu}(\vec{x},t)+
\hat{E}_{+}^{\alpha}\;\mscr{P}_{\alpha}^{\nu}(\vec{x},t)+\hat{E}_{-}^{\alpha}\;\mscr{Q}_{\alpha}^{\nu}(\vec{x},t) \;;  \\  \no
\mbox{Matrix-potentials }\;\mfrak{A}^{\nu}&=&\mfrak{A}^{\nu}(\mscr{Q},\mscr{P};k)\;\mbox{ of canonical fields }\;
\phi_{\alpha}^{\nu}(\vec{x},t)=(\mscr{Q}_{\alpha}^{\nu}(\vec{x},t;k)\,,\,\mscr{P}_{\alpha}^{\nu}(\vec{x},t;k)\,)\;.
\eeq
In comparison to previous sections \ref{s2}-\ref{s4}, we assign to the time-like matrix
potential \(\mfrak{A}^{0}(\vec{x},t)\) the Lax matrix \(\mfrak{M}(\vec{x},t)\) and to the
(N-1) spatial matrix potential components \(\mfrak{A}^{i}(\vec{x},t)\), '\(\vec{\mfrak{A}}(\vec{x},t)\)'
the spatial Lax "vector" \(\vec{\mfrak{L}}(\vec{x},t)\) with corresponding (N-1) spatial Lax matrix components
\(\mfrak{L}^{i}(\vec{x},t)\) (\ref{s5_3}). The equivalence of mixed and exchanged partial
derivatives \(\pp^{\mu}\pp^{\nu}\Xi(\vec{x},t)\stackrel{!}{=}\pp^{\nu}\pp^{\mu}\Xi(\vec{x},t)\)
of (\ref{s5_1}) enforces \(N(N-1)/2\) zero-curvature equations (\ref{s5_4})
instead of the single zero-curvature condition in previous sections for (1+1) dimensions.
In analogy to the electrodynamic case, one can introduce a matrix-valued field strength tensor \(\mfrak{F}^{\mu\nu}\)
and a matrix-valued current \(\mfrak{J}^{\mu\nu}\) (\ref{s5_5}),
given by the commutator of two matrix potentials \(\mfrak{A}^{\mu}\), \(\mfrak{A}^{\nu}\)
\beq\lb{s5_3}
\mfrak{A}^{0}(\vec{x},t)&=&\mfrak{M}(\vec{x},t)\;\;,\;\;\;\mfrak{A}^{i}(\vec{x},t)=\mfrak{L}^{i}(\vec{x},t)\;;\;\;\;
(i=1,\ldots,N-1)\;; \\ \lb{s5_4}
0&=&\pp^{\mu}\mfrak{A}^{\nu}-\pp^{\nu}\mfrak{A}^{\mu}+
\boldsymbol{\big[}\mfrak{A}^{\nu}\;\boldsymbol{,}\;\mfrak{A}^{\mu}\boldsymbol{\big]_{-}}\;;\;\;\;
\mfrak{A}^{\nu}\in\mbox{sub-algebra}\subseteq\mbox{sl}(n,\mathbb{C})\;; \\  \lb{s5_5}
\mfrak{F}^{\mu\nu} &=&\pp^{\mu}\mfrak{A}^{\nu}-\pp^{\nu}\mfrak{A}^{\mu}=\mfrak{J}^{\mu\nu}=
\boldsymbol{\big[}\mfrak{A}^{\mu}\;\boldsymbol{,}\;\mfrak{A}^{\nu}\boldsymbol{\big]_{-}} \;.
\eeq
By using the Maurer-Cartan structure equations for the gauge matrices \(\mscr{G}_{0}\)
\be\lb{s5_6}
\left(\pp^{\mu}\big(\pp^{\nu}\mscr{G}_{0}\big)\:\mscr{G}_{0}^{-1}\right)-
\left(\pp^{\nu}\big(\pp^{\mu}\mscr{G}_{0}\big)\:\mscr{G}_{0}^{-1}\right)+
\boldsymbol{\big[}\big(\pp^{\nu}\mscr{G}_{0}\big)\:\mscr{G}_{0}^{-1}\;\boldsymbol{,}\;
\big(\pp^{\mu}\mscr{G}_{0}\big)\:\mscr{G}_{0}^{-1}\boldsymbol{\big]_{-}}=0\;,
\ee
we can straightforwardly derive a gauge invariance of the zero-curvature eqs. (\ref{s5_4},\ref{s5_5}) under the
gauge transformation (\ref{s5_7}) for the matrix potentials \(\mfrak{A}^{\nu}\rightarrow \mfrak{A}^{\bprime\nu}\)
\beq\lb{s5_7}
\mfrak{A}^{\nu}\; \rightarrow \;\mfrak{A}^{\bprime\nu}&=&\mscr{G}_{0}\;\mfrak{A}^{\nu}\;\mscr{G}_{0}^{-1} +
\big(\pp^{\nu}\mscr{G}_{0}\big)\;\mscr{G}_{0}^{-1} \;; \\ \lb{s5_8}
0&=&\pp^{\mu}\mfrak{A}^{\bprime\nu}-\pp^{\nu}\mfrak{A}^{\bprime\mu}+
\boldsymbol{\big[}\mfrak{A}^{\bprime\nu}\;\boldsymbol{,}\;\mfrak{A}^{\bprime\mu}\boldsymbol{\big]_{-}}\;;\;\;\;
\mscr{G}_{0}\in\mbox{sub-group}\subseteq\mbox{SL}(n,\mathbb{C}) \;.
\eeq
In later sections we have to perform a gauge transformation in order to prove the involution of conserved
quantities. One has to take a gauge transformation for the scalar product
\(\wt{\mfrak{L}}(\vec{x}(\zeta),t)=(d\vec{x}(\zeta)\,/\,d\zeta)\cdot\vec{\mfrak{L}}(\vec{x}(\zeta),t)\) along
chosen 'loops' \(\vec{x}(\zeta=0)=\vec{x}(\zeta=2\pi)\) to the Cartan sub-algebra elements
\(\wt{\mfrak{L}}_{{\scrscr\hat{H}}}(\vec{x}(\zeta),t)=\hat{H}^{i}\;
\vec{\lambda}_{i}^{{\scrscr(\hat{H}^{i})}}(\vec{x}(\zeta),t)\cdot(d\vec{x}(\zeta)\,/\,d\zeta)\).

The nonlinear equations from (\ref{s5_3}-\ref{s5_5}) can be transferred to the structure constants
\(\im\:c^{jk}_{\ph{jk}l}\) of the underlying algebra from the matrix potentials \(\mfrak{A}^{\mu}(\vec{x},t)\) which
are defined with the generators \(\mfrak{g}^{j}\) and
corresponding fields \(\Psi_{j}^{\mu}(\mscr{Q},\mscr{P};k)\) (\ref{s5_9}) as an ansatz
for nonlinear equations. As we resolve relations (\ref{s5_4},\ref{s5_10}) in terms of the generator
basis \(\mfrak{g}^{j}\) and structure constants \(\im\:c^{jk}_{\ph{jk}l}\) (\ref{s5_9}),
we finally attain the set of nonlinear equations
(\ref{s5_11}) with the ansatz of the \(\Psi_{j}^{\mu}(\mscr{Q},\mscr{P};k)\) fields which can also be related to the more
fundamental fields \(\mscr{Q}_{\alpha}^{\nu}(\vec{x},t)\), \(\mscr{P}_{\alpha}^{\nu}(\vec{x},t)\) in eq. (\ref{s5_2})
\beq \lb{s5_9}
\boldsymbol{\big[}\mfrak{g}^{j}\;\boldsymbol{,}\;\mfrak{g}^{k}\boldsymbol{\big]_{-}} &=&\im\;c^{jk}_{\ph{jk}l}\;\mfrak{g}^{l}\;;\;\;\;
\mfrak{A}^{\mu}(\vec{x},t)=\mfrak{g}^{j}\;\Psi^{\mu}_{j}(\mscr{Q},\mscr{P};k) \;;\;\;\;  \\  \lb{s5_10}
0 &=& \bigg(\mfrak{g}^{l}\Big((\pp^{\mu}\Psi_{l}^{\nu}\big)-\big(\pp^{\nu}\Psi_{l}^{\mu}\big)\Big)+
\boldsymbol{\big[}\mfrak{g}^{j}\;\boldsymbol{,}\;\mfrak{g}^{k}\boldsymbol{\big]_{-}}\;\:
\Psi^{\nu}_{j}\;\Psi^{\mu}_{k}\bigg)\;; \\  \lb{s5_11}
0&=&\big(\pp^{\mu}\Psi_{l}^{\nu}(\mscr{Q},\mscr{P};k)\big)-\big(\pp^{\nu}\Psi_{l}^{\mu}(\mscr{Q},\mscr{P};k)\big)+\im\;c^{jk}_{\ph{jk}l}\;\:
\Psi^{\nu}_{j}(\mscr{Q},\mscr{P};k)\;\Psi^{\mu}_{k}(\mscr{Q},\mscr{P};k)\;.
\eeq

\subsection{Conserved quantities of monodromy matrix paths with nontrivial homotopy of fields}\lb{s52}

In the following we consider a closed 'loop' \(\vec{x}(\xi)=\vec{x}_{\xi}\), \(\xi\in[0,2\pi)\) (\ref{s5_19}) within the
group manifold of the chosen Lie algebra (\(\mbox{sl}(n,\mathbb{C})\) or a corresponding sub-group) for the matrix potentials
\(\mfrak{A}^{\nu}=(\mfrak{M},\vec{\mfrak{L}})\) where a reference point \(\vec{x}_{P_{0}}\) defines the beginning and the end
of the loop
\be \lb{s5_19}
\vec{x}(\xi)=\vec{x}_{\xi}\;;\;\;\;\;\xi\in[0,2\pi)\;;\;\;\;\;
\vec{x}(\xi=0)=\vec{x}_{P_{0}}\;;\;\;\;\;\vec{x}(\xi=2\pi)=\vec{x}_{P_{0}}\;.
\ee
It is of crucial importance that the fibre space, given by the N-1 spatial coordinates for the base space and by the mapping to
the group manifold of the generators \(\mfrak{A}^{i}=\mfrak{L}^{i}\), does not allow a continuous contraction of the loop to a
trivial, single point as e.\ g.\ the reference point \(\vec{x}_{P_{0}}\). This supposition yields a straightforward
derivation of conserved quantities along nontrivial, non-contractable loops in analogy to section \ref{s41}.
One begins with the matrix path from the reference point \(\vec{x}_{P_{0}}\) to a point \(\vec{x}(\xi)=\vec{x}_{\xi}\),
which is specified by a parametrization \(\vec{x}(\zeta)\), (\(\zeta\in[0,\xi)\), \(\xi\in[0,2\pi)\)), and performs
a spatial ordering of exponentials \(\exp\{{\scr\Delta \zeta}\:(d\vec{x}_{\zeta}\,/\,d\zeta)\cdot\vec{\mfrak{L}}(\vec{x}_{\zeta},t;k)\}\)
along a part of the loop
\beq\lb{s5_20}
\mfrak{T}(\vec{x}_{\xi},\vec{x}_{P_{0}};t;k) &=&\overrightarrow{\exp}\bigg\{\int_{0}^{\xi}d\zeta\;\;
\frac{d\vec{x}_{\zeta}}{d\zeta}\cdot\vec{\mfrak{L}}(\vec{x}_{\zeta},t;k)\bigg\} \;.
\eeq
In correspondence to section \ref{s41}, we take the time derivative of (\ref{s5_20}) and have to regard the product rule
following from the spatial ordering of the exponential step operators along the part \(\zeta\in[0,\xi)\) of the loop
\(\vec{x}_{\zeta}=\vec{x}(\zeta)\)
\beq\lb{s5_21}
\pp_{t}\mfrak{T}(\vec{x}_{\xi},\vec{x}_{P_{0}};t;k) &=&\int_{0}^{\xi}d\zeta\;\;
\overrightarrow{\exp}\bigg\{\int_{\zeta}^{\xi}d\zeta_{1}\;\;
\frac{d\vec{x}_{\zeta_{1}}}{d\zeta_{1}}\cdot\vec{\mfrak{L}}(\vec{x}_{\zeta_{1}},t;k)\bigg\}\;\times \\ \no &\times&
\Big(\frac{d\vec{x}_{\zeta}}{d\zeta}\cdot\big(\pp_{t}\vec{\mfrak{L}}(\vec{x}_{\zeta},t;k)\,\big)\Big)\;
\overrightarrow{\exp}\bigg\{\int_{0}^{\zeta}d\zeta_{2}\;\;
\frac{d\vec{x}_{\zeta_{2}}}{d\zeta_{2}}\cdot\vec{\mfrak{L}}(\vec{x}_{\zeta_{2}},t;k)\bigg\}\;.
\eeq
We can replace the scalar product \((d\vec{x}_{\zeta}\,/\,d\zeta)\cdot(\pp_{t}\vec{\mfrak{L}}(\vec{x}_{\zeta},t;k)\,)\) in (\ref{s5_21})
by the zero-curvature relations (\ref{s5_22},\ref{s5_23})
\beq\lb{s5_22}
\pp_{t}\vec{\mfrak{L}}-\vec{\pp}\mfrak{M}=
\boldsymbol{\big[}\mfrak{M}\;\boldsymbol{,}\;\vec{\mfrak{L}}\boldsymbol{\big]_{-}} &\Longrightarrow&
\pp_{t}\vec{\mfrak{L}} = \vec{\pp}\mfrak{M}+
\boldsymbol{\big[}\mfrak{M}\;\boldsymbol{,}\;\vec{\mfrak{L}}\boldsymbol{\big]_{-}} \;; \\  \lb{s5_23}
\frac{d\vec{x}_{\zeta}}{d\zeta}\cdot\pp_{t}\vec{\mfrak{L}} &=&\frac{d\vec{x}_{\zeta}}{d\zeta}\cdot\vec{\pp}\mfrak{M}+
\boldsymbol{\Big[}\mfrak{M}\;\boldsymbol{,}\;\frac{d\vec{x}_{\zeta}}{d\zeta}\cdot\vec{\mfrak{L}}\boldsymbol{\Big]_{-}}  \;,
\eeq
in order to transform the integrand of (\ref{s5_21}) to a total derivative of the loop parameter \(\zeta\in[0,\xi)\)
(cf.\ section \ref{s41} eqs.\ (\ref{s4_5}-\ref{s4_7}))
\beq\lb{s5_24}
\pp_{t}\mfrak{T}(\vec{x}_{\xi},\vec{x}_{P_{0}};t;k) &=&\int_{0}^{\xi}d\zeta\;\;
\overrightarrow{\exp}\bigg\{\int_{\zeta}^{\xi}d\zeta_{1}\;\;
\frac{d\vec{x}_{\zeta_{1}}}{d\zeta_{1}}\cdot\vec{\mfrak{L}}(\vec{x}_{\zeta_{1}},t;k)\bigg\}\;\times \\ \no &\times&
\bigg(\frac{d\vec{x}_{\zeta}}{d\zeta}\cdot\vec{\pp}\mfrak{M}+
\boldsymbol{\Big[}\mfrak{M}\;\boldsymbol{,}\;\frac{d\vec{x}_{\zeta}}{d\zeta}\cdot\vec{\mfrak{L}}\boldsymbol{\Big]_{-}}\bigg)\;
\overrightarrow{\exp}\bigg\{\int_{0}^{\zeta}d\zeta_{2}\;\;
\frac{d\vec{x}_{\zeta_{2}}}{d\zeta_{2}}\cdot\vec{\mfrak{L}}(\vec{x}_{\zeta_{2}},t;k)\bigg\} \\ \no &=&
\int_{0}^{\xi}d\zeta\;\;\frac{\pp}{\pp\zeta}\bigg(\overrightarrow{\exp}\bigg\{\int_{\zeta}^{\xi}d\zeta_{1}\;\;
\frac{d\vec{x}_{\zeta_{1}}}{d\zeta_{1}}\cdot\vec{\mfrak{L}}(\vec{x}_{\zeta_{1}},t;k)\bigg\}\;\times \\ \no &\times&
\mfrak{M}(\vec{x}_{\zeta},t;k)\;\;
\overrightarrow{\exp}\bigg\{\int_{0}^{\zeta}d\zeta_{2}\;\;
\frac{d\vec{x}_{\zeta_{2}}}{d\zeta_{2}}\cdot\vec{\mfrak{L}}(\vec{x}_{\zeta_{2}},t;k)\bigg\} \bigg) \\  \no  &=&
\mfrak{M}(\vec{x}(\xi),t;k)\;\;\mfrak{T}(\vec{x}_{\xi},\vec{x}_{P_{0}};t;k)-
\mfrak{T}(\vec{x}_{\xi},\vec{x}_{P_{0}};t;k)\;\;\mfrak{M}(\vec{x}_{P_{0}},t;k)\;\;\;.
\eeq
As one takes \(\xi=2\pi\) so that \(\vec{x}(\xi=2\pi)=\vec{x}(\xi=0)=\vec{x}_{P_{0}}\) are the beginning and the end
of a closed loop '\(\odot\)' in the base space for a nontrivial, (non-contractable) homotopic mapping of the fields, one transforms
the time derivative of monodromy matrix (\ref{s5_26}) to a commutator which finally gives
the conserved quantities (\ref{s5_27}) of traces of powers of the monodromy matrix \(\mfrak{T}(\odot,t;k)\)
\beq \lb{s5_25}
\mfrak{T}(\vec{x}(\xi=2\pi),\vec{x}_{P_{0}};t;k) &=&\mfrak{T}(\odot,t;k) \;; \\  \lb{s5_26}
\pp_{t}\mfrak{T}(\odot,t;k) &=& \boldsymbol{\big[}\mfrak{M}(\vec{x}_{P_{0}},t;k)\;
\boldsymbol{,}\;\mfrak{T}(\odot,t;k)\boldsymbol{\big]_{-}} \;; \\  \lb{s5_27}
\pp_{t}\mbox{Tr}\big[\mfrak{T}^{n}(\odot,t;k)\big] &=&\big(\pp_{t}C^{(n)}(\odot,t;k)\big)=
n\;\mbox{Tr}\Big[\big(\pp_{t}\mfrak{T}(\odot,t;k)\,\big)\;
\mfrak{T}^{n-1}(\odot,t;k)\Big] \\ \no &=& n\;\mbox{Tr}\Big[\boldsymbol{\big[}\mfrak{M}(\vec{x}_{P_{0}},t;k)\;\boldsymbol{,}\;
\mfrak{T}(\odot,t;k)\boldsymbol{\big]_{-}}\;\mfrak{T}^{n-1}(\odot,t;k)\Big]=0\;\;\;.
\eeq

\subsection{Involution of conserved quantities and the classical $\boldsymbol{\mfrak{r}}$-matrix in arbitrary spacetime dimensions}\lb{s53}

As we have generalized the derivation of conserved quantities from a circle \(x\in[0,2\pi)\) within (1+1) dimensions
to a closed loop \(\vec{x}(\xi)\), \(\xi\in[0,2\pi)\) within a (N-1) dimensional base space for a nontrivial, homotopic mapping
of a Lie group \(\mbox{SL}(n,\mathbb{C})\) (or a sub-group as \(\mbox{SU}(n)\), etc.\ ), we can also extend the involution
properties of section \ref{s42} for the (1+1) dimensional case to those of the ((N-1)+1) spacetime. One can repeat the calculations
of section \ref{s42} by replacing the spatial Lax matrix \(\mfrak{L}(x,t)\) by the scalar product
\(\wt{\mfrak{L}}(\vec{x}_{\zeta},t;k)=\frac{d\vec{x}_{\zeta}}{d\zeta}\cdot\vec{\mfrak{L}}(\vec{x}_{\zeta},t;k)\)
where the 'loop' parameter \(\zeta\in[0,2\pi)\) substitutes the spatial coordinate '\(x\)' or '\(\xi\)' of the circle
for the (1+1) dimensional case in section \ref{s42}
\beq\lb{s5_28}
\mbox{Apply }\wt{\mfrak{L}}(\vec{x}_{\zeta},t;k)&=&\frac{d\vec{x}_{\zeta}}{d\zeta}\cdot\vec{\mfrak{L}}(\vec{x}_{\zeta},t;k)\;
\mbox{ instead of }\;\mfrak{L}(x,t;k)\;\mbox{ or }\;\mfrak{L}(\xi,t;k)\;\mbox{ in section \ref{s42} and } \\ \no &&
\mbox{ repeat calculation of involution properties }  \;; \\  \lb{s5_29}
\lefteqn{\hspace*{-2.8cm}\mbox{Diagonalize by a gauge transformation :} }  \\ \no
\wt{\mfrak{L}}(\vec{x}_{\zeta},t;k)&\rightarrow& \wt{\mfrak{L}}_{\hat{H}}(\vec{x}_{\zeta},t;k)=
\hat{H}^{i}\;\vec{\lambda}_{i}^{{\scrscr(\hat{H}^{i})}}(\vec{x}_{\zeta},t;k)\cdot\frac{d\vec{x}_{\zeta}}{d\zeta}\;\;\;.
\eeq
Similarly, the classical $\mfrak{r}$-matrix approach can be conveyed from appendix \ref{sb} of the (1+1) dimensional case
to ((N-1)+1) dimensions with the replacement (\ref{s5_28}) under restriction of
spatially constant matrices \(\mfrak{r}(k_{i}^{\nu},k_{j}^{\mu};t)\).
However, we emphasize again that the analogous, corresponding transformations in place of the (1+1)
dimensional case of section \ref{s42} and appendix \ref{sb}
only hold for {\it non-contractable, nontrivial, homotopic mappings} from the loop within the (N-1) dimensional base space
to the chosen Lie group manifold as \(\mbox{SL}(n,\mathbb{C})\) or one of its sub-groups.

\section{Summary and conclusion} \lb{s6}

\subsection{Lax pairs and chaotic behaviour of (1+1) GP-type equations} \lb{s61}

This article has been initiated by the notion whether any Lax pair construction can only lead to
a completely integrable behaviour. As we have verified in sections \ref{s21} and \ref{s22} for Lax pairs
of the (1+1) GP-equations as generators of the $\mbox{sl}(2,\mathbb{C})$ algebra, one can even determine
Lax pairs for arbitrary external potentials \(V(x,t)\) without changing the spatial Lax matrix component
\(\mfrak{L}(x,t)\). Since the conserved quantities and their involution only depend on the exponential
step operators with the spatial Lax matrix \(\mfrak{L}(\xi,t)\), \(\xi\in[0,2\pi)\), one can directly
conclude for a Liouville integrability as in the cases without an external potential, either from
a spatially constant \(\mfrak{r}_{12}\)-matrix approach according to appendix \ref{sb} or from a gauge
transformation to the diagonal, commuting Cartan sub-algebra elements as the eigenvalues of \(\mfrak{L}(\xi,t)\)
within the tensorial Poisson bracket relation (A symmetric dependence of the eigenvalues on the
canonical fields within the Poisson bracket has to be presupposed.). As we reduce the generators of
\(\mfrak{L}(x,t)\), \(\mfrak{M}(x,t)\), either to the sub-algebra \(\mbox{su}(2)\) or to \(\mbox{sl}(2,\mathbb{R})\)
(of the most general, nontrivial $\mbox{sl}(2,\mathbb{C})$ algebra, cf.\ appendix \ref{sa}) for an attractive or
repulsive interaction, the hermitian property of the prevailing Hamiltonian, following from \(\mbox{su}(2)\)
or \(\mbox{sl}(2,\mathbb{R})\) Lax pairs, prevents any chaotic behaviour, due to the chosen compactness with
a spatial circle \(x\in[0,2\pi)\). However, as we combine the
\(\psi_{\mbox{\scz su}(2)}(x,t)\) and \(\psi_{\mbox{\scz sp}(2,\mathbb{R})}(x,t)\)
fields of the two integrable, (1+1) GP-equations with attractive and repulsive interaction to the complex-valued
parameter fields within the most general, non-compact $\mbox{sl}(2,\mathbb{C})$ algebra, probability or density
of the \(\psi_{\mbox{\scz su}(2)}(x,t)\) and \(\psi_{\mbox{\scz sp}(2,\mathbb{R})}(x,t)\)
fields can flow and change between
the coupled GP-equations which separately contain incoherent, non-hermitian terms for a chaotic behaviour.
This chaotic behaviour from Lax pair construction is even possible for \(\mbox{sl}(n>2,\mathbb{C})\) algebras
where one has to select a sub-algebra for a compact sub-group as \(\mbox{SU}(n)\subset\mbox{SL}(n,\mathbb{C})\)
so that coupled nonlinear equations of physical fields are also composed of incoherent terms, giving rise to
unlimited increase of the \(\psi_{\mbox{\scz su}(n)}(x,t)\) fields and corresponding chaotic behaviour.

\subsection{Lax pair construction in arbitrary spacetime} \lb{s62}

The given construction of Lax pairs for the (1+1) GP-equations as generators of \(\mbox{sl}(n,\mathbb{C})\)
(or of a sub-algebra as \(\mbox{su}(n)\)) straightforwardly generalize to arbitrary spacetime dimensions.
However, we emphasize again that this extension beyond (1+1) dimensions necessarily has to involve a nontrivial
homotopic mapping from the loop within the (N-1) dimensional base space to the considered group manifold,
following from the Lax pair generators. In absence of a nontrivial homotopy, it is possible to contract the
loop within the fibre space to trivial point mappings so that the construction of the conserved quantities and
their involution becomes trivial and meaningless. Therefore, Lax pair constructions for a Liouville integrability
or a possible chaotic behaviour beyond (1+1) dimensions have to be accompanied by an investigation for a
nontrivial homotopy of the underlying fibre space (\cite{mietopy1}).

\begin{appendix}

\section{Reduction of $\boldsymbol{\mbox{gl}(n,\mathbb{C})}$ to
$\boldsymbol{\mbox{sl}(n,\mathbb{C})}$ Lax pairs by separating the trivial trace
parts} \lb{sa}

In this part \ref{sa} of the appendix we assume that Lax matrices \(\mfrak{L}(x,t)\), \(\mfrak{M}(x,t)\)
are not traceless and therefore belong to the $\mbox{gl}(n,\mathbb{C})$ algebra as the most general case of
\(n\times n\) matrices. The general Lax matrices \(\mfrak{L}(x,t)\), \(\mfrak{M}(x,t)\) are
separated into diagonal unity parts \(\hat{1}\;\Delta\mfrak{L}(x,t)\), \(\hat{1}\;\Delta\mfrak{M}(x,t)\) and
remaining traceless parts \(\mfrak{L}_{0}(x,t)\), \(\mfrak{M}_{0}(x,t)\) (\ref{sa_1},\ref{sa_2}) of the
\(\mbox{sl}(n,\mathbb{C})\) algebra (\(n=N_{\mfrak{L}}=N_{\mfrak{M}}\)).
This trace separation is also performed for the initial matrix \(\mfrak{M}_{ini}(x=0,t)\) at the
coordinate origin within the solution of the zero-curvature relation for
\(\mfrak{M}(x,t)=\mfrak{M}_{0}(x,t)+\hat{1}\;\Delta\mfrak{M}(x,t)\) where explicit use is made for the
trace splitting of \(\mfrak{L}(x,t)= \mfrak{L}_{0}(x,t)+\hat{1}\;\Delta\mfrak{L}(x,t)\)
with the '\(\mfrak{ad}\)'-operator
\beq\lb{sa_1}
\mfrak{L}(x,t) &=&
\mfrak{L}_{0}(x,t)+{\ts\frac{\hat{1}}{N_{\mfrak{L}}}}\cdot\mbox{Tr}\big[\mfrak{L}(x,t)\big]
= \mfrak{L}_{0}(x,t)+\hat{1}\;\Delta\mfrak{L}(x,t)\;; \;\;\;
\Big(\mbox{Tr}\big[\mfrak{L}_{0}(x,t)\big] \equiv 0\Big) \;; \\  \lb{sa_2}
\mfrak{M}(x,t) &=&
\mfrak{M}_{0}(x,t)+{\ts\frac{\hat{1}}{N_{\mfrak{M}}}}\cdot\mbox{Tr}\big[\mfrak{M}(x,t)\big]
= \mfrak{M}_{0}(x,t)+\hat{1}\;\Delta\mfrak{M}(x,t)\;;\;\;\;
\Big(\mbox{Tr}\big[\mfrak{M}_{0}(x,t)\big]\equiv 0\Big) \;;  \\  \lb{sa_3}
\mfrak{M}_{ini}(x=0,t) &=& \mfrak{M}_{ini}^{(0)}(x=0,t)+
\hat{1}\;\Delta\mfrak{M}_{ini}(x=0,t)\;;\;\;\;
\Big(\mbox{Tr}\big[\mfrak{M}_{ini}^{(0)}(x=0,t)\big]= 0\Big) \;; \\  \lb{sa_4}
\mfrak{M}(x,t) &=& \mfrak{M}_{0}(x,t)+\hat{1}\;\Delta\mfrak{M}(x,t)
= \\ \no &=& \overleftarrow{\exp}\Big\{\int_{0}^{x}d\xi\:
\overrightarrow{\boldsymbol{[}\mfrak{L}_{0}(\xi,t)+\hat{1}\;\Delta\mfrak{L}(\xi,t)\:\boldsymbol{,}\:
\ldots\boldsymbol{]_{-}}}\Big\}\Big(\mfrak{M}_{ini}^{(0)}(x=0,t)+
\hat{1}\;\Delta\mfrak{M}_{ini}(x=0,t)\Big)+  \\ \no &+&
\int_{0}^{x}dy\:\overleftarrow{\exp}\Big\{\int_{y}^{x}d\xi\:
\overrightarrow{\boldsymbol{[}\mfrak{L}_{0}(\xi,t)+\hat{1}\;\Delta\mfrak{L}(\xi,t)
\:\boldsymbol{,}\:\ldots\boldsymbol{]_{-}}}\Big\}\:\Big(\big(\pp_{t}\mfrak{L}_{0}(y,t)\,\big)+
\hat{1}\;\big(\pp_{t}\Delta\mfrak{L}(y,t)\,\big)\Big)\;.
\eeq
As we regard the complete vanishing of the '\(\mfrak{ad}\)'-operator part
{\small\(\overrightarrow{\boldsymbol{[}\hat{1}\;\Delta\mfrak{L}(\xi,t)\:\boldsymbol{,}\:\ldots\boldsymbol{]_{-}}}\)}
in (\ref{sa_4}), one can conclude for the separation of the solution of the zero-curvature relation into
the two independent parts (\ref{sa_5},\ref{sa_6}) where equation (\ref{sa_5}) only consists of the total traceless
\(\mbox{sl}(n,\mathbb{C})\) generators \(\mfrak{L}_{0}(x,t)\), \(\mfrak{M}_{0}(x,t)\),
\(\mfrak{M}_{ini}^{(0)}(x=0,t)\), and where equation (\ref{sa_6}) separately has the remaining fields
\(\Delta\mfrak{L}(x,t)\), \(\Delta\mfrak{M}(x,t)\), \(\Delta\mfrak{M}_{ini}(x=0,t)\) from the diagonal
unity part '\(\hat{1}\)'
\beq\lb{sa_5}
\mfrak{M}_{0}(x,t)&=&\overleftarrow{\exp}\Big\{\int_{0}^{x}d\xi\:
\overrightarrow{\boldsymbol{[}\mfrak{L}_{0}(\xi,t)\:\boldsymbol{,}\:\ldots\boldsymbol{]_{-}}}\Big\}
\mfrak{M}_{ini}^{(0)}(x=0,t)+   \\ \no &+&
\int_{0}^{x}dy\:\overleftarrow{\exp}\Big\{\int_{y}^{x}d\xi\:
\overrightarrow{\boldsymbol{[}\mfrak{L}_{0}(\xi,t)\:\boldsymbol{,}\:\ldots\boldsymbol{]_{-}}}\Big\}\:
\big(\pp_{t}\mfrak{L}_{0}(y,t)\big)\;\;;  \\  \lb{sa_6}
\Delta\mfrak{M}(x,t)&=&\Delta\mfrak{M}_{ini}(x=0,t)+
\int_{0}^{x}dy\;\big(\pp_{t}\Delta\mfrak{L}(y,t)\,\big)  \\ \no &=&
\Delta\mfrak{M}_{ini}(x=0,t)+
\pp_{t}\Big(\int_{0}^{x}dy\;\mbox{Tr}\big[\mfrak{L}(y,t)\big]/N_{\mfrak{L}}\Big)\;\;.
\eeq
We can also verify from the zero-curvature relation (\ref{sa_7}) the separation property into traceless
\(\mbox{sl}(n,\mathbb{C})\) matrices and diagonal unit parts. Since the diagonal unit parts
\(\Delta\mfrak{L}(x,t)\), \(\Delta\mfrak{M}(x,t)\) only cause vanishing commutators within the general
zero-curvature condition, the latter relation splits into the solely traceless part of \(\mbox{sl}(n,\mathbb{C})\)
generators \(\mfrak{L}_{0}(x,t)\), \(\mfrak{M}_{0}(x,t)\) and a simple part for the completely diagonal terms
with the fields \(\Delta\mfrak{L}(x,t)\), \(\Delta\mfrak{M}(x,t)\)
\beq\lb{sa_7}
\lefteqn{\big(\pp_{t}\mfrak{L}\big)-\big(\pp_{x}\mfrak{M}\big)+\boldsymbol{\big[}\mfrak{L}\;\boldsymbol{,}\;
\mfrak{M}\boldsymbol{\big]_{-}}=\overbrace{\big(\pp_{t}\mfrak{L}_{0}\big)-
\big(\pp_{x}\mfrak{M}_{0}\big)+\boldsymbol{\big[}\mfrak{L}_{0}\;\boldsymbol{,}\;
\mfrak{M}_{0}\boldsymbol{\big]_{-}} }^{=0} + } \\ \no
&+&\frac{\hat{1}}{N_{\mfrak{L}}}
\underbrace{\bigg[\pp_{t}\Big(\mbox{Tr}\big[\mfrak{L}(x,t)\big]\Big)-\pp_{x}\:\pp_{t}
\Big(\int_{0}^{x}dy\;\mbox{Tr}\big[\mfrak{L}(y,t)\big]\Big)\bigg]}_{\equiv0}=0
\;;\;\;\; (N_{\mfrak{L}}=N_{\mfrak{M}})\;.
\eeq
Therefore, the purely traceless generators with \(\mfrak{L}_{0}(x,t)\) (and the derived or traceless chosen
generators for \(\mfrak{M}_{0}(x,t)\)) can give rise to nontrivial, nonlinear equations within a relevant
Lax pair construction.

\section{Involution of monodromy matrices for spatially constant $\boldsymbol{\mfrak{r}}$-matrix}\lb{sb}

A standard proof of conserved quantities from the spatial Lax matrix \(\mfrak{L}(x,t)\subseteq\mbox{sl}(n,\mathbb{C})\)
begins with the assumption of the fundamental, tensorial Poisson bracket relation (\ref{sb_1}), having a delta function
\(\delta(x_{1}-x_{2})\) for the 'ultralocal' case of fields. We note that this relation transforms the quadratic
term of \(\mfrak{L}(x,t)\) to a linear part within a commutator of the so-called '\(\mfrak{r}_{12}\)'-matrix
which is supposed to be independent on the space coordinate '\(x\)'
\beq\lb{sb_1}
\boldsymbol{\big\{}\mfrak{L}_{1}(x_{1},t;k_{1})\;\stackrel{\otimes}{\boldsymbol{,}}\;
\mfrak{L}_{2}(x_{2},t;k_{2})\boldsymbol{\big\}} &=&
\boldsymbol{\big[}\mfrak{r}_{12}(k_{1},k_{2};t)\;\boldsymbol{,}\;\mfrak{L}_{1}(x_{1},t;k_{1})+
\mfrak{L}_{2}(x_{2},t;k_{2})\,\boldsymbol{\big]_{-}}\;\delta(x_{1}-x_{2})\;.
\eeq
We consider the spatial evolution eqs.\ (\ref{sb_3},\ref{sb_4}) for the initial coordinate '\(y\)' and end
point '\(x\)' of the defined type (\ref{sb_2}) of monodromy matrix \(\mfrak{T}(x,y;t;k)\) which follows
from subsequent spatial ordering of exponential step operators with Lax matrix \(\mfrak{L}(\xi,t;k)\).
As one applies the Leibniz and product rule for the tensorial Poisson bracket (\ref{sb_5}) of two
monodromy matrices (\ref{sb_2}), we accomplish the fundamental Poisson bracket (\ref{sb_1},\ref{sb_6}) within
a double spatial integral \(\xi_{1}\in[y_{1},x_{1}]\), \(\xi_{2}\in[y_{2},x_{2}]\)
\beq \lb{sb_2}
\mfrak{T}(x,y;t;k) &=& \overleftarrow{\exp}\Big\{\int_{y}^{x}d\xi\;\mfrak{L}(\xi,t;k)\Big\}\;;  \\ \lb{sb_3}
\big(\pp_{x}\mfrak{T}(x,y;t;k)\,\big) &=& \mfrak{L}(x,t;k)\;\mfrak{T}(x,y;t;k)\;;  \\  \lb{sb_4}
\big(\pp_{y}\mfrak{T}(x,y;t;k)\,\big) &=& -\:\mfrak{T}(x,y;t;k)\;\mfrak{L}(y,t;k)\;;  \\   \lb{sb_5}
\boldsymbol{\big\{}\mfrak{T}_{1}(x_{1},y_{1};t;k_{1})\;\stackrel{\otimes}{\boldsymbol{,}}\;
\mfrak{T}_{2}(x_{2},y_{2};t;k_{2})\boldsymbol{\big\}} &=&
\int_{y_{1}}^{x_{1}}d\xi_{1}\int_{y_{2}}^{x_{2}}d\xi_{2}\;\;\mfrak{T}_{1}(x_{1},\xi_{1};t;k_{1})\;\;
\mfrak{T}_{2}(x_{2},\xi_{2};t;k_{2})\;\;\times  \\ \no &\times&
\boldsymbol{\big\{}\mfrak{L}_{1}(\xi_{1},t;k_{1})\;\stackrel{\otimes}{\boldsymbol{,}}\;
\mfrak{L}_{2}(\xi_{2},t;k_{2})\boldsymbol{\big\}}\;\;
\mfrak{T}_{1}(\xi_{1},y_{1};t;k_{1})\;\;\mfrak{T}_{2}(\xi_{2},y_{2};t;k_{2})\;;  \\   \lb{sb_6}
\boldsymbol{\big\{}\mfrak{L}_{1}(\xi_{1},t;k_{1})\;\stackrel{\otimes}{\boldsymbol{,}}\;
\mfrak{L}_{2}(\xi_{2},t;k_{2})\boldsymbol{\big\}} &=&
\boldsymbol{\big[}\mfrak{r}_{12}(k_{1},k_{2};t)\;\boldsymbol{,}\;\mfrak{L}_{1}(\xi_{1},t;k_{1})+\mfrak{L}_{2}(\xi_{2},t;k_{2})\,
\boldsymbol{\big]_{-}}\;\delta(\xi_{1}-\xi_{2})\;.
\eeq
After substitution and insertion of (\ref{sb_1},\ref{sb_6},\ref{sb_3},\ref{sb_4}) into (\ref{sb_5}), the tensorial
Poisson bracket of monodromy matrices reduces to simple integrals (\ref{sb_7},\ref{sb_8}) because the corresponding
integrands only consist of total derivatives \(\delta(\xi_{1}-\xi_{2})\;(\pp_{\xi_{1}}+\pp_{\xi_{2}})\) and
\(\pp_{\xi}\), respectively
\beq \lb{sb_7}
\lefteqn{\boldsymbol{\big\{}\mfrak{T}_{1}(x_{1},y_{1};t;k_{1})\;\stackrel{\otimes}{\boldsymbol{,}}\;
\mfrak{T}_{2}(x_{2},y_{2};t;k_{2})\boldsymbol{\big\}} =\mfrak{T}_{1}(x_{1},x_{min};t;k_{1})\;\;
\mfrak{T}_{2}(x_{2},x_{min};t;k_{2})\;\;\times }  \\ \no &\times&
\int_{y_{max}}^{x_{min}}d\xi_{1}\;d\xi_{2}\;\delta(\xi_{1}-\xi_{2})\;
\bigg[\mfrak{T}_{1}(x_{min},\xi_{1};t;k_{1})\;\;
\mfrak{T}_{2}(x_{min},\xi_{2};t;k_{2})\;\;\times \\ \no &\times&\mfrak{r}_{12}(k_{1},k_{2};t)\;
\Big(\big(\pp_{\xi_{1}}\mfrak{T}_{1}(\xi_{1},y_{max};t;k_{1})\,\big)+\big(\pp_{\xi_{2}}\mfrak{T}_{2}(\xi_{2},y_{max};t;k_{2})\,\big)\Big)+
\\ \no &+&
\Big(\big(\pp_{\xi_{1}}\mfrak{T}_{1}(x_{min},\xi_{1};t;k_{1})\,\big)+\big(\pp_{\xi_{2}}\mfrak{T}_{2}(x_{min},\xi_{2};t;k_{2})\,\big)\Big)\;
\mfrak{r}_{12}(k_{1},k_{2};t)\;\times  \\ \no &\times&
\mfrak{T}_{1}(\xi_{1},y_{max};t;k_{1})\;\;\mfrak{T}_{2}(\xi_{2},y_{max};t;k_{2})\bigg]\;\;
\mfrak{T}_{1}(y_{max},y_{1};t;k_{1})\;\;\mfrak{T}_{2}(y_{max},y_{2};t;k_{2})  \\ \no &=&
\mfrak{T}_{1}(x_{1},x_{min};t;k_{1})\;\;\mfrak{T}_{2}(x_{2},x_{min};t;k_{2})\;\;
\int_{y_{max}}^{x_{min}}d\xi_{1}\;d\xi_{2}\;\delta(\xi_{1}-\xi_{2})\;\big(\pp_{\xi_{1}}+\pp_{\xi_{2}}\big)\;\times  \\  \no &\times&
\bigg[\mfrak{T}_{1}(x_{min},\xi_{1};t;k_{1})\;\;\mfrak{T}_{2}(x_{min},\xi_{2};t;k_{2})\;\;\mfrak{r}_{12}(k_{1},k_{2};t)\;
\mfrak{T}_{1}(\xi_{1},y_{max};t;k_{1})\;\;\mfrak{T}_{2}(\xi_{2},y_{max};t;k_{2})\bigg]\;\;\times  \\ \no &\times&
\mfrak{T}_{1}(y_{max},y_{1};t;k_{1})\;\;\mfrak{T}_{2}(y_{max},y_{2};t;k_{2}) \;;  \\ \lb{sb_8}
\lefteqn{\boldsymbol{\big\{}\mfrak{T}_{1}(x_{1},y_{1};t;k_{1})\;\stackrel{\otimes}{\boldsymbol{,}}\;
\mfrak{T}_{2}(x_{2},y_{2};t;k_{2})\boldsymbol{\big\}} =}   \\ \no  &=&
\mfrak{T}_{1}(x_{1},x_{min};t;k_{1})\;\;\mfrak{T}_{2}(x_{2},x_{min};t;k_{2})\;\;
\int_{y_{max}}^{x_{min}}d\xi\;\;\times  \\  \no &\times&\pp_{\xi}
\bigg(\mfrak{T}_{1}(x_{min},\xi;t;k_{1})\;\;\mfrak{T}_{2}(x_{min},\xi;t;k_{2})\;\;\mfrak{r}_{12}(k_{1},k_{2};t)\;
\mfrak{T}_{1}(\xi,y_{max};t;k_{1})\;\;\mfrak{T}_{2}(\xi,y_{max};t;k_{2})\bigg)\;\;\times  \\ \no &\times&
\mfrak{T}_{1}(y_{max},y_{1};t;k_{1})\;\;\mfrak{T}_{2}(y_{max},y_{2};t;k_{2}) \;.
\eeq
After performing the spatial integration in (\ref{sb_8}), we achieve the tensorial Poisson bracket (\ref{sb_9})
of evoluton matrices with the remaining integration boundaries for the spatial \(\xi\) integration.
Let us assume the integer relation (\(n_{1}\geq n_{2}\)) so that taking the integration boundaries transforms the
right-hand side of (\ref{sb_9}) to a commutator-like relation (\ref{sb_10}) with the classical '\(\mfrak{r}_{12}\)'-matrix
\beq\lb{sb_9}
\lefteqn{\boldsymbol{\big\{}\mfrak{T}_{1}(2\pi\,n_{1},0;t;k_{1})\;\stackrel{\otimes}{\boldsymbol{,}}\;
\mfrak{T}_{2}(2\pi\,n_{2},0;t;k_{2})\boldsymbol{\big\}} =}   \\ \no &=&
\mfrak{T}_{1}(2\pi\,n_{1},\xi;t;k_{1})\;\;\mfrak{T}_{2}(2\pi\,n_{2},\xi;t;k_{2})\;\;\mfrak{r}_{12}(k_{1},k_{2};t)\;\;
\mfrak{T}_{1}(\xi,0;t;k_{1})\;\;\mfrak{T}_{2}(\xi,0;t;k_{2})\bigg|_{\xi=0}^{\xi=2\pi\,Min(n_{1},n_{2})} \; \;; \\ \lb{sb_10}
\lefteqn{\boldsymbol{\big\{}\mfrak{T}_{1}(2\pi\,n_{1},0;t;k_{1})\;\stackrel{\otimes}{\boldsymbol{,}}\;
\mfrak{T}_{2}(2\pi\,n_{2},0;t;k_{2})\boldsymbol{\big\}} =}   \\ \no &=&
\mfrak{T}_{1}(2\pi\,n_{1},2\pi\,n_{2};t;k_{1})\;\;\mfrak{r}_{12}(k_{1},k_{2};t)\;\;\mfrak{T}_{1}(2\pi\,n_{2},0;t;k_{1})\;\;
\mfrak{T}_{2}(2\pi\,n_{2},0;t;k_{2}) +  \\ \no &-&
\mfrak{T}_{1}(2\pi\,n_{1},0;t;k_{1})\;\;\mfrak{T}_{2}(2\pi\,n_{2},0;t;k_{2})\;\;\mfrak{r}_{12}(k_{1},k_{2};t)\;\;.
\eeq
As we perform the total trace \(\mbox{Tr}_{12}\) in order to obtain the conserved quantities
\(C^{(n_{1})}(k_{1})\), \(C^{(n_{2})}(k_{2})\) within the Poisson bracket of physical, canonical fields,
the right-hand side of (\ref{sb_10}) yields the trace \(\mbox{Tr}_{12}\) of a commutator with the
'\(\mfrak{r}_{12}\)'-matrix which results into zero and therefore demonstrates the independence of the conserved
quantities \(C^{(n_{1})}(k_{1})\), \(C^{(n_{2})}(k_{2})\)
\beq\lb{sb_11}
\lefteqn{\mbox{Tr}_{12}\Big[\boldsymbol{\big\{}\mfrak{T}_{1}(2\pi\,n_{1},0;t;k_{1})\;\stackrel{\otimes}{\boldsymbol{,}}\;
\mfrak{T}_{2}(2\pi\,n_{2},0;t;k_{2})\boldsymbol{\big\}}\Big] =
\boldsymbol{\big\{}C^{(n_{1})}(k_{1})\;\boldsymbol{,}\;C^{(n_{2})}(k_{2})\boldsymbol{\big\}}= }   \\ \no &=&
\mbox{Tr}_{12}\Big[\overbrace{\mfrak{T}_{1}(2\pi\,n_{2},0;t;k_{1})\;\;\mfrak{T}_{1}(2\pi\,n_{1},2\pi\,n_{2};t;k_{1})}^{
\mfrak{T}_{1}(2\pi\,n_{1},0;t;k_{1})}\;\;\mfrak{T}_{2}(2\pi\,n_{2},0;t;k_{2})\;\;\mfrak{r}_{12}(k_{1},k_{2};t) +  \\ \no &-&
\mfrak{T}_{1}(2\pi\,n_{1},0;t;k_{1})\;\;\mfrak{T}_{2}(2\pi\,n_{2},0;t;k_{2})\;\;\mfrak{r}_{12}(k_{1},k_{2};t)\Big] = \\ \no &=&
\mbox{Tr}_{12}\bigg[\boldsymbol{\Big[}\mfrak{T}_{1}(2\pi\,n_{1},0;t;k_{1})\;\;\mfrak{T}_{2}(2\pi\,n_{2},0;t;k_{2})\;\;
\boldsymbol{,}\;\;\mfrak{r}_{12}(k_{1},k_{2};t)\boldsymbol{\Big]_{-}}\bigg] \equiv 0 \;\;.
\eeq
The given proof of this appendix \ref{sb} relies on the spatial independence of the '\(\mfrak{r}_{12}\)'-matrix,
\(\mfrak{r}_{12}(k_{1},k_{2};t)\) with the ultralocal property \(\delta(x_{1}-x_{2})\) in the assumed fundamental
Poisson bracket relation (\ref{sb_1},\ref{sb_6}). The spatial independence of '\(\mfrak{r}_{12}\)'-matrix
has to be incorporated in order to transform the spatial integrations in (\ref{sb_7},\ref{sb_8}) to total
spatial derivatives so that the integrands simplify to remaining integration boundaries in (\ref{sb_9},\ref{sb_10}).

\end{appendix}


\end{document}